\documentclass[aps,
prd,
10pt,
twocolumn,
nofootinbib,
amsmath,
amssymb,
longbibliography,
superscriptaddress]{revtex4-2}

\usepackage[utf8]{inputenc}
\usepackage{lmodern}
\usepackage[T1]{fontenc}
\usepackage[dvipsnames,table]{xcolor}
\usepackage[pdftex,breaklinks,colorlinks,
linkcolor=Blue,
citecolor=teal,
anchorcolor=red,
urlcolor=cyan,
pdfencoding=auto]{hyperref}
\usepackage{upgreek}
\usepackage{graphicx,amsfonts}
\usepackage{multirow}
\usepackage{tabularx}
\usepackage{booktabs}
\usepackage{array}
\usepackage{color,colortbl}

\usepackage{epsfig}
\usepackage{url}
\usepackage{bm}
\usepackage{mathrsfs}
\usepackage{enumerate}
\usepackage{amsthm}
\usepackage{bbm}
\usepackage[normalem]{ulem}
\usepackage{tensor}
\usepackage{siunitx}
\usepackage{orcidlink}
\usepackage[caption=false]{subfig}
\usepackage{float}

\font\tenscr=rsfs10 scaled1100
\font\sevenscr=rsfs7 
\font\fivescr=rsfs5 
\skewchar\tenscr='177
\skewchar\sevenscr='177
\skewchar\fivescr='177
\newfam\scrfam
\textfont\scrfam=\tenscr
\scriptfont\scrfam=\sevenscr
\scriptscriptfont\scrfam=\fivescr

\def\scri{{\fam\scrfam I}}

\makeatletter
\newcommand{\subalign}[1]{%
  \vcenter{%
    \Let@ \restore@math@cr \default@tag
    \baselineskip\fontdimen10 \scriptfont\tw@
    \advance\baselineskip\fontdimen12 \scriptfont\tw@
    \lineskip\thr@@\fontdimen8 \scriptfont\thr@@
    \lineskiplimit\lineskip
    \ialign{\hfil$\m@th\scriptstyle##$&$\m@th\scriptstyle{}##$\hfil\crcr
      #1\crcr
    }%
  }%
}
\makeatother

\newcommand{\beq}{\begin{equation}}
\newcommand{\eeq}{\end{equation}}
\newcommand{\bea}{\begin{eqnarray}}
\newcommand{\eea}{\end{eqnarray}}
\newcommand{\bit}{\begin{itemize}}
\newcommand{\eit}{\end{itemize}}
\newcommand{\ben}{\begin{enumerate}}
\newcommand{\een}{\end{enumerate}}
\newcommand{\nn}{\nonumber}

\newcommand{\D}{{\cal D}}
\newcommand{\id}{i_{\rm d}}
\newcommand{\iF}{i_{\rm f}}
\newcommand{\polarR}{{\varrho}}
\newcommand{\pfR}{R}
\newcommand{\pftheta}{\bar{\theta}}
\newcommand{\pfphi}{\bar{\varphi}}
\newcommand{\pfC}{{\bar{C}}}
\newcommand{\pfl}{{\bar{l}}}
\newcommand{\pfm}{{\bar{m}}}
\newcommand{\pfn}{{n}}

\begin{document}

\title{Multi-domain spectral method for self-force calculations}

\author{Rodrigo Panosso Macedo\,\orcidlink{0000-0003-2942-5080}}
\affiliation{Niels Bohr International Academy, Niels Bohr Institute, Blegdamsvej 17, 2100 Copenhagen, Denmark}
\author{Patrick Bourg\,\orcidlink{0000-0003-0015-0861}}
\affiliation{Institute for Mathematics, Astrophysics and Particle Physics, Radboud University, Heyendaalseweg 135, 6525 AJ Nijmegen, The Netherlands}
\affiliation{School of Mathematical Sciences and STAG Research Centre, University of Southampton, Southampton, SO17 1BJ, United Kingdom}
\author{Adam Pound\,\orcidlink{0000-0001-9446-0638}}
\affiliation{School of Mathematical Sciences and STAG Research Centre, University of Southampton, Southampton, SO17 1BJ, United Kingdom}
\author{Samuel D.\ Upton\,\orcidlink{0000-0003-2965-7674}}
\affiliation{School of Mathematical Sciences and STAG Research Centre, University of Southampton, Southampton, SO17 1BJ, United Kingdom}
\affiliation{Astronomical Institute of the Czech Academy of Sciences, Boční II 1401/1a, CZ-141 00 Prague, Czech Republic}
\date{\today}

\begin{abstract}
Second-order self-force calculations will be critical for modelling extreme-mass-ratio inspirals, and they are now known to have high accuracy even for binaries with mass ratios $\sim 1:10$. Many of the challenges facing these calculations are related to slow convergence of spherical-harmonic (or spheroidal harmonic) mode sums in a region containing the small companion. In this paper, we begin to develop a multi-domain framework that can evade those problems. Building on recent work by Osburn and Nishimura, in the problematic region of spacetime we use a puncture scheme and decompose the punctured field equations into a basis of Fourier and azimuthal $m$ modes, avoiding a harmonic decomposition in the $\theta$ direction. Outside the problematic region, we allow for a complete spherical- or spheroidal-harmonic decomposition. As a demonstration, we implement this framework in the simple context of a scalar charge in circular orbit around a Schwarzschild black hole. Our implementation utilizes several recent advances: a spectral method in each region, hyperboloidal compactification, and an extremely high-order puncture.
\end{abstract}

\maketitle
\tableofcontents

\section{Introduction}\label{sec:Introduction}
Gravitational-wave astronomy has historically focused on comparable-mass compact binaries, in which the two compact objects are roughly equal in mass. However, as detectors become more sensitive and new detectors come online that are able to access significantly lower frequencies, we will observe a much broader range of systems. In particular, space-based detectors will observe extreme- and intermediate-mass-ratio inspirals (EMRIs and IMRIs) in which the larger body is hundreds to millions of times more massive than the smaller body~\cite{LISA:2022yao}

To model these systems, the gravitational-wave community primarily uses self-force theory~\cite{Barack:2018yvs,Pound:2021qin}, in which one expands the field equations in powers of a small parameter \(\varepsilon\) that characterizes post-test-particle corrections. In the gravitational case, $\varepsilon$ is taken to be the mass ratio between the two bodies, characterizing the effect of the smaller body on the spacetime of the larger. One then iteratively solves the field equations order by order, using solutions of the lower-order equations as input into the source terms in higher-order equations. Currently, this is done through second order in the mass ratio, a requirement to track the EMRI waveform phase throughout the entirety of the inspiral~\cite{Hinderer:2008dm,Burke:2023lno}. To date, at second order, calculations have been performed of the binding energy~\cite{Pound:2019lzj}, energy flux~\cite{Warburton:2021kwk}, and gravitational waveform~\cite{Wardell:2021fyy} for a spinning~\cite{Mathews:2021rod,Mathews-Capra} smaller object orbiting a Schwarzschild (or slowly spinning~\cite{Mathews-Capra}) black hole.

Traditionally, nearly all self-force calculations at first and second order have utilized a complete $(\ell,m)$ mode decomposition in a basis of spherical or spheroidal harmonics. This separates the field equations, reducing them to ordinary differential equations (ODEs, in the frequency domain) or to $1+1$D partial differential equations (PDEs, in the time domain)~\cite{Pound:2021qin}. In this paper, alongside the companion paper~\cite{Bourg2023}, we instead elect to perform a decomposition into azimuthal $m$ modes $e^{im\varphi}$ \emph{without} expanding in associated Legendre or spheroidal basis functions of the polar angle $\theta$. By also expanding in Fourier harmonics $e^{-i\omega t}$, we then work with two-dimensional PDEs in $(r,\theta)$. Such $m$-mode decompositions have been used previously in self-force calculations in both the time domain~\cite{Barack:2007we,Dolan:2010mt,Dolan:2011dx,Dolan:2012jg,Thornburg:2016msc} and frequency domain~\cite{Osburn:2022bby}, and our approach is specifically a direct successor to that of~\citet{Osburn:2022bby}. However, all of those previous works were motivated by the fact that the Lorenz-gauge linearized Einstein equations are nonseparable in a Kerr background.\footnote{Though see Refs.~\cite{Dolan:2021ijg,Dolan:2023enf} for a recent method of obtaining Lorenz-gauge metric perturbations from the solutions of separable equations.} We instead aim to use $m$ modes to tackle the major bottleneck in second-order self-force calculations: constructing the source term in the second-order Einstein equations.

Current second-order calculations are based on the approach laid out in Refs.~\cite{Miller:2020bft,Spiers:2023mor,Miller:2023ers}, in which the metric perturbations and field equations are decomposed into $(\ell,m)$ modes in the Barack--Lousto--Sago (BLS) basis of tensor spherical harmonics $Y^{i\ell m}_{\mu\nu}$ ($1 \leq i \leq 10$)~\cite{Barack:2005nr,Barack:2007tm}. Modes of the second-order metric perturbation $h^{(2)}_{\mu\nu}$ are sourced by modes of the second-order Einstein tensor, $\delta^{2}G_{\mu\nu}[h^{(1)},h^{(1)}]$. To compute these source modes, one must sum up quadratic combinations of first-order modes and their derivatives, using a mode-coupling formula of the form
\begin{multline}    
    \delta^{2}G_{i\ell m}[h^{(1)},h^{(1)}] \\ = \sum_{\substack{i_{1}\ell_{1}m_{1}\\i_{2}\ell_{2}m_{2}}} \mathcal{D}^{i\ell m}_{i_{1}\ell_{1}m_{1}i_{2}\ell_{2}m_{2}}\bigl[h^{(1)}_{i_{1}\ell_{1}m_{1}},h^{(1)}_{i_{2}\ell_{2}m_{2}}\bigr], \label{eq:d2Gilm}
\end{multline}
where $\mathcal{D}^{ilm}_{i_{1}\ell_{1}m_{1}i_{2}\ell_{2}m_{2}}$ is a bilinear differential operator explicitly given in Ref.~\cite{Spiers:2023mor}. Near the particle, in order for this sum to numerically converge and to find a single mode of the second-order source, one must sum an arbitrarily large number of first-order $\ell$-modes~\cite{Miller:2016hjv}. Obtaining enough modes of the first-order field to attain numerical convergence in this manner is unfeasible. Instead, the method used has been to directly decompose the most singular part of the four-dimensional \(\delta^{2}G_{\mu\nu}[h^{(1)},h^{(1)}]\) into \((\ell,m)\) modes by numerically integrating it against the tensor spherical harmonics over two-spheres of fixed areal radius $r$. The modes of the less singular pieces of \(\delta^{2}G_{\mu\nu}[h^{(1)},h^{(1)}]\) are then obtained using the coupling formula~\eqref{eq:d2Gilm}~\cite{Miller:2016hjv}. More concretely, the first-order field is split into a singular puncture field and more regular residual field, $h^{(1)}_{\mu\nu}=h^{(1)\cal P}_{\mu\nu}+h^{(1)\cal R}_{\mu\nu}$~\cite{Barack:2018yvs,Pound:2021qin}, with the puncture known analytically in four dimensions and the residual field calculated numerically at the level of $\ell m$ modes; the modes of $\delta^{2}G_{\mu\nu}[h^{(1)\cal P},h^{(1)\cal P}]$ are then obtained through numerical integration over two-spheres, and the remaining pieces are obtained through the mode-coupling formula. While this approach successfully provides modes of the second-order source, it is a major computational bottleneck even in the simplest case of quasicircular orbits in Schwarzschild spacetime.\footnote{The use of an alternate gauge with a weaker singularity structure to circumvent this problem has been suggested~\cite{Pound:2017psq}, but the development of this is still in its infancy~\cite{Upton:2021oxf,Upton:2023tcv} and will require significant work to bring it to the same status as the Lorenz gauge.}

By solely decomposing the first-order field into $m$ modes, we can alleviate the slow convergence that appears when summing over the $\ell$ modes. There are at least two reasons for this:
\begin{enumerate}
    \item As demonstrated in the companion paper~\cite{Bourg2023} (and earlier literature~\cite{Dolan:2012jg,Thornburg:2016msc}), it is relatively straightforward to obtain analytical expressions for the $m$ modes of the puncture field to any order in distance from the particle; the companion paper shows this for the toy model of a scalar charge in a quasicircular orbit around a Schwarzschild black hole, but we expect the method to extend to gravity and to Kerr. The $m$ modes of $\delta^{2}G_{\mu\nu}[h^{(1)\cal P},h^{(1)\cal P}]$ should be obtainable analytically in the same manner, avoiding the need for a burdensome two-dimensional numerical integration.
    \item The slow decay of $h^{(1)}_{\mu\nu}$'s $\ell$-modes stems from the fact that they are defined through integration over spheres centered on the central black hole. For a large region of spacetime, these spheres pass through the region near the particle, such that each $\ell$ mode receives a large contribution from the singularity. This suggests that spherical harmonics on these spheres are ill-adapted to representing the field near the particle. A tailored basis of functions around the particle (or an adaptive finite-difference grid) should be able to more efficiently represent the local field, leading to more rapid convergence.  
\end{enumerate}

Of course, these potential advantages come at the cost of solving two-dimensional PDEs rather than decoupled ODEs. Our aim in this paper is to develop a scheme to rapidly solve $m$-mode field equations using a multi-domain spectral solver, based on methods from~\cite{Meinel:2008kpy} and code developed for Ref.~\cite{PanossoMacedo:2022fdi}. We focus on the same scalar-charge toy problem as the companion paper~\cite{Bourg2023}, taking advantage of the puncture derived there. Like Osburn and Nishimura~\cite{Osburn:2022bby}, we utilize a puncture scheme, in which the puncture field is moved to the right-hand side of the field equations as a so-called ``effective source'', and the field equations are then solved for the residual field; see the companion paper~\cite{Bourg2023} and references therein for a more comprehensive description of the puncture scheme and its use in self-force calculations. 

Our approach builds on Osburn and Nishimura's but differs in several ways. First, we use a spectral method rather than a finite-difference scheme. Spectral methods have proved to be a successful approach to self-force computations, but applications were mainly restricted to time-domain calculations~\cite{Field:2009kk,Canizares:2009ay,Canizares:2010yx,Canizares:2011kw,Diener2019,DaSilva:2023xif,OBoyle:2023jqo,Wittek:2023nyi}. This work extends the frequency-domain strategy initiated in Ref.~\cite{PanossoMacedo:2022fdi}. Second, we only solve the $m$-mode equations within a spherical shell around the central black hole, using a basis of functions adapted to the behavior of the puncture field near the particle. Outside that shell, we obtain the field from $\ell m$ modes rather than $m$ modes, exploiting the fact that the $\ell$-mode sum converges sufficiently rapidly at points far from the particle's orbital radius; this should ultimately enable a modular approach in which the choice of basis functions can be optimized in each region. In the regions outside the shell, we use Ref.~\cite{PanossoMacedo:2022fdi}'s one-dimensional $\ell m$-mode spectral solver on compactified hyperboloidal slices. Finally, we accelerate convergence using the very-high-order puncture derived in the companion paper~\cite{Bourg2023}. While our demonstration is for the toy problem of a scalar charge circling around a non-spinning black hole, it provides a proof of concept of these methods, which should (by design) extend straightforwardly to the gravitational case and to orbits around a Kerr black hole.

We begin in Sec.~\ref{sec:GeomSetup} by outlining the geometrical setup of our scheme: firstly, by recapping the standard formalism used for the scalar self-force and, secondly, describing the transformation to the local, comoving coordinates detailed in the companion paper~\cite{Bourg2023}. We then, in Sec.~\ref{sec:AsympFrame}, detail the compactified hyperboloidal coordinates that will be used to capture the asymptotic behavior far from the particle. Sections~\ref{sec:coord_domain_decomposition} and~\ref{sec:BCs by domain} then describe how the various individual coordinate systems are combined together to form the multi-domain grid, along with the appropriate boundary conditions required. The details of the numerical scheme and its implementation are described in Sec.~\ref{sec:NumericalMethods}, with Sec.~\ref{sec:External Data} illustrating how the puncture field of Ref.~\cite{Bourg2023} and the $\ell m$ modes of Ref.~\cite{PanossoMacedo:2022fdi} are incorporated into the numerical solver. Section~\ref{sec:NumericalSolutions} examines the convergence properties of our numerical solution and of a scalar toy second-order source. The key outcome is that one can rapidly construct the $m$ modes of the second-order source using the outputs of our scheme, thereby avoiding the costly source-construction in the $(\ell,m)$-mode approach.

\subsection{Notation, conventions and definitions}\label{sec:Notation}

The imaginary unit is represented by ${\rm i}$, whereas the symbol $i$ is mostly employed to describe discrete grid points $\{x_i\}_{i=0}^N$ in Sec.~\ref{sec:NumericalMethods}. The particle's proper time is represented by $\uptau$, whereas $\tau$ is the hyperboloidal time coordinate. We use geometrized units with $G=c=1$.

\section{Geometrical setup}\label{sec:GeomSetup}

\subsection{Schwarzschild coordinates}

We first review the scalar self-force formalism in the standard Schwarzschild coordinates $x^a = \{t,r,\theta,\varphi\}$. 

\subsubsection{Spacetime}

The line element reads
\beq
\label{eq:SchwarzschildMetric}
ds^2 = -f(r) dt^2 +\dfrac{1}{f(r)} dr^2 + r^2 \left(d\theta^2 + \sin^2\theta d\varphi \right), 
\eeq
with $f(r) = 1-\dfrac{r_h}{r}$, where $r_h=2M$ is the Schwarzschild radius of the black hole. It will be convenient to define the tortoise coordinate $r_*$ via
\beq
\dfrac{d r_*}{dr} = \dfrac{1}{f(r)} \Rightarrow r_* = r + r_h \ln \left( \dfrac{r}{r_h} - 1\right).
\eeq

\subsubsection{Particle's trajectory}

The particle with mass $\mu$ and scalar charge $q$ follows a geodesic with worldline  
\beq
x^a(\uptau) = \{t_p(\uptau), r_p(\uptau), \theta_p(\uptau), \varphi_p(\uptau)\}. 
\eeq
We focus on a circular orbit of radius $r_p$ around the black hole. Given the background symmetry, we place the trajectory on the equatorial plane without loss of generality. Thus, the particle follows the trajectory 
\beq
x^a_p(t) = \{t, r_p, \pi/2, \varphi_p(t)\}
\eeq
(reparametrized in terms of coordinate time) with
\beq
\varphi_p(t) = \Omega_p \, t, \quad \Omega_p = \sqrt{\dfrac{M}{r_p^3}}.
\eeq
From the timelike and axial Killing vectors $\partial_t$ and $\partial_\varphi$, respectively, we derive the constants of motion
\beq
E_p = \dfrac{f_p}{\sqrt{1-3M/r_p}}, \quad L_p = \dfrac{\sqrt{r_p M}}{\sqrt{1-3M/r_p}}
\eeq
with $f_p = f(r_p)$. They represent the particle's energy and angular momentum per unit mass $\mu$, respectively.

\subsubsection{Scalar field}

The particle generates a scalar field $\Phi(t,r,\theta,\varphi)$ satisfying the wave equation
\bea
\label{eq:4D_ScalarWaveEq}
\square_g \Phi = \dfrac{1}{\sqrt{-\boldsymbol{g}}} \partial_a \left(\sqrt{-\boldsymbol{g}}\, g^{ab} \partial_b \right) \Phi = S_\Phi, \\ \nn
\eea
with $\boldsymbol{g} = -r^4 \sin^2\theta$ the metric's determinant, and 
\bea
\label{eq:4D_source}
S_\Phi &=& - 4\pi q \int d\uptau \, \dfrac{\delta^4(x^a - x_p^a(\uptau))}{\sqrt{-\boldsymbol{g}}} \nonumber\\
&=& -\frac{4 \pi q f_p}{E_p r_p^2} \delta(r-r_p) \delta(\theta-\pi/2) \delta(\varphi-\varphi_p(t))
\eea
the source term supported on the particle's worldline. Here, $\delta^4(x^a)$ is the four-dimensional Dirac delta function. 

Off the worldline, the problem reduces to solving the homogeneous equation $\square_g \Phi = 0$. We are interested in the retarded solution $\Phi^{\rm ret}(t,r,\theta,\varphi)$, i.e., the solution satisfying the boundary conditions
\beq
\label{eq:4D_BC}
\Phi^{\rm ret} \sim \dfrac{e^{-{\rm i}\omega \left(t\mp r_* \right)}}{r}, \quad r_* \rightarrow \pm \infty
\eeq
for each frequency $\omega$ in the solution's spectrum. These conditions ensure that energy propagates in the form of gravitational waves out to the wave zone at $r \rightarrow \infty$ ($r_* \rightarrow \infty$), as well as into the black hole at $r=r_h$ ($r_* \rightarrow -\infty$).

\subsubsection{Frequency domain, \texorpdfstring{$m$}{m}-mode decomposition}
Because the source term only depends on $t$ and $\varphi$ in the combination $(\varphi-\Omega_p t)$, it can be decomposed into $m$ modes in the form
\bea
S_\Phi = \sum_{m=-\infty}^\infty S_{\phi_m}(r,\theta) e^{{\rm i}m(\varphi-\Omega_p t)}
\eea
where 
\bea
\label{eq:Source_m}
S_{\phi_m} &=& \frac{1}{2 \pi} \int_{0}^{2\pi} S_{\Phi}\, e^{-{\rm i}m(\varphi-\Omega_p t)} d\varphi \nn \\
\nn \\
&=& - \frac{2 q f_{p}}{E_{p} r_{p}^2}\delta(r-r_p)\delta(\theta - \pi/2).
\eea
With the ansatz
\beq
\label{eq:Freq_m_decomposition_2}
\Phi(t,r,\theta,\varphi) = \sum_{m=-\infty}^{\infty} \phi_m(r,\theta) e^{{\rm i} m \left(\varphi - \Omega_p t\right)},
\eeq
the wave equation~\eqref{eq:4D_ScalarWaveEq} reduces to
\beq
\label{eq:EllipticEq_SchwarzschildCoord}
\Delta_{m} \phi_m = S_{\phi_m},
\eeq
with the elliptic operator
\bea
\Delta_{m}  &=& f(r)\partial^2_{rr} + \dfrac{1+f(r)}{r} \partial_r  \nn \\
 &+& \dfrac{1}{r^2}\left(  \partial^2_{\theta \theta} + \cot\theta \partial_\theta\right) + \dfrac{\omega^2}{f(r)} - \dfrac{m^2}{r^2\,\sin^2\theta}.
 \label{eq:EllipticOperator}
\eea
Here we have introduced $\omega:=m\Omega_p$.

The fact that $S_\Phi$ is real implies that $\Phi$ is as well. This in turn implies that its $m$ modes are related by the symmetry $\phi_{-m}(r,\theta) = \phi^*_m(r,\theta)$, with ${}^*$ representing the complex conjugate. Thus, we can concentrate on finding solutions for $m\geq 0$.

\subsubsection{Boundary conditions}\label{sec:BC}
Equation \eqref{eq:EllipticEq_SchwarzschildCoord} is defined in the domain $(r,\theta)\in[r_h,\infty)\times[0,\pi]$, and a unique solution follows from specifying appropriate boundary/regularity conditions at the domain boundaries. 

As before, away from the particle's position $(r_p,\theta_p)$, Eq.~\eqref{eq:EllipticEq_SchwarzschildCoord} reduces to the homogeneous elliptic equation $\Delta_m \phi_m=0$.
The retarded solution $\phi_m^{\rm ret}(r,\theta)$ follows from imposing boundary conditions equivalent to Eq.~\eqref{eq:4D_BC}, i.e.,
\beq\label{eq:2D_BC}
\phi^{\rm ret}_m \sim \dfrac{e^{\pm {\rm i}\omega  r_*}}{r}, \quad r_* \rightarrow \pm \infty.
\eeq
For the angular coordinate, we demand regularity for the solution at $\theta=0$ and $\theta=\pi$, which yields\footnote{This regularity condition follows from imposing smoothness in Cartesian coordinates $({\sf x},{\sf y},{\sf z})$ at the ${\sf z}$ axis:  $\phi_m(\theta)e^{{\rm i}m\phi}=\frac{\phi_m(\theta)}{r^m\sin^m\theta}({\sf x}+{\rm i}\,{\sf y})^m$ can only be smooth if $\phi_m(\theta)\sim \sin^m\theta$ (times an even function of $\sin\theta$). Alternatively, one can use an Ansatz $\phi_m^{\rm ret}(x) = (1-x^2)^{k/2} \hat \phi_m^{\rm ret}(x)$, with $x=\cos\theta$. The choice $k^2 = m^2$ eliminates the singular piece $\sim m^2/(1-x^2)$ in the field equation, yielding a regular equation for the auxiliary field $\hat \phi_m^{\rm ret}(x)$. For instance, this condition corresponds to the behaviour of the Associated Legendre Polynomials $P_\ell^m(x)$.  
}
\beq
\label{eq:Axis_RegCond}
\phi^{\rm ret}_m \sim \sin^{m}\theta, \quad \sin\theta \rightarrow 0
\eeq
(or $\sim\sin^{|m|}\theta$ if we consider  negative $m$). The system also possesses the equatorial symmetry $\phi_m(r,\theta) = \phi_m(r,\pi-\theta)$, which can be used to reduce the computational domain to $(r,\theta)\in[r_h,\infty)\times[0,\pi/2]$. In this case, apart from the boundary condition \eqref{eq:Axis_RegCond} at $\theta = 0$, we must also ensure the equatorial symmetry conditions
\beq
\label{eq:Eq_Sym_Condition}
\left. \partial^{k}_\theta \phi^{\rm ret}_m \right|_{\theta = \pi/2} = 0, \quad k = 1, 3, \ldots 
\eeq
The equatorial symmetry condition is equivalent to stating that $\phi^{\rm ret}_m$ is an even function of $x=\cos\theta$. Thus, a Taylor expansion in the angular direction only contains terms proportional to $y=\cos^2\theta$. This property will justify the choice of coordinates in Sec.~\ref{sec:hyp_coord}.

However, the unique solution $\phi^{\rm ret}_m(r,\theta)$ resulting from fixing the boundary conditions is not regular at the particle's location $r=r_p$ and $\theta=\theta_p$. In the next section, we describe the regularization scheme to obtain a bounded solution at the particle's location. 

\subsubsection{Effective-source approach}
To regularise the solution at the particle's world line, we follow the effective source approach~\cite{Barack:2007we,Vega:2007mc,Dolan:2010mt,Wardell:2015kea}. We decompose the retarded field into the Detweiler-Whiting regular $\Phi^{\rm R}$ and singular $\Phi^{\rm S}$ components~\cite{Detweiler:2002mi}
\beq
\Phi^{\rm ret} = \Phi^{\rm R} + \Phi^{\rm S}.
\eeq
Then, we introduce a puncture field $\Phi^{\mathcal{P}} \approx \Phi^{\rm S}$ that captures the singularity structure of $\Phi^{\rm S}$. Concretely, given a local expansion in powers of distance $\hat R$ from the particle, we define an $n$th-order puncture as any field satisfying 
\beq
\Phi^{\mathcal{P}}=\Phi^{\rm S}+O(\hat R^{n+1}), 
\eeq
along with a corresponding residual field 
\beq
\Phi^{\mathcal{R}}:= \Phi^{\rm ret} -\Phi^{\mathcal{P}}, 
\eeq
which satisfies $\Phi^{\mathcal{R}}=\Phi^{\rm R}+O(\hat R^{n+1})$. 

At the level of $m$ modes, we define $\phi^{\mathcal{P}}_m$ as the mode coefficients of $\Phi^{\mathcal{P}}$ in a decomposition of the form~\eqref{eq:Freq_m_decomposition_2}, and the residual field mode coefficients
\beq
\label{eq:Res_Ret_Punc}
\phi^{\mathcal{R}}_m = \phi^{\rm ret}_m - \phi^{\mathcal{P}}_m.
\eeq
Each $\phi^{\mathcal{R}}_m$ satisfies the equation
\beq
\label{eq:Eq_effectiveSource}
\Delta_m \phi^{\mathcal{R}}_m = S^{\rm eff}, 
\eeq
with the effective source given by
\beq
S^{\rm eff} = S_{\phi_m} - \Delta_m \phi^{\mathcal{P}}_m.
\eeq
We employ this approach in a subdomain $\D_{\mathcal{R}}\subset [r_h,\infty)\times[0,\pi/2]$, which will be fixed in Sec.~\ref{sec:coord_domain_decomposition}. Outside the domain $\D_{\mathcal{R}}$, the retarded field satisfies the homogeneous equation $\Delta_m \phi^{\rm ret}_m = 0$.

The next section summarises the method we use to obtain the puncture field and effective source. 

\subsection{Particle's frame}\label{sec:ParticlesFrame}
To efficiently calculate the puncture field $\phi_m^{\mathcal{P}}$, we follow the strategy put forward in Ref.~\cite{Bourg2023}. The basic idea consists of introducing a coordinate system comoving with the particle and exploiting a spherical harmonic decomposition on a sphere defined around the particle's location (a method outlined previously in Ref.~\cite{Pound:2012dk}). 

\subsubsection{Coordinate transformation}

We introduce a locally Cartesian coordinate system $X^a=\{T,X,Y,Z\}$ around the particle's worldline, defined by the following transformation from the original coordinates $x^a$: 
\bea
&T=E_p\,  t  - L_p \, \varphi , \quad X = \dfrac{\delta r}{\sqrt{f_p}}, \quad Y = - r_p \cos\theta, \nn  \\
&  Z = -  z_c \sin\left(\dfrac{\varphi - \Omega_p \,t}{2} \right),
\label{eq:CoordParticleFrame}
\eea
where 
\beq
\delta r = r-r_p\quad\text{and}\quad z_c = 2\dfrac{E_p \, r_p}{\sqrt{f_p}}. 
\eeq
This choice leads to a line element which is locally flat, i.e.,
\beq
\left. ds^2 \right|_{X^a=X^a_p}  = -dT^2 + dX^2 + dY^2 + dZ^2.
\eeq
The source term~\eqref{eq:4D_source} in these coordinates reduces to the flat-space form
\beq
S_\Phi=-4\pi q\,\delta(X) \delta(Y)\delta(Z).
\eeq

We abuse the notation and identify the scalar $\Phi(T,X,Y,Z)$ directly with $\Phi(x^a(X^b))$. The decomposition \eqref{eq:Freq_m_decomposition_2} then yields
\beq
\label{eq:phi_partilcle_frame}
\Phi(T,X,Y,Z) = \sum_{m=-\infty}^{\infty} \phi_m(X,Y) e^{ -2 {\rm i} m \sin^{-1}(Z/z_c) }.
\eeq
We observe that the scalar field $\Phi$ is static (independent of $T$) in the particle's frame. 

It will also be useful to define cylindrical coordinates $\{\polarR,\pfphi,Z\}$, where $\{\polarR,\pfphi\}$ are polar coordinates in the $XY$ plane, defined in the natural way as 
\beq
\label{eq:ParticlePolarCoordinates}
X = {\polarR} \cos \pfphi, \quad Y = \polarR \sin\pfphi.
\eeq
In terms of the original coordinates $\{r,\theta\}$, they read
\bea
\label{eq:ParticlePolarCoordinates2}
{\polarR^2} = \dfrac{\delta r^2}{f_p} + r_p^2 \cos\theta^2, \quad
\tan \pfphi = -\dfrac{ r_p \sqrt{f_p}\cos\theta}{\delta r}.
\eea

\subsubsection{Wave operator}
Since the metric reduces to the Minkowski metric at the origin, we can express the scalar wave equation \eqref{eq:4D_ScalarWaveEq} as a Poisson-like equation,
\beq
\label{eq:4D_PoissonEq}
\Delta_{\rm flat} \Phi = -4\pi q\,\delta(X)\delta(Y)\delta(Z)+\pfC(\Phi).
\eeq
where
\beq
\Delta_{\rm flat} = \partial^2_{X}  + \partial^2_{Y} + \partial^2_{Z}
\eeq
is the flat-space Laplacian, and
\beq
\pfC = g^{\mu \nu} \Gamma_{\mu \nu}^\rho \partial_\rho -(g^{\mu \nu}-\eta^{\mu \nu})\partial_\mu \partial_\nu.
\eeq
Reference~\cite{Bourg2023} details the techniques to solve the above equation. Here, we review the most important features impacting our $m$-mode scheme.
\subsubsection{Puncture field and effective source}
One can solve \eqref{eq:4D_PoissonEq} for $\Phi$ by decomposing the field into spherical harmonics centered on the particle and expanding in powers of radial distance $\hat R$ from the particle:
\beq
\label{eq:PhiDecomp}
\Phi = \sum_{\pfn=-1}^\infty \sum_{\pfl=0}^{\pfl_{max}} \sum_{\pfm=-\pfl}^\pfl \Phi_{\pfl\pfm\pfn} \pfR^\pfn Y_{\pfl\pfm} (\pftheta,\pfphi),
\eeq
where $\Phi_{\pfl\pfm\pfn}$ are constants, and $(\pfR,\pftheta,\pfphi)$ are the flat-space spherical coordinates\footnote{Alternatively, defined from the polar coordinates $(\polarR,\pfphi,Z)$ in Eq.~\eqref{eq:ParticlePolarCoordinates2} via $\polarR = R \sin \pftheta$. } defined from $(X,Y,Z)$,
\begin{align}
X &=: \pfR \sin \pftheta \cos \pfphi,\\
Y &=: \pfR \sin \pftheta \sin \pfphi,\\
Z &=: \pfR \cos \pftheta.
\end{align}
Here we have followed the notation from Ref.~\cite{Bourg2023}, with the barred quantities referring to a representation based on the coordinate system centred at the particle's frame.  

With the substitution of the ansatz~\eqref{eq:PhiDecomp} into Eq.~\eqref{eq:4D_PoissonEq} and by expanding $\hat{C}(\Phi)$ in a similar fashion, one obtains an algebraic system of equations for the coefficients $\Phi_{\pfl\pfm\pfn}$.

At each given order $\pfn$, the constants $\Phi_{\pfl\pfm\pfn}$ are uniquely specified by the field equation except the specific constant with $\pfl=\pfn$ ($\geq0$), which is undetermined because $\Delta_{\rm flat}(\Phi_{\pfn\pfm\pfn}\pfR^\pfn Y_{\pfn\pfm})=0$ for any value of $\Phi_{\pfn\pfm\pfn}$. These coefficients $\Phi_{\pfn\pfm\pfn}$ represent the freedom in the general solution; they can only be determined by boundary conditions. The singular field $\Phi^{\rm S}$ is the particular solution with all of these special coefficients set to zero: $\Phi_{\pfn\pfm\pfn}=0$. We then define a puncture of order $\pfn_{\rm max}$ as the expansion \eqref{eq:PhiDecomp} truncated at $\pfn = \pfn_{\rm max}$ and with all free coefficients $\Phi_{\pfn\pfm\pfn}$ set to zero. We denote this as $\Phi^{\mathcal{P}}_{\pfn_{\rm max}}$, though we will generally omit the label $\pfn_{\rm max}$. From these analytical expressions, we obtain the $m$-mode decomposition by analytically evaluating the integrals over $\varphi$,
\bea
\label{eqn:PhiPm}
\phi^{\mathcal{P}}_{\pfn_{\rm max},m} &=& \frac{1}{2 \pi} \int_{0}^{2\pi} \Phi^{\mathcal{P}}_{\pfn_{\rm max}}\, e^{-{\rm i}m(\varphi-\Omega_p t)} d\varphi,
\eea
as described in Ref.~\cite{Bourg2023}.

\subsection{Compactified hyperboloidal coordinates}\label{sec:AsympFrame}
We now turn our attention to the coordinate system best adapted to account for the asymptotic behavior as $r_*\rightarrow \pm \infty$. We introduce hyperboloidal coordinates $\bar x^a = \{\tau, \sigma, y, \varphi\}$ such that: (i) the hypersurfaces $\tau=$ constant penetrate the black-hole horizon and extend towards future null infinity~\cite{Zenginoglu:2007jw,Zenginoglu:2011jz}; (ii) the radial direction is compact, i.e. $\sigma\in [0,1]$ with $\sigma=0$ corresponding to $\scri^+$ and $\sigma=1$ the black-hole horizon; and (iii) the angular coordinate $y$ incorporates the equatorial symmetry as discussed in Sec.~\ref{sec:BC}. The hyperboloidal time coordinate $\tau$ is not to be confused with the particle's proper time, which we have denoted $\uptau$.

\subsubsection{Hyperboloidal transformation}\label{sec:hyp_coord}
The transformation to our new coordinates reads~\cite{Zenginoglu:2007jw}
\beq
\label{eq:HypCoord}
t = \lambda\bigg( \tau - H(\sigma)\bigg), \quad r = \dfrac{r_h}{\sigma}, \quad y = \cos^2(\theta),
\eeq
with $\lambda$ a given length scale. The hyperboloidal degrees of freedom are fixed by choosing the minimal gauge~\cite{Ansorg:2016ztf,PanossoMacedo:2023qzp}, for which the height function $H(\sigma)$ reads
\begin{equation}
    \label{eq:height_function}
    H(\sigma) = \frac{r_h}{\lambda}\bigg[ \ln(1-\sigma) - \frac{1}{\sigma} + \ln(\sigma) \bigg].
\end{equation}
In such coordinates, the line element \eqref{eq:SchwarzschildMetric} conformally re-scales into 
\bea
\label{eq:conformal_rescale}
ds^2 &=& \Omega^{-2} d\bar{s}^2, \quad \Omega(\sigma)=\sigma/\lambda,
\eea
 with the conformal line element
\bea
\label{eq:conformal_metric}
d\bar{s}^2 &=& - \sigma^2 (1-\sigma) d\tau^2 \\
&&+ \dfrac{2 r_h}{\lambda}(1-2\sigma^2) d\tau d\sigma +\left( \dfrac{2r_h}{\lambda}\right)^2 (1+\sigma) d\sigma^2 \nn \\
&&+\left( \dfrac{r_h}{\lambda} \right)^2 \bigg( \dfrac{dy^2}{4y(1-y)} +(1-y)d\varphi^2\bigg). \nn
\eea
Fixing  $\lambda= 2 r_h$ simplifies the $(\tau,\sigma)$ sector of the metric.

In the hyperboloidal coordinates, obtaining the fields $\phi_m(r,\theta)$ reduces to solving a boundary value problem in the compact (square) region $(\sigma, y)\in[0,1]^2$, with $\sigma=0$ corresponding to future null infinity and $\sigma=1$ the black-hole horizon. In the angular direction, $y=0$ is the equator and $y=1$ the axis of symmetry. 

In the next two sections, we introduce the elliptic equations defined in this compact domain and the corresponding boundary conditions.

\subsubsection{Elliptic equation and effective source}\label{sec:hyp_elliptic_eq}
To account for the boundary behaviors in the frequency domain via the hyperboloidal approach, we re-scale the scalar field $\phi_m(r,\theta)$ into $\bar \phi_m(\sigma,y)$ via
\beq
\label{eq:phi_hyperboloidal_rescaling}
\phi_m  = {\cal Z} \, \bar \phi_m, \quad {\cal Z}(\sigma, y) = \Omega(\sigma)\, e^{s H(\sigma) } (1-y)^{-|m|/2},
\eeq
with 
\beq
s= -{\rm i} \lambda \omega
\eeq
the dimensionless frequency parameter. 

The function ${\cal Z}$ is chosen such that if the rescaled field $\bar \phi_m$ is smooth at the boundaries $\sigma=0$ (future null infinity), $\sigma =1$ (future horizon) and $y=1$ (symmetry axis), then the physical field $\phi_m$ satisfies the desired boundary conditions \eqref{eq:2D_BC} in the original coordinates~\cite{Zenginoglu:2011jz, PanossoMacedo:2023qzp}. 

The factor $\Omega(\sigma) e^{s H(\sigma)}$ in ${\cal Z}$ is standard in hyperboloidal calculations~\cite{Zenginoglu:2011jz,PanossoMacedo:2023qzp}. It incorporates the ingoing behavior at the horizon and outgoing behavior at future null infinity from Eq.~\eqref{eq:2D_BC} directly from the height function. The factor $\Omega(\sigma)\propto 1/ r$ results from the peeling properties, and it ensures that $\bar \phi_m$ at $\sigma=0$ represents the wave amplitude at future null infinity.

The factor $(1-y)^{-|m|/2}$ incorporates the regularity condition at the axis $\theta=0$ ($y=1$). According to Eq.~\eqref{eq:Axis_RegCond}, this choice introduce a behavior $\bar \phi_m \sim (1-y)^{|m|} $, enforcing $\left. \bar \phi_m\right|_{y=1} = 0$. An alternative choice derived directly from Eq.~\eqref{eq:Axis_RegCond} is to set ${\cal Z} \sim (1-y)^{|m|/2}$, which would yield a regular, nonvanishing $\left. \bar \phi_m\right|_{y=1}$. This latter option would be analogous to our choice to make the rescaled field go to a nonvanishing finite function at future null infinity; however, we find it yields highly oscillating functions with steep gradient as $y\rightarrow 1$. We therefore opt for the rescaling shown in Eq.~\eqref{eq:phi_hyperboloidal_rescaling} as it enhances the numerical accuracy for the boundary conditions. We discuss this further in Appendix~\ref{app:LegChebPoly}. 

Given our rescalings, the elliptic operator \eqref{eq:EllipticOperator} transforms as
\beq
\label{eq:operator_rescaling}
\Delta_m \phi_m = {\cal F} {\boldsymbol A} \bar \phi_m, \quad {\cal F}(\sigma, y) = \dfrac{{\cal Z}(\sigma,y)}{r(\sigma)^2}.
\eeq
We express the operator ${\boldsymbol A}$ in its generic form
\bea
\label{eq:operator_A}
{\boldsymbol A} = \alpha_2 \partial^2_{\sigma\sigma} + \alpha_1 \partial_{\sigma} + \alpha_0 + \Gamma_2\partial^2_{yy} + \Gamma_1\partial_{y},
\eea
with the functions (for $m\geq0$)
\bea
\label{eq:alpha_2}
\alpha_2(\sigma) &=& \sigma^2(1-\sigma), \\
\alpha_1(\sigma) &=& \sigma(2-3\sigma) + (1-2\sigma^2)s, \\
\alpha_0(\sigma) &=& -(1+\sigma)s^2 - 2\sigma s - \left[m\left(m-1\right) + \sigma \right], \\
\label{eq:Gamma_2}
\Gamma_2(y) &=&  4 y (1-y), \\
\label{eq:Gamma_1}
\Gamma_1(y) &=&  2 - 2y(3-2m).
\eea

The rescaling \eqref{eq:phi_hyperboloidal_rescaling} and transformation \eqref{eq:operator_rescaling} applies for all fields considered, i.e., $\phi_m^{\rm ret}$, $\phi_m^{\mathcal{R}}$ and $\phi_m^{\mathcal{P}}$. Thus, the hyperboloidal equivalent of Eq.~\eqref{eq:Eq_effectiveSource} reads
\beq
\label{eq:EqRes}
{\boldsymbol A} \bar \phi_m^{\mathcal{R}} = \bar S^{\rm eff}, \quad \bar S^{\rm eff}= \dfrac{S^{\rm eff}}{\cal F}.
\eeq
We recall that $S^{\rm eff}$ is defined in a subdomain $\D_{\mathcal{R}}$ of the hyperboloidal compact region that includes the particle location, meaning $\D_{\mathcal{R}}\subset [0,1]^2$, with $(\sigma_p, 0) \in \D_{\mathcal{R}}$. For $(\sigma, y)\notin \D_{\mathcal{R}}$, the retarded field $\bar \phi_m^{\rm ret}$ satisfies the homogenous equation 
\beq
\label{eq:EqHomo} 
{\boldsymbol A} \bar \phi_m^{\rm ret} = 0.
\eeq 

\subsubsection{Regularity conditions}\label{sec:reg_conditions}

The operator ${\boldsymbol A}$'s principal part is described by the functions $\alpha_2(\sigma)$ and $\Gamma_2(y)$, given in Eqs.~\eqref{eq:alpha_2} and~\eqref{eq:Gamma_2}, respectively. These functions vanish at the domain boundaries, such that for any regular $\bar{\phi}_m$, the field equation~\eqref{eq:EqHomo} on the boundaries reduces to a relationship between the field and its first normal derivative. Therefore, we can impose the desired regularity condition by imposing the field equation directly on the boundary.

\subsection{Multi-domain square mapping}\label{sec:coord_domain_decomposition}
So far, we have introduced coordinate systems adapted to capture the key features of the solution near the particle (Sec.~\ref{sec:ParticlesFrame}) and at the boundaries (Sec.~\ref{sec:AsympFrame}). This section brings these mappings together so that we can exploit the properties of each region individually, while seeking a global solution. 

We work with the hyperboloidal parametrisation of the exterior black-hole region, and we divide the plane $(\sigma,y)\in[0,1]^2$ into four subdomains $\D_{\id}$, $\id = 0,\ldots, 3$. Each subdomain is parametrised by a square grid $(x^1,x^2)\in[-1,1]^2$, which will be discretised into a Chebyshev spectral grid in Sec.~\ref{sec:NumericalMethods}. Figure~\ref{fig:multi_domain} illustrates the multi-domain decomposition for a particle in a circular orbit with radius $r_p=10M$ on the equatorial plane. Further features of this figure will be detailed in the next sections.

\subsubsection{Outer source-free region: Domain \texorpdfstring{$0$}{0}}
This domain describes the region extending to future null infinity, comprising an outer spherical ``shell''  $\{r,\theta\}\in[r_+, \infty)\times[0,\pi/2]$. The radius $r_+ > r_p$ is a free parameter marking the transition into the near-particle region. This domain has the label $\id = 0$ in Fig.~\ref{fig:multi_domain}.

In terms of the hyperboloidal coordinates, this region corresponds to the rectangle $\D_0 = [0,\sigma_-]\times[0,1]$, with $\sigma_-= r_h/r_+$; cf.~Eq.~\eqref{eq:HypCoord}. The mapping into the square region $\{x^1,x^2\}\in[-1,1]^2$ then follows straightforwardly via
\beq
\sigma = \sigma_- \dfrac{1+x^1}{2}, \quad y = \dfrac{1+x^2}{2}.
\eeq

\subsubsection{Inner source-free region: Domain \texorpdfstring{$3$}{3}}
In analogy with the previous case, this domain comprises an inner shell $(r,\theta)\in[r_h, r_-]\times[0,\pi/2]$ between the black-hole future horizon and the beginning of a near-particle region at a free parameter $r_-<r_p$. In Fig.~\ref{fig:multi_domain}, this domain is labeled by $\id = 3$.

In the hyperboloidal coordinates, this region is the rectangle $\D_3 = [\sigma_+,1]\times[0,1]$, with $\sigma_+= r_h/r_-$. It is easily mapped into the square region $(x^1,x^2)\in[-1,1]^2$ by
\bea
\sigma = \sigma_+\dfrac{1-x^1}{2} + \dfrac{1+x^1}{2}, \quad y = \dfrac{1+x^2}{2}.
\eea

\subsubsection{Effective-source region: Domain \texorpdfstring{$2$}{2}}\label{sec:Map_D2}
We now turn our attention to the region in which we will employ the effective source formalism. We specify the domain's boundary to be a ball around the particle, which we fix at a radius $\polarR_o$ in the particle's co-moving frame of Sec.~\ref{sec:ParticlesFrame}. That is, we take it to be a domain $\D_{\mathcal{R}}=\D_2$ defined by $(\polarR, \pfphi) \in [0,\polarR_o]\times[0,\pi]$. In Fig.~\ref{fig:multi_domain}, this domain is labeled by $\id = 2$.

Then, the map into spectral coordinates $(x^1,x^2)\in[-1,1]^2$ follows directly from the polar representation for the coordinate centered at the particle via
\beq
\label{eq:x1_x2_PolarParticle}
x^1 = -\cos\pfphi, \quad x^2 = 2\rho -1,
\eeq
with $\rho={\polarR}/{\polarR_o}\in[0,1]$. In particular, the particle's location in those spectral coordinates is stretched into the line $x^2=-1$, $x^1 \in[-1,1]$. Finally, expressions \eqref{eq:CoordParticleFrame} and \eqref{eq:ParticlePolarCoordinates} for the particle's co-moving frame fix the relation between Eqs.~\eqref{eq:x1_x2_PolarParticle} and the hyperboloidal coordinates $(\sigma,y)$ from Eq.~\eqref{eq:HypCoord} to be
\bea
\label{eq:map_dom2}
\sigma &=& \dfrac{\sigma_p}{1 - \eta\, \sigma_p \sqrt{f_p}\, \rho(x^2) x^1} , \\
y &=& \sigma_p^{2}\, \eta^2\,  \rho(x^2)^2 [1-(x^1)^2].
\eea
The parameter 
\beq
\eta:=\polarR_o/r_h 
\eeq
is a dimensionless measure for the size of the ball $\polarR_o$ in units of the horizon radius $r_h$, and it specifies a shell $\D_{1,2}=\D_1\cup\D_2$ surrounding the particle. This world tube has thickness $\sigma\in[\sigma_-, \sigma_+]$, where $\sigma_\pm$ are the extreme values of $\sigma$ in ${\cal D}_2$. This thickness is determined by the points at which the ball's boundary intersects the equatorial plane $\cos \pfphi = \mp 1 \leftrightarrow x^1 =\pm 1$, where the coordinate $\sigma$ assumes its maximal and minimal values at the equatorial points. These extreme values are given by
\beq
\label{eq:sigma_plusminus}
\sigma_\pm = \dfrac{\sigma_p}{1 \mp \eta\, \sigma_p \sqrt{f_p}} \Longleftrightarrow r_\mp = r_p \mp \eta\, \sqrt{f_p}\, r_h.
\eeq
The upper boundary ($x^2 = 1$) of this region is, by definition, given by $\polarR = \polarR_o = \eta\, r_h$. In terms of the hyperboloidal coordinates $(\sigma,y)$, Eq.~\eqref{eq:map_dom2} then gives
\beq
\label{eq:D1-D2_parameter_x1}
\sigma_o(x^1) = \dfrac{\sigma_p}{1 - \eta\, \sigma_p\, \sqrt{f_p}\, x^1}, \quad
y_o(x^1) = \eta^2\, \sigma_p^2\, (1-(x^1)^2).
\eeq

Appendix \ref{App:alternative_map_D2} introduces an alternative map for domain $\D_2$, which still fixes $x^2 = 1$ to be a ball of radius $\polarR_o=\eta\, r_h$ around the particle in its co-moving frame, but relaxes this condition for the surfaces $x^2<1$.

\subsubsection{Intermediate source-free  region: Domain \texorpdfstring{$1$}{1}} The last domain, $\D_1$, covers the remainder of the shell $\sigma\in[\sigma_-, \sigma_+]$ (i.e., the portion of $\D_{1,2}$ outside the effective source region $\D_2$). In Fig.~\ref{fig:multi_domain}, this domain is labeled by $\id = 1$. Here we solve the source-free equation for $\phi_m$. 

The mapping into the square region $\{x^1, x^2\} \in [-1,1]^2$ must interpolate between the shell boundaries $\sigma\in\sigma_{\pm}$, as well as between the axis $y=0$ and the effective-source boundary $(\sigma,y) = (\sigma_o,y_o)$ as given in Eq.~\eqref{eq:D1-D2_parameter_x1}.

For practical purposes when discretising the problem, it is useful to define the spectral grid $(x^1, x^2)$ in $\D_1$ such that the mapping $\{\sigma_o(x^1),y_o(x^1)\}$ refers to the same grid points irrespective of whether $x^1$ is the spectral coordinate in region $\D_1$ or $\D_2$.
The whole domain of $\D_2$ is then given by$\{\sigma,y\}\in[\sigma_-,\sigma_+]\times[y_o(\sigma), 1]$, where
\beq
\label{eq:Map_Dom1}
\sigma = \sigma_o(x^1),  \quad
y = y_o(x^1)\,\dfrac{1-x^2}{2} + \dfrac{1+x^2}{2}.
\eeq
When visualising this mapping, cf. fig.~\ref{fig:multi_domain}, an apparent drastic grid stretch develops around $y=0$ and $\sigma=\sigma_-$. However, this visual effect does not impact the numerical code.

\subsection{Boundary conditions}\label{sec:BCs by domain}
In this section, we describe the boundary conditions imposed on each of the subdomains $\D_{\id}$ ($\id = 0,\ldots, 3$) to enforce the uniqueness of the numerical solution. The boundary conditions are divided into three types: regularity, transition and external conditions.

\subsubsection{Regularity conditions}\label{sec:RegCond}
The regularity conditions have already been mentioned in Sec.~\ref{sec:reg_conditions}. The operator $\boldsymbol{A}$ in Eq.~\eqref{eq:operator_A} is singular at the surfaces $\sigma = 0$ (future null infinity) and $\sigma=1$ (horizon), as well as $y=0$ (equator)  and $y=1$ (symmetry axis). Because we seek the regular solutions to the elliptic equations, one can directly impose the differential equations at these surfaces.

Conceptually, this argument also extends to the particle's location at $(\sigma, y) = (\sigma_p, 0)$. The coordinate mapping~\eqref{eq:map_dom2} in domain $\D_{2}$ is based on the polar representation of local Cartesian coordinates around the particle. As in any polar chart, the origin blows up into a line parametrised by the angular coordinate.  Here, the origin corresponds to the particle's point $(\sigma_p, 0)$, which is parametrised in terms of the coordinate line $x^2=-1$, $x^1\in[-1,1]$. With the help of Eqs.~\eqref{eq:map_dsigma}--\eqref{eq:map_d2sigmay}, the mapping
\eqref{eq:map_dom2} imposes the following conditions for the regularity of first derivatives in $\sigma$ and $y$:
\beq
\label{eq:cond_first_der}
\phi_{,1} \dot= 0, \quad \phi_{,2} \dot= x^1 \, \phi_{,12}. 
\eeq
The notation $\dot = $ indicates that the equality holds only at $x^2=-1$. In particular, one can directly differentiate Eqs.~\eqref{eq:cond_first_der} along $x^1$ to obtain
\beq
\label{eq:cond_first_der_implications}
\phi_{,11} \dot = 0, \quad \phi_{,112} \dot= 0.
\eeq
If conditions \eqref{eq:cond_first_der} and \eqref{eq:cond_first_der_implications} are satisfied, then we verify that 
\beq
{\boldsymbol A} \bar\phi \sim {\cal O}(1) \quad {\rm as} \quad x^2 \rightarrow -1. 
\eeq
Thus, it remains only to ensure that the effective source is also finite at the particle. This is guaranteed for punctures with orders $\hat n_{\rm max} \geq 1$. 

At the practical level, however, exactly evaluating the effective source at the particle with the algorithm from Ref.~\cite{Bourg2023} requires a cumbersome limiting procedure using l'Hopital's rule. Section~\ref{sec:NumericalMethods} will describe a solution to circumvent this process at the numerical level. It involves choosing a grid that avoids the particle altogether, while still keeping the convergence properties of the spectral solver. Quantities exactly at the particle's position are then obtained by an interpolation at $x^2=-1$.

Even though the numerical approach guarantees an accurate solution for the field and its first $\sigma$ and $y$ derivatives, the algorithm faces limitations if higher derivatives in $\sigma$ and $y$ are necessary to construct further quantities. Indeed, the regularity of second $(\sigma,y)$-derivatives as $x^2\rightarrow -1$ is more intricate as it requires higher order derivatives in $(x^1, x^2)$ via
\bea
\label{eq:cond_second_der}
\phi_{,122} \dot= x^1\, \phi_{,22}, \quad \phi_{,222} \dot= x^1\, \phi_{,1222} - \dfrac{1}{3} (x^1)^2 \, \phi_{,11222}.
\eea
The accuracy loss for numerical calculations up to $5^{\rm th}$ order $(x^1, x^2)$-derivatives spoils the evaluation of division by small numbers when interpolation to $x^2=-1$.

Appendix \ref{App:alternative_map_D2} introduces an alternative polar mapping around the particle that has better regularity properties for high order $\sigma$ and $y$ derivatives at particle. The downside of such alternative map is that it jeopardises the behavior of the puncture field and effective source at the surfaces $x^2=$ constant, to be discussed in Sec.~\ref{sec:results_EffSource}

All in all,  regularity conditions follow from directly imposing the differential equations at the following boundaries:
\bit

\item Inner source-free region (Domain $0$): 
\bea
&&\text{horizon:}\, \sigma=1 \leftrightarrow x^1=1, \nn \\ 
&&\text{equator:}\, y=0 \leftrightarrow x^2=-1 , \nn \\
&&\text{symmetry axis:}\, y=1 \leftrightarrow x^2=1 .  \nn
\eea

\item Intermediate source-free  region (Domain $1$): 
\bea
&&\text{symmetry axis:}\, y=1 \leftrightarrow x^2=1.  \nn
\eea

\item Effective-source region (Domain $2$): 
\bea
&&\text{equator (left of particle):}\, y=0  \leftrightarrow x^1=-1, \nn \\
&&\text{equator (right of particle):}\,y=0 \leftrightarrow x^1= 1, \nn \\
&&\text{particle:}\, (\sigma,y)=(\sigma_p,0) \leftrightarrow x^2= -1. \nn
\eea

\item Outer source-free region (Domain $3$): 
\bea
&&\text{future null infinity:}\, \sigma=0 \leftrightarrow x^1=-1, \nn \\ 
&&\text{equator:}\, y=0 \leftrightarrow x^2=-1 , \nn \\
&&\text{symmetry axis:}\, y=1 \leftrightarrow x^2=1 .  \nn
\eea

\eit 

\subsubsection{Transition conditions}\label{sec:transition_conds}
This section discusses the transition conditions between domains $\D_0$-$\D_1$, $\D_1$-$\D_3$, and $\D_1$-$\D_2$. These conditions result from imposing specific jumps on the corresponding fields and their normal derivative across the domains' boundaries. 

We first consider the source-free region comprised by the domains $\D_0$, $\D_1$ and $\D_3$. As already mentioned, in this region the fields $\bar \phi_m^{(\id)}$ ($\id=0,1,3$) represent the retarded field $\bar \phi^{\rm ret}_m$ satisfying the homogeneous equation ${\boldsymbol A} \bar \phi_m^{(\id)} = 0$. Transition conditions at the boundaries $\D_0$-$\D_1$, $\D_1$-$\D_3$ follow from imposing the continuity of the retarded field and its derivative along the $\sigma$-direction:
\bit
\item Interface $\D_0$-$\D_1:$
\bea
\label{eq:TransitionCondition_D0_D1_field}
\left . \left . \bar \phi^{(\id = 1)}_m \right |_{x^1 = -1} - \bar \phi_m^{(\id = 0)} \right |_{x^1 = 1} = 0,\\
\nn \\
\label{eq:TransitionCondition_D0_D1_der}
\left .  \partial_\sigma \left . \bar \phi^{(\id = 1)}_m \right |_{x^1 = -1} - \partial_\sigma \bar \phi_m^{(\id = 0)} \right |_{x^1 = 1} = 0,
\eea
\item Interface $\D_1$-$\D_3:$
\bea
\label{eq:TransitionCondition_D1_D3_field}
\left . \left . \bar \phi^{(\id = 3)}_m \right |_{x^1 = -1}-\bar \phi_m^{(\id = 1)} \right |_{x^1 = 1} =0, \\
\nn \\
\label{eq:TransitionCondition_D1_D3_der}
\left .  \partial_\sigma \left . \bar \phi^{(\id = 3)}_m \right |_{x^1 = -1} - \partial_\sigma \bar \phi_m^{(\id = 1)} \right |_{x^1 = 1} = 0.
\eea
\eit

On the interface $\D_1$-$\D_2$, the transition conditions follow from the identification $\bar \phi^{(1)}_m=\bar \phi^{\rm ret}_m$ and $\bar \phi^{(2)}_m=\bar \phi^{\mathcal{R}}_m$. Thus, they are fixed by the puncture field via Eq.~\eqref{eq:Res_Ret_Punc}:
\bit
\item Interface $\D_1$-$\D_2:$
\begin{align}
\label{eq:TransitionCondition_D1_D2_field}
 \left .  \bar \phi^{(\id = 2)}_m \right |_{x^2 = 1}- \left. \bar \phi_m^{(\id = 1)} \right |_{x^2 = -1} &= - \left. \bar \phi^{\mathcal{P}}_m \right |_{x^2 = 1}\\
\nn \\
\label{eq:TransitionCondition_D1_D2_der}
 \left .  \partial_{\vec n_{2}}  \bar \phi^{(\id = 2)}_m \right |_{x^2 = 1} - \left . \partial_{\vec n_{2}} \bar \phi_m^{(\id = 1)} \right |_{x^2 = -1} &= - \left.\partial_{\vec n_{2}} \bar \phi^{\mathcal{P}}_m \right |_{x^2 = 1},
\end{align}
\eit
with $\partial_{\vec n_{2}}$ the normal derivative to $x^2 =$ constant,
\beq
\partial_{\vec n_{2}} = \dfrac{\sigma_{,x^1} \partial_y - y_{,x^1} \partial_\sigma}{\sqrt{\sigma_{,x^1} ^2 + y_{,x^1}^2}}.
\eeq

\subsubsection{Multi-basis configuration}\label{sec:ext_conditions}

So far, we have assumed that the same spectral basis of functions $e^{im(\varphi-\Omega t)}$ is used in all four domains, and we have adopted square coordinates in each domain in order to use a Chebyshev basis in the sections below. However, this uniformity is not a requirement. The $m$-mode scheme we use is specifically intended to bypass difficulties with an $\ell m$-mode scheme in the shell $\D_{1,2}$ containing the particle; nothing prevents us from using an $\ell m$-mode decomposition in the regions $\D_0$ and $\D_3$.

In fact, as alluded to in the Introduction, we expect there are significant advantages to using a full decomposition into spherical harmonics $Y_{\ell m}(\theta, \varphi)$ in the regions $\D_0$ and $\D_3$. In our context, this amounts to using associated Legendre functions $P_\ell^m(\cos\theta)$ as basis functions in the polar direction: 
\beq
\bar{\phi}_m(\sigma,y) = \sum_{\ell = |m|}^{\infty}\phi_{\ell m}(\sigma)(1-y)^{|m|/2} P^m_{\ell}(\sqrt{y}).
\eeq
For a consistent representation in terms of $P_\ell^m(\cos\theta)$, the factor $(1-y)^{|m|/2}$ is introduced to undo the scaling in Eq.~\eqref{eq:phi_hyperboloidal_rescaling}.

The 2D elliptic equations in $\D_0$ and $\D_3$ then reduce to a set of decoupled ODEs in $\sigma$, one for each $\phi_{\ell m}$. Because the sum over $\ell$ converges rapidly in these domains, relatively few $\ell$ modes are required. Regularity conditions at the axis are not needed, and the interface conditions~\eqref{eq:TransitionCondition_D0_D1_field}--\eqref{eq:TransitionCondition_D1_D3_der} remain as written. We note that although the different $\ell$ modes decouple in the field equations, they do couple to one another through these transition conditions to the intermediate domain~$\D_1$. 

In this paper, we do not fully implement such a scheme. However, we take a step toward it by using external $\ell m$-mode data in the regions $\D_0$ and $\D_3$, which then provide boundary data for our $m$-mode calculation in $\D_{1,2}$. Such data can be computed using the code from Ref.~\cite{PanossoMacedo:2022fdi} or other readily available codes~\cite{BHPToolkit}, which work with a global $\ell m$ mode decomposition in all regions of the problem. In some sense this means we solve the field equations in $\D_{1,2}$ twice: once using an $\ell m$ decomposition in order to compute data in $\D_0$ and $\D_3$, and then again using an $m$-mode scheme. However, in reality the $\ell m$ scheme captures relatively little of the physical, 4D field in $\D_{1,2}$ because of the slow convergence with $\ell$ in that region; in contrast, our $m$-mode scheme is able to capture the 4D field with high accuracy throughout $\D_{1,2}$. Moreover, reducing our numerical domain to $\D_{1,2}$ allows us to most easily explore the features of the problem in that region, which is our region of interest here. 

To concretely construct boundary conditions from external $\ell m$-mode data, assume the data is given in the typical form
\beq
\Phi^{\rm ret} = \sum_{m=-\infty}^{m=\infty} \sum_{\ell = |m|}^{\infty} \psi^{\rm ret}_{\ell,m}(r) Y_{\ell m}(\theta, \varphi) e^{-im\Omega_p t}.
\eeq
Comparing this with the $m$-mode decomposition \eqref{eq:Freq_m_decomposition_2} after the hyperboloidal mapping \eqref{eq:phi_hyperboloidal_rescaling}, we see that the boundary conditions for our field in $\D_1$ are given by
\bea
\label{eq:ExtData_lm_mode_minus}
\left.\bar\phi^{(1)}_m(y)\right|_{x^1=- 1} &=& \sum_{\ell = m}^{\infty} c_{\ell m}\, \dfrac{\psi^{\rm ret}_{\ell,m}(r_+) P_{\ell}^m(\sqrt{y})}{{\cal Z}(\sigma_-, y)}, \\
\nn \\
\label{eq:ExtData_lm_mode_plus}
\left.\bar\phi^{(1)}_m(y)\right|_{x^1=+1} &=& \sum_{\ell = m}^{\infty} c_{\ell m}\, \dfrac{\psi^{\rm ret}_{\ell,m}(r_-) P_{\ell}^m(\sqrt{y})}{{\cal Z}(\sigma_+, y)},
\eea
where
\beq
c_{\ell m}=\sqrt{\dfrac{2\ell+1}{4\pi} \dfrac{(\ell-m)!}{(\ell+m)!}}.
\eeq 
Given this boundary data, we then employ the 2D elliptic solver only in the shell $\D_1$ and $\D_2$.

\subsection{Remarks}
We observe that the decomposition presented in this section is not restricted to a spherically symmetric background given by the Schwarzschild metric. From the numerical, technical point of view, the resulting elliptic equation captures the main structures present in more complicated scenarios, such as a rotating central black hole and the second-order self-force formalism in the two-timescale approach. When going from the scalar field model to the gravity case, we anticipate the need of solving the equations for two coupled complex fields, so that the equatorial symmetry is properly accounted for. 

\begin{figure*}[t!]
	\centering
    \includegraphics[width=\textwidth]{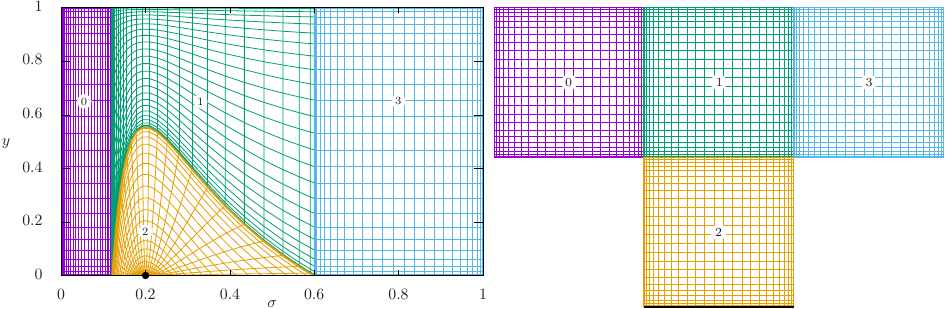}
	\caption{{\em Left panel:} 2D space in hyperboloidal coordinates $\{\sigma,y\}$. The compact radial coordinate $\sigma=r_h/r$ includes future null infinity at $\sigma=0$ and the black hole's future horizon at $\sigma=1$. The angular coordinate $y=\cos^2\theta$ has the symmetry axis at $y=1$ and the equator at $y=0$. The black dot at $(\sigma,y) = (1/5, 0)$ represents the position of a particle with orbital radius $r_p = 10M$ on the equatorial plane. The black hole's exterior is divided into three nested spherical shells, $\D_{0}$, $\D_{1,2}$, and $\D_3$. We refer to domains $\D_{0}$ and $\D_{3}$, respectively, as the {\it outer source-free region} and {\it inner source-free region}. The intermediate shell, $\D_{1,2}$, is divided into domains $\D_{1}$ and $\D_2$. Domain $\D_2$ is the {\it effective-source region}, in which we use a puncture method. We refer to the domain $\D_1=\D_{1,2}-\D_2$ as the {\it intermediate source-free region}. {\em Right panel:} The same four domains $\D_{\id}$ ($\id=0,\ldots, 3$), each parametrised with square coordinates $(x^1, x^2)\in[-1,1]^2$ adapted to a spectral solver for elliptic equations. Transition conditions at the interfaces between $\D_0$ and $\D_1$, and between $\D_1$ and $\D_3$, follow from the continuity of the retarded field and its first normal derivative at these boundaries. At the interface between $\D_1$ and $\D_2$, the puncture field fixes the jumps between the retarded solution in Domain $1$ and the residual field in Domain $2$. The polar character of the coordinate chart in $\D_2$ stretches the particle's location into the line $x^2=-1$, $x^1\in[-1,1]$ (shown as a black line).}
	\label{fig:multi_domain}
\end{figure*}

A bigger challenge may arise for a particle following  eccentric and inclined orbits. We expect a domain decomposition for eccentric orbits at the equatorial plane similar to fig.~\ref{fig:multi_domain}, with the world-tube boundaries $\sigma_\pm$ coinciding with the orbit's radial parameters. The challenges appear in the construction of the puncture field and the presence of Gibbs phenomena in the frequency domain (e.g.~\cite{Leather:2023dzj}). For inclined orbits on the other hand, one may need to elaborate further on the domain decomposition.

\section{Multi-domain spectral method}\label{sec:NumericalMethods}
As anticipated, we solve the elliptic equation \eqref{eq:EqRes} (and its homogenous version) with a multi-domain, spectral code. This section introduces the main concepts of the code, which is based on the approach from~\cite{Meinel:2008kpy} (and references therein). In the context of self-force calculations, this algorithm extends the code from Ref.~\cite{PanossoMacedo:2022fdi} into 2D problems.

We first overview the generic formulation for the elliptic spectral solver. Then, we concentrate on the specific features of the problem under consideration. 

Before proceeding, we remark on the notation used in the next sections: for a given discrete quantity (e.g., fields, domain, grid points), upper case $N $ labels the last value in the corresponding discretisation index $i = 0, \ldots, N$, whereas lower case $n = N+1$ denotes the total number of discrete elements.

\subsection{Spectral solver: collocation method}\label{sec:SpecSolver}
Let $f^{(\iF,\id)}(x^1,x^2)$ represent a real-valued function defined on the domain $\D_{\id} = [-1,1]^2$. As in the previous section, the index $\id=0,\ldots, N_{\rm d}$ labels one of the $n_{\rm d} = N_{\rm d}+1$ domains considered. The label $\iF=0,\ldots, N_{\rm f}$ accounts for one in a total of $n_{\rm f} = N_{\rm f} +1$ fields. 

In the specific case of this work, the total number of domains is either $n_{\rm d} = 4$, or $n_{\rm d} = 2$, depending on whether one exploits the multi-basis configuration from Sec.~\eqref{sec:ext_conditions} to set boundary conditions at $\sigma=\sigma\pm$ and solve the problem only within domains $\D_1$ and $\D_2$. In each domain, one must solve for $n_{\rm f} = 2$ real fields, representing the real and imaginary parts of the complex field~$\bar \phi^{(\id)}$. 

For a fixed numerical truncation $N_1^{\id}$ and $N_2^{\id}$ within each domain, the numerical scheme approximates the exact function $f^{(\iF,\id)}(x^1,x^2)$ via the expansion
\beq
\label{eq:f_approximation}
f^{(\iF,\id)}_{N_1^{\id}, N_2^{\id }}(x^1,x^2) = \sum_{k_1=0}^{N_1^{\id}} \sum_{k_2=0}^{N_2^{\id}} c^{(\id, \iF)}_{k_1, k_2}  T_{k_1}(x^1) T_{k_2}(x^2)
\eeq
with $T_k(x)=\cos[k \arccos(x)]$ the Chebyshev polynomials of the first kind. The Chebyshev coefficients $c^{(\id, \iF)}_{k_1, k_2}$ are fixed by a collocation method. For this purpose, we introduce discrete grids $\{x^1_{i_1} \}_{i_1 = 0}^{N_1^{\id}}$ and $\{x^2_{i_2} \}_{i_2 = 0}^{N_2^{\id}}$, with a total of $n_1^{\id}=N_1^{\id}+1$ and $n_2^{\id}=N_1^{\id}+1$ points along the $x^1$ and $x^2$ directions in a given domain. The Chebyshev coefficients $c^{(\id, \iF)}_{k_1, k_2}$ are uniquely fixed by imposing that the approximate representation \eqref{eq:f_approximation} coincides with the exact function $f^{(\iF,\id)}(x^1,x^2)$ at the grid points. Explicitly, let us define the discrete value from each $\iF$ field, in each $\id$ domain, at each grid point $(x^1_{i_1}, x^2_{i_2})$ with the notation
\beq
\label{eq:discrete_functions}
f^{(\iF,\id)}_{i_1, i_2} := f^{(\iF,\id)}_{N_1^{\id}, N_2^{\id }}(x^1_{i_1},x^2_{i_2}).
\eeq
Then, the collocation method imposes that these discrete set of values correspond to the exact solution, i.e. $f^{(\iF,\id)}_{i_1, i_2}= f^{(\iF,\id)}(x^1_{i_1},x^2_{i_2})$.

From the discrete values $f^{(\iF,\id)}_{i_1, i_2}$, spectral derivatives in the directions $x^1$ and $x^2$ follows by applying the appropriate spectral derivative matrices ${\rm D}^{x^{\rm A}}_{{i_{\rm i}}, {j_{\rm i}}}$: 
\beq
\label{eq:Disc_Derv}
\begin{array}{c}
\displaystyle
\left( f_{,x^{1}} \right)^{(\iF,\id)}_{{i_1}, {i_2}}= \sum_{j_1=0}^{N_1^{\rm id}} {\rm D}^{x^1}_{{i_1}, {j_1}}\, f^{(\iF,\id)}_{{j_1}, {i_2}}, \\ 
\\
\displaystyle
\left( f_{,x^{2}} \right)^{(\iF,\id)}_{{i_1}, {i_2}}= \sum_{j_2=0}^{N_2^{\rm id}} {\rm D}^{x^2}_{{i_2}, {j_2}}\, f^{(\iF,\id)}_{{i_1}, {j_2}},
\end{array}
\eeq
where we employed the notation
\beq
\left( f_{,x^{\rm A}} \right)^{(\iF, \id)}_{{i_1}, {i_2}} := \left. \dfrac{\partial}{\partial_{x^{\rm A}}} {f^{(\iF,\id)}_{N_1^{\id}, N_2^{\id }}(x^1,x^2)} \right|_{(x^1_{i_1},x^2_{i_2})}
\eeq
to represent the differentiation operation along a direction $x^{\rm A}$ (${\rm A}=1,2$). Higher derivatives follow from the same procedure. Appendix \ref{sec:SpecMatrix} shows the matrices ${\rm D}^{x^{\rm A}}_{{i_{\rm i}}, {j_{\rm i}}}$ associated with the grids employed in this work, to be introduced in Sec.~\ref{sec:Chebyshev grids}. 

This discretisation procedure allows us to cast the differential equations \eqref{eq:EqRes} and \eqref{eq:EqHomo}, with their corresponding boundary/transition conditions from Sec.~\ref{sec:BCs by domain} into a system of linear algebraic equations for the unknowns $f^{(\iF,\id)}_{i_1, i_2}$, as detailed in Appendix~\ref{sec:LinAlgSystem}. In a compact notation, this linear system assumes the form $\vec F(\vec X) = 0$, with
\beq
\label{eq:linear_system}
\vec F(\vec X) = \hat J\, \cdot \, \vec{X} - \vec{S}.
\eeq
In the above expression,  $\vec X$ collects all components $f^{(\iF,\id)}_{i_1, i_2}$ into a single vector
\beq
\label{eq:vec_X}
\vec X = \left( f^{(\iF,\id)}_{i_1, i_2} \right) \quad {\rm for} \quad \left\{ 
\begin{array}{l}
\id = 0,\ldots, N_{\rm d} \\  
\iF = 0,\ldots, N_{\rm f}    \\ 
i_1 = 0,\ldots, N_1^{\id} \\
i_2 = 0,\ldots, N_2^{\id}
\end{array}
\right. ,
\eeq
with size 
\beq
n_{\rm total} = N_{\rm total} +1 = n_{\rm f} \sum_{\id=0}^{N_{\rm d}} n_1^{\rm id} n_2^{\rm id}. 
\eeq
The matrix $\hat J$ contains the discrete representation of the operator $\eqref{eq:operator_A}$ and the transition conditions expressed by the left-hand side of Eqs.~\eqref{eq:TransitionCondition_D1_D2_field}--\eqref{eq:TransitionCondition_D1_D2_der}. The non-vanishing components of the source vector $\vec{S}$ collect the right-hand side of Eqs.~\eqref{eq:EqRes}, \eqref{eq:TransitionCondition_D1_D2_field} and \eqref{eq:TransitionCondition_D1_D2_der}, i.e. the effective source and puncture field at the relevant grid points. When solving the problem in the multi-basis configuration with external data at $
\sigma_\pm$, the source vector $\vec{S}$ also contains information from the boundary data resulting from the $\ell m$-scheme in Eqs.~\eqref{eq:ExtData_lm_mode_minus} and \eqref{eq:ExtData_lm_mode_plus}.

\begin{figure}[t!]
	\centering
     \includegraphics[scale=1.]{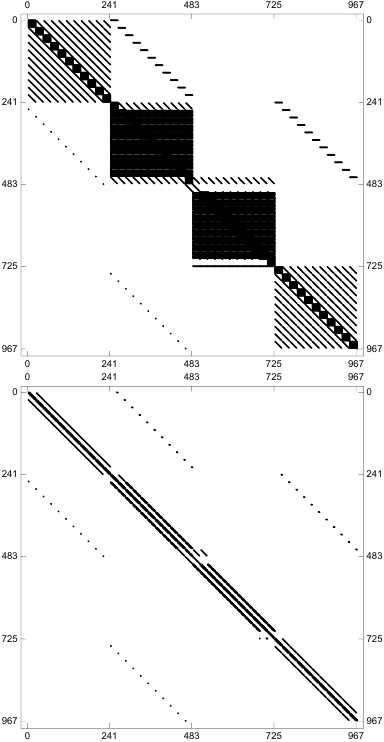}
	\caption{Visualisation of the Jacobian matrix $\hat J$ for the linear system including all domains $\D_{\id}$, $\id=0,\ldots, 3$ ($n_{d}=4$). The grids have parameters $N_1 = N_2 = 10$ in all considered domains. {\em Top Panel:} The spectral representation for the derivative matrices yields dense block matrices within each domain. Inversion with LU decomposition algorithms becomes prohibitive for large resolutions. {\em Bottom Panel:} Finite difference representation of the derivative matrices used as preconditioner to the iterative solver BiCGSTAB. Despite the sparse structure, the coupling of domain $\D_1$ to $\D_0$ and $\D_3$ yields a broad band, which spoils the speed-up efficiency of the preconditioner in the current implementation of the code.	
}
	\label{fig:Matrices_4domain}
\end{figure}

\begin{figure}[t!]
	\centering
    \includegraphics[scale=1.]{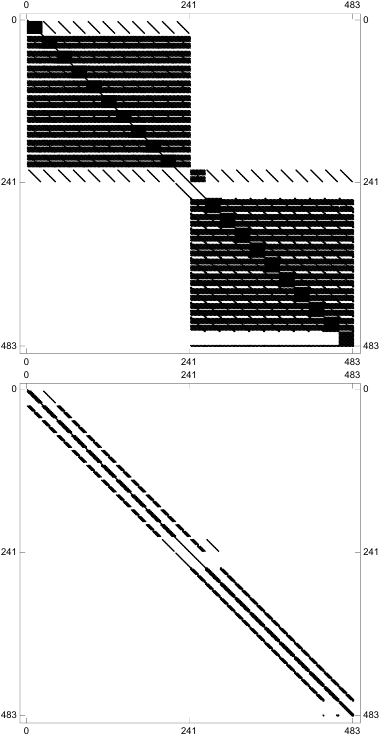}
	\caption{Visualisation of the Jacobian matrix $\hat J$ for the linear system restricted to domains $\D_1$ and $\D_2$ ($n_{d}=2$). The grids have parameters $N_1 = N_2 = 10$ in all considered domains. {\em Top Panel:} As in Fig.~\ref{fig:Matrices_4domain}, the spectral representation for the derivative matrices yields dense block matrices. Even though the matrix size is reduced by a half when comparing the $n_d=2$ and $n_d=4$ cases, inversion with LU decomposition algorithms is still prohibitive for large resolutions. {\em Bottom Panel:} The finite difference representation for the derivative matrices provides a very narrow band matrix, which is an efficient preconditioner to the iterative solver BiCGSTAB.
}
	\label{fig:Matrices_2domain}
\end{figure}

The components $X_{\rm I}$ for the vector $\vec X$, and consequently the rows and columns of the matrix $J_{\rm I \rm J}$, are arranged such that the even entries ($i_{\rm f}=0$) hold the real part of the complex field, while the odd entries ($i_{\rm f} = 1$) hold the imaginary part (recall that, here, $n_{\rm f}=2$). In any given domain $\id^*$, the vector $(f^{(\iF,\id^*)}_{i_1, i_2})$ is arranged so that the first $n_1^{\id^*}$ entries are the components with $i_2 = 0$, $i_1\in[0,N^{\id^*}_1]$, followed by entries with $i_2 = 1$, $i_1\in[0,N^{\id^*}_1]$, and so on. Put together, the vector $(f^{(\iF,\id^*)}_{i_1, i_2})$ for a given $\id^*$, has length $n_{\rm f} \times n_1^{\id^*} \times n_2^{\id^*}$ and encodes the information of the complex field in that domain. Considering the subsequent domains, one finds $n_d$ such vectors, which are put together into a single vector $\vec X$ in indexical order $\id^*$. This sequential indexing is summarised by the mapping
\beq
\label{eq:vec_index}
{\rm I}(\id^*, \iF, i_1, i_2) = \iF + n_{\rm f} \left( i_1 + i_2 \, n^{\id^*}_1 + \sum_{j_{\rm d}=0}^{\id^* -1} n_1^{j_{\rm d}} \, n_2^{j_{\rm d}}  \right).
\eeq 
Note that in the above expression, the index $\id^*$ always has the range $0,\ldots, N_d$. This notation is needed to avoid confusion with the labelling of the several domains $\D_{\id}$ introduced in Sec.~\ref{sec:coord_domain_decomposition}. For example, in a setup involving all $n_d=4$ domains, the index $\id^* = 0,\ldots, 3$ does coincide with the labels in $\D_{\id}$ ($\id = 0,\ldots, 3$). However, if we exclude the outer ($\D_0$) and inner ($\D_3$) source-free regions, one has an $n_d=2$ setup with $\id^*=0$ and $\id^*=1$ corresponding to $\D_1$ and $\D_2$, respectively.

The top panels of Figs.~\ref{fig:Matrices_4domain} and~\ref{fig:Matrices_2domain} display the structure of $\hat J$ for the setups with $n_d=4$ and $n_d=2$ domains, with a relative small number of grid points ($n_1 = n_2 = 11$) in all domains. One observes the diagonal structure of dense matrices resulting from discretising the operator ${\boldsymbol A}$ in terms of spectral differentiation matrices. The off diagonal elements illustrate the coupling between the domains via the transition conditions on the fields and normal derivatives. One option to solve the linear system is via a direct inversion of the matrix $\hat {J}$ with a LU decomposition algorithm. Though very robust, this approach becomes prohibitively expensive for moderate to large resolutions $n_1 \sim n_2 \sim n$, as the LU inversion scales as $n_{\rm total}^3 \sim n^6$. 

Alternatively, it is common to solve the system \eqref{eq:linear_system} with iterative linear solvers~\cite{Barrett1994,Saad_2003}. In particular, we employ the Bi-conjugate stabilized method (BiCGStab)~\cite{Vorst1992} with pre-conditioner based on the finite difference representation for the differentiation operators. Such finite difference (fd) representation yields a sparse band structure for the matrix $\hat J_{\rm fd}$. The bottom panels of Figs.~\ref{fig:Matrices_4domain} and~\ref{fig:Matrices_2domain} display the finite difference structure of $\hat J_{\rm fd}$ for the setups with $n_d=4$ and $n_d=2$ domains, respectively. The sparse nature in the form of a band matrix is apparent in the plots.

The inversion of band matrices at the preconditioner level in the BiCGStab algorithm scales as $\sim n_{\rm total} \, n_{\rm band}$, with $n_{\rm band}$ the band size. For narrow band matrices, this step is much faster than the full LU decomposition of the original matrices, and it ensures that the iterative BiCGStab algorithm converges after just a few steps.  In particular, the grid points ordering according to Eq.~\eqref{eq:vec_index} yields a very narrow band in the $n_d=2$ domains setup. The same conclusion is not valid in $n_d=4$ domains setup, as the transition conditions from $\D_1$ into $\D_0$ and $\D_3$ significantly enlarges the band structure. This complication provides additional motivation for concentrating on the $n_d=2$ domains setup in the present paper. As explained in the Introduction and Sec.~\ref{sec:ext_conditions}, an $\ell m$-mode decomposition with coordinates centred on the black hole is most natural in the inner and outer shell, and there are multiple readily available codes that can provide $\ell m$-mode data in those regions. From the numerical perspective, our current code is significantly more efficient when restricted to $n_d=2$ domains, taking boundary data from one of those existing codes. 

We do not discard the possibility of solving the full $n_d=4$ domains system. For that purpose, a Schur complement domain decomposition method~\cite{Saad_2003,Rashti:2021ihv} should be the best option to enhance the code's efficiency. However, implementing that algorithm and studying the $n_d=4$ case is beyond the scope of this work.

We finish this section with a discussion on the grid points employed in each domain, with particular focus on the choice for the region around the particle.

\subsection{Chebyshev grids}\label{sec:Chebyshev grids}
\begin{figure}[bt!]
	\centering
	\includegraphics[width=\columnwidth]{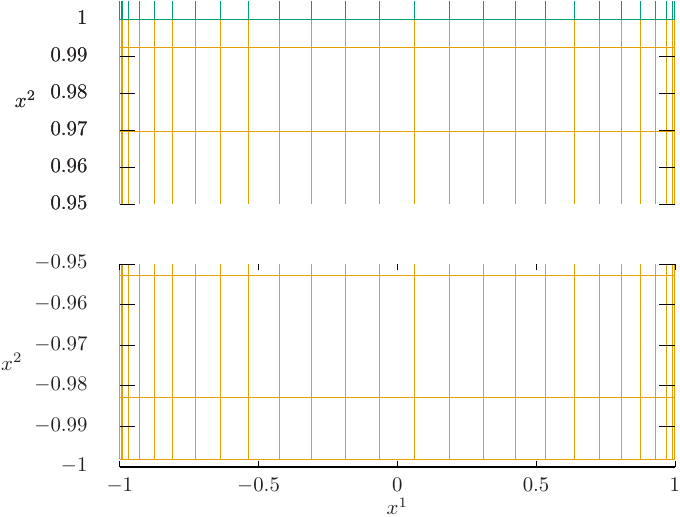}
	\caption{Square coordinates $(x^1, x^2)\in[-1,1]^2$ parametrising domain $\D_2$. The vertical axis zooms in on the particle's location at $x^2=-1$ (bottom panel, black line), and the transition into domain $\D_1$ at $x^2=1$ (top panel). The discrete grid along the $x^2$ direction is based on Radau-Chebyshev collocation points (yellow grid), as defined in Eq.~\eqref{eq:ChebRadau_grid}. This choice excludes the surface $x^2=-1$ from the numerical grid and includes the transition surface at $x^2=1$ where $\D_2$ connects with $\D_1$ (green grid).}
	\label{fig:specgrid_zoom_particle}
\end{figure}

We employ two different grids in this work. The first one is the well-known Chebyshev-Lobatto grid, which is obtained by considering the extremes of the Chebyshev polynomial $T_{N}(x)$ of order $N$~\cite{trefethen2000spectral,Boyd,canuto2007spectral,GraNov07,Ansorg2013}. The Chebyshev-Lobatto grid reads
\beq
\label{eq:ChebLobatto_grid}
x_i = \cos \left(\dfrac{ \pi i}{N}\right), \quad i=0,\ldots, N.
\eeq
This grid includes the interval ends $x=\pm 1$ as grid points with labels $i=0$ and $i=N$. Therefore, it is ideal to solve boundary-value problems and to impose transition conditions between the domains.

We employ the Chebyshev-Lobatto grid \eqref{eq:ChebLobatto_grid} along the $x^1$ direction in {\em all} domains $\D_{\id}$ ($\id = 0,\ldots, N_{\rm d}$). This grid is also the best choice along the $x^2$ direction in domains  $\D_{0}$, $\D_{1}$ and $\D_{3}$, i.e., in the domains describing the source-free region, where we solve the homogeneous elliptic equation for the retarded field. 

In the domain $\D_{2}$, which contains the particle, we opt for a less widely known grid. The so-called Chebyshev-Radau grid considers half of the Fourier discretisation for the interval $\theta=[0,2\pi]$ into $\theta_i = 2\pi i/(2N+1)$ ($i=0,\ldots, 2N$)~\cite{trefethen2000spectral,Boyd,canuto2007spectral,GraNov07,Ansorg2013}. Specifically, the Chebyshev-Radau grid reads
\beq\label{eq:ChebRadau_grid}
x_i = \cos\left(\dfrac{2\pi i}{2N+1}\right), \quad i = 0, \ldots, N.
\eeq
This grid has the peculiar property of including the interval end-value $x=1$ as a grid point with label $i=0$, but at the same time excluding the value $x=-1$ from the numerical grid. Indeed, the grid point with label $i=N$ assumes the value
\beq
x_N = \cos\left(\dfrac{2\pi N}{2N+1}\right) \approx -1 + \dfrac{\pi^2}{8 N^2} > -1.
\eeq
We employ the Chebyshev-Radau grid along the $x^2$ direction in domain $\D_{2}$. This choice allows us to impose the condition \eqref{eq:TransitionCondition_D1_D2_der} at the surface $x^2_0=1$ when transitioning from domain $\D_{2}$ into $\D_{1}$. At the same time, we avoid the surface $x^2=-1$ where the particle is located, and at which cumbersome regularity conditions would need to be carefully derived. By imposing the elliptic equation directly at the last grid point $x^2_N\approx -1$, we ensure that such regularity conditions are implicitly taken care of, thus retaining the solver's spectral properties. 

These grids are displayed in Fig.~\ref{fig:multi_domain} for a discretisation with parameters $N_1 = N_2 = 25$ in all domains. The left panel in Fig.~\ref{fig:multi_domain} shows the grid structure in the physical domain with hyperboloidal coordinates $\{\sigma,y\}\in [0,1]^2$, whereas the right panel depicts the spectral grid parametrised by square coordinates $\{x^1, x^2\}\in[-1,1]^2$ within each domain $\D_{\id}$. It becomes evident how the particle, represented by a point in the physical domain (black dot, left panel Fig.~\ref{fig:multi_domain}), stretches into a line in the spectral domain (black line, right panel Fig.~\ref{fig:multi_domain}).

Figure~\ref{fig:specgrid_zoom_particle} focuses on the domain $\D_2$ to illustrate the properties of the Chebyshev-Radau grid. It zooms in on the region around the coordinates $x^2=-1$, where one observes the discrepancy between the last grid point $x^2_{N_2}\approx -1$ and the particle at $x^2 = -1$. The figure also zooms in around $x^2=1$ to show that this surface is the grid point $x^2_0$, and that it provides the transition between domain $\D_2$ (yellow lines) into $\D_1$ (green line).

\section{External Data}\label{sec:External Data}
\begin{figure}[tb!]
	\centering
	\includegraphics[width=\columnwidth]{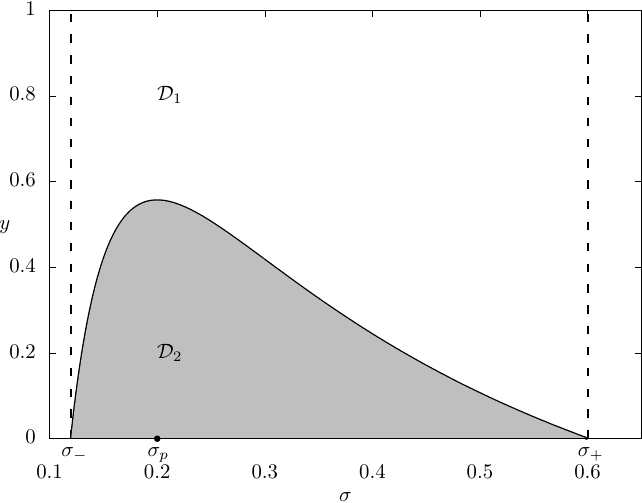}
	\caption{Regions in the hyperboloidal coordinates $\{\sigma,y\}$ where external data is required for a 2-domain $m$-mode elliptic solver: (i) The black solid line delimits the domain boundary $\D_1$-$\D_2$. Boundary conditions follow from the puncture field $\bar \phi^{\mathcal{P}}_m$, $\partial_\sigma \bar \phi^{\mathcal{P}}_m$ and $\partial_y \bar \phi^{\mathcal{P}}_m$; (ii) In the grey region (domain $\D_2$), one needs the effective source $\bar S^{\rm eff}$; (iii) At the dashed lines at $\sigma_\pm$, cf.~\eqref{eq:sigma_plusminus}, boundary conditions follow from a $\ell$-mode-sum over the retarded field modes $\bar \phi^{\rm ret}_{\ell m}$. Here, the particle's orbit is at $r_p = 10 M$ with $\D_2$ excision parameter $\eta \approx 3.727$ --- cf.~\eqref{eq:eta_plus}.
}
	\label{fig:external_data_region}
\end{figure}

Our first numerical studies concentrate on the external data needed in the multi-domain elliptic solver. Figure~\ref{fig:external_data_region} depicts the regions where external data is needed: (i)~the puncture field at the domain boundary $\D_1$-$\D_2$ (black solid line); (ii) the effective source in domain $\D_2$ (grey region); and (iii) the retarded $\ell m$-mode sum at the domain boundaries $\D_0$-$\D_2$, and $\D_2$-$\D_3$ (dashed lines). 

We obtain the puncture field and effective source by adapting the \texttt{Mathematica} notebook developed in Ref.~\cite{Bourg2023} to the geometrical setup presented in Sec.~\ref{sec:GeomSetup}. The retarded field $\psi_{\ell, m}$ entering the boundary data in Eqs.~\eqref{eq:ExtData_lm_mode_minus}--\eqref{eq:ExtData_lm_mode_plus} is generated with the code from Ref.~\cite{PanossoMacedo:2022fdi}. While Ref.~\cite{PanossoMacedo:2022fdi} extensively discusses the accuracy of individual $(\ell, m)$ modes, this section benchmarks the numerical implementation of the analytical strategy introduced in Ref.~\cite{Bourg2023}. We concentrate our studies on the configuration $r_p=10M$, and we discuss the role played by the parameters $\eta$, $n_{\rm max}$ and $m$. 

\subsection{Boundary conditions for domain \texorpdfstring{$\D_1$}{D₁}: \texorpdfstring{\\}{} external retarded field from \texorpdfstring{$(\ell,m)$}{(ℓ,m)} modes}\label{sec:external data}

\begin{figure*}[t!]
	\centering
    \includegraphics[width=\textwidth]{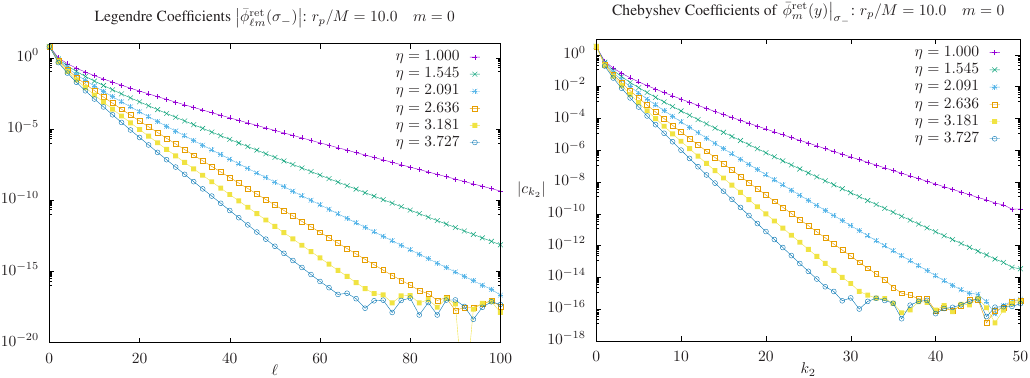}
	\caption{
	{\em Left Panel}: Hyperboloidal retarded field modes $\bar \phi_{\ell m}^{\rm ret}$ obtained with the code from Ref.~\cite{PanossoMacedo:2022fdi} to construct the boundary conditions \eqref{eq:ExtData_lm_mode_minus} for domain $\D_1$ at $\sigma_-$. For a fixed $m$, these modes are equivalent to the coefficients of a spectral decomposition in a Legendre polynomial basis. The position of the boundary $\sigma_-$ is controlled by the parameter $\eta$ according to Eq.~\eqref{eq:sigma_plusminus}, and the exponential decay rate improves as $\sigma_-$ moves further away from $\sigma_p$. {\em Right panel:} In the current code, the resulting boundary condition $\bar \phi^{(1)}_m(y)\bigr|_{\sigma_-}$ in Eq.~\eqref{eq:ExtData_lm_mode_minus} is decomposed into a Chebyshev polynomial basis. The corresponding Chebyshev coefficients also decay exponentially, with the decay rate improving for larger $\eta$. Here, $m=0$ and the particle is at $r_p=10 M$. The results are qualitatively the same at the boundary $\sigma_+$, and for different values of $m$ and $r_p$.
}
	\label{fig:bound_lm}
\end{figure*}

When solving the elliptic equation in domains $\D_1$ and $\D_2$, we need to impose boundary conditions at the surfaces $\sigma_\pm$, $y\in[0,1]$. These surfaces delimit the worldtube around the particle, and the boundary conditions communicate the information from the outer ($\D_0$) and inner ($\D_3$) source-free regions into the intermediate source-free region ($\D_1$). As discussed in Sec.~\ref{sec:ext_conditions}, Eqs.~\eqref{eq:ExtData_lm_mode_minus} and \eqref{eq:ExtData_lm_mode_plus} give the boundary conditions at $\sigma_\pm$, with the retarded fields $\psi^{\rm ret}_{\ell m}(r_{\mp})$ obtained from the available codes. We employ the hyperboloidal spectral solver for ODEs (ODE-HypSpecGSF code) introduced by Ref.~\cite{PanossoMacedo:2022fdi} to calculate $\psi^{\rm ret}_{\ell m}(r_{\mp})$.

The size of the worldtube should be large enough so that $\ell$-mode contributions decay sufficiently fast in Eqs.~\eqref{eq:ExtData_lm_mode_minus} and \eqref{eq:ExtData_lm_mode_plus}. At the same time, this size should not compromise the validity of the puncture and effective source approach within the particle's excision domain $\D_2$. In practical terms, the worldtube size should also adapt to a generic value of particle's orbit $r_p$ in order to avoid a tedious fine-tuning parametrisation exercise at each new orbital radius configuration.

In the ODE-HypSpecGSF, the worldtube boundaries $\sigma_\pm$ are free parameters. Ref.~\cite{PanossoMacedo:2022fdi} points out that a convenient choice satisfying the above requirements is to place the inner surface at
$\sigma_- =\sigma_p/2$, i.e. halfway between future null infinity ($\sigma=0$) and the particle ($\sigma=\sigma_p$). Similarly, the outer surface is fixed at $\sigma_+ = (1+\sigma_p)/2$, i.e. halfway between the particle $\sigma=\sigma_p$ and the black-hole horizon $\sigma=1$. 

In the current setup, $\sigma_-$ and $\sigma_+$ are not independent quantities. Rather, Eq.~\eqref{eq:sigma_plusminus} fixes their values in terms of the parameter $\eta$, which controls the size of the radius (in units of horizon length) delimiting the effective-source region. Here, we devise a strategy to fix $\eta$ in terms of $\sigma_p$ motivated by the one employed in Ref.~\cite{PanossoMacedo:2022fdi}.

For a given value of particle's orbital radius $r_p$, the worldtube size cannot exceed the $\sigma$-coordinate distance between the particle and the horizon. This constraint introduces a maximum value $\eta_{\rm max} = r_p \sqrt{f_p}/r_h$, achieved when the outer boundary is exactly at the horizon, $\sigma_+=1$. This bound prevents us from from placing the worldtube inner boundary at $\sigma_- = \sigma_p/2$ as suggested in Ref.~\cite{PanossoMacedo:2022fdi}. Indeed, this choice would yield $\eta_- = r_p/(r_h\,\sqrt{f_p}) > \eta_{\rm max}$, with the corresponding outer boundary $\left.\sigma_+\right|_{\eta_-}>1$ exceeding the horizon coordinate position. In fact, it is interesting to observe that $\left.\sigma_+\right|_{\eta_-} \rightarrow \infty$, i.e., the circle around the particle would extend all the way until the singularity at $r=0 \Leftrightarrow \sigma\rightarrow \infty$.

On the other hand, it is perfectly possible to fix the worldtube's outer boundary halfway between the particle at $\sigma=\sigma_p$ and the black-hole horizon at $\sigma=1$ by choosing 
\beq
\label{eq:eta_plus}
\sigma_+ = \dfrac{1+\sigma_p}{2} \Longleftrightarrow \eta_+ =  \dfrac{\sqrt{f_p}}{\sigma_p\left( 1 + \sigma_p \right)}.
\eeq
 With this choice the inner boundary assumes the value
\beq
\left. \sigma_-\right|_{\eta_+} = \dfrac{\sigma_p}{2 }\left(1 + \sigma_p \right).
\eeq
In this way, we can adjust the worldtube size directly from the particle's orbit, ensuring that the boundaries are placed sufficiently far from the particle for an efficient decay of the $\ell$-mode contributions.

To demonstrate this property, the left panel of Fig.~\ref{fig:bound_lm} shows the conformal retarded field at the inner boundary, $\bar \phi_{\ell m}(\sigma_-)$ for different values $\eta\in [1, \eta_+ ]$. The results were calculated with the code from Ref.~\cite{PanossoMacedo:2022fdi} for $r_p = 10 M$ and $m=0$. An exponential decay for $\bar \phi_{\ell m}(\sigma_-)$ is apparent in Fig.~\ref{fig:bound_lm} for all values of $\eta$. However, the decay rate for a small worldtube is very slow despite being exponential. Indeed, for $
\eta=1$, the retarded field is still of order $\bar \phi_{\ell m}(\sigma_-)\sim 10^{-8}$ at $\ell = 100$. For $\eta=\eta_+\sim 3.727$, on the other hand, machine round off error is achieved around $\bar \phi_{\ell m}(\sigma_-)\sim 10^{-18}$ for $\ell \sim 60$. At the outer boundary $\sigma_+$, the decay rates are better than at $\sigma_-$. As a rule of thumb, we observe that the modes $\bar \phi_{\ell m}(\sigma_+)$ decay at least twice as fast as $\bar \phi_{\ell m}(\sigma_-)$. Also, we have verified that these results are qualitatively the same for different $m$ modes and orbital radii as large as $r_p = 10^5 M$.

Recall that the coefficients $\bar \phi_{\ell m}(\sigma_-)$ result from a decomposition of the spacetime retarded field $\Phi$ into a spherical harmonic basis with the angular coordinates defined in the black-hole frame. For a given $m$-mode, this is nothing more then a spectral decomposition having the associated Legendre polynomials as basis functions. As for any spectral representation, the corresponding (Legendre) spectral coefficients should decay exponentially for analytic functions, as we indeed observe.  

When summing these coefficients over $\ell$, the boundary conditions in Eqs.~\eqref{eq:ExtData_lm_mode_minus} and \eqref{eq:ExtData_lm_mode_plus} reconstruct the angular dependence by taking into account not only the Legendre polynomial basis $P_\ell^m(y)$, but also a factor $\sim (1-y)^{|m|/2}$ coming from the regularisation at the symmetry axis $y=1$ in Eq.~\eqref{eq:phi_hyperboloidal_rescaling}. The current code then employs an overall Chebyshev spectral scheme, so the resulting boundary data $\bar \phi^{(1)}_m(y)\bigr|_{\sigma_\pm}$ is re-expanded in terms of the Chebyshev polynomial basis. The right panel of Fig.~\ref{fig:bound_lm} depicts the Chebyshev coefficients $c_{k_2}$ resulting from this re-expansion\footnote{\label{fn:appCheb}See Appendix \ref{sec:SpecApprox_error_conv} for a discussion of the Chebyshev representation of one-dimensional functions.}. 

Since the two sets of modes are just representations in two different polynomial bases, it comes as no surprise that the Chebyshev coefficients $c_{k_2}$ associated with the function $\bar \phi^{(1)}_m(y)\bigr|_{\sigma_\pm}$ behave in the same way as the corresponding Legendre coefficients $\bar \phi_{\ell m}(\sigma_-)$. The apparent faster decay of the Chebyshev coefficients in Fig.~\ref{fig:bound_lm} is only a scaling effect. Due to the symmetries of the problem, the non-vanishing $\bar \phi_{\ell m}(\sigma_-)$ modes occur when $\ell+m=$ even. For $m=0$, this property implies that we only have $\ell = 0, 2, 4, \ldots$ Even though the top panel of Fig.~\ref{fig:bound_lm} shows modes up to $\ell_{\rm max} = 100$, they amount only to a total of $50$ non-vanishing $\bar \phi_{\ell m}(\sigma_-)$ modes. The Chebyshev coefficients $c_{k_2}$, on the other hand, are present for all $k_2 = 0, 1, 2, 3 \ldots$. The bottom panel of Fig.~\ref{fig:bound_lm} shows modes up to $k_2^{\rm max} = 50$, equivalent to the total number of non-vanishing $\bar \phi_{\ell m}(\sigma_-)$ modes.

As a final remark, we comment that the resemblance between the Legendre and Chebyshev decompositions is a consequence of the particular choice of angular dependence of the function ${\cal Z}(\sigma,y)$ in Eq.~\eqref{eq:phi_hyperboloidal_rescaling}. As discussed in Sec.~\ref{sec:hyp_elliptic_eq}, this choice ensures that $\bar \phi_m(\sigma,y)=0$ when $y=1$ ($m\neq0$). An alternative re-scaling directly taking into account the behavior  in Eq. \eqref{eq:Axis_RegCond} is also possible, but it yields a worse behavior for the Chebyshev coefficients due to the presence of highly oscillating functions with steep gradients around $y=1$. We discuss this issue in Appendix \ref{app:LegChebPoly}.

\subsection{Domain transition \texorpdfstring{$\D_1$}{D₁}-\texorpdfstring{$\D_2$}{D₂}: the puncture field}\label{ref:results_punc_field}

As discussed in Sec.~\ref{sec:transition_conds} --- see Eqs.~\eqref{eq:TransitionCondition_D1_D2_field} and \eqref{eq:TransitionCondition_D1_D2_der} --- the puncture field $\bar \phi^{\mathcal{P}}_m$, together with its derivatives $\partial_\sigma \bar \phi^{\mathcal{P}}_m$ and $\partial_y \bar \phi^{\mathcal{P}}_m$, fixes the transition conditions at the interface between domains $\D_1$-$\D_2$. These transition conditions are defined along the surface $x^2=-1$ in domain $\D_1$ and $x^2=1$ in $\D_2$. Therefore, they are one-dimensional functions parameterised by the coordinate $x^1\in[-1,1]$, and the decay rate for the Chebyshev coefficients $c_{k_1}$ resulting from a spectral decomposition in the Chebyshev polynomials is a strong indication of the underlying function's regularity\footref{fn:appCheb}. 

\begin{figure}[t!]
	\centering
	\includegraphics[width=\columnwidth]{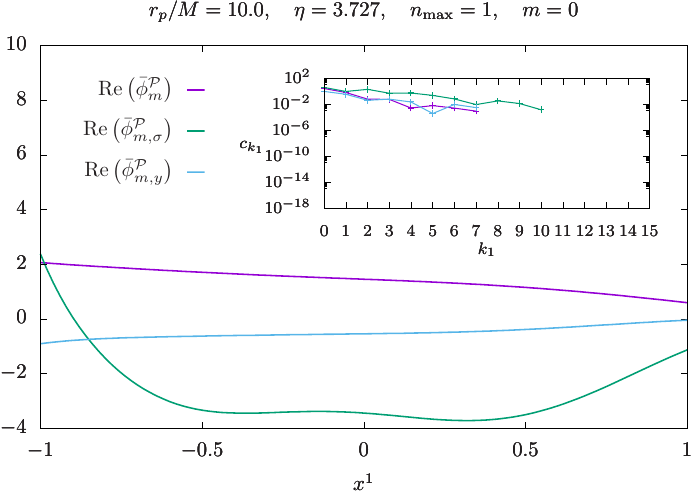}
	\caption{
The puncture field $\bar \phi^{\mathcal{P}}_m$ and its $\sigma$- and $y$-derivatives fix the boundary data at the domain interface $\D_1$-$\D_2$. These functions are parametrised by the coordinate $x^1\in[-1,1]$, and the inset shows their respective Chebyshev coefficients as a measure of the functions' accuracy and analytical properties. The vanishing of $c_{k_1}$ for $k_1>k_1^{\rm max}$ indicates the polynomial character of these functions in the coordinate $x^1$.
Here, the mode number is $m=0$ and the particle's orbital radius is $r_p = 10 M$. The algorithm's configuration has $n_{\rm max}=1$ and $\eta \approx 3.727$ --- cf.~\eqref{eq:eta_plus}.
}
	\label{fig:punc_data_interface}
\end{figure}

Figure \ref{fig:punc_data_interface}'s left panel shows an example for the set of boundary data for a configuration with $r_p=10M$, $\eta\approx 3.727$, $n_{\rm max}=1$, and $m=0$. The inset in the middle panel displays the corresponding Chebyshev coefficients for the functions ${\rm Re}(\bar \phi^{\mathcal{P}}_m)$ (purple), ${\rm Re}(\partial_\sigma\bar \phi^{\mathcal{P}}_m)$ (green) and ${\rm Re}(\partial_y\bar \phi^{\mathcal{P}}_m)$ (blue).

For this configuration, the Chebyshev coefficients vanish after an index $k_1^{\rm max}=7$ ($\bar \phi^{\mathcal{P}}_m$ and $\partial_y\bar \phi^{\mathcal{P}}_m$) or $k_1^{\rm max}=10$ ($\partial_\sigma\bar \phi^{\mathcal{P}}_m$), indicating that the functions are polynomials of order $7$ and $10$ in the coordinate $x^1$, respectively. We also observe the polynomial behavior for the mode $m=0$ at higher orders $n_{\rm max}>1$ for the puncture field. For higher angular modes $m>0$, however, the behavior is not directly observed because the Chebyshev coefficients decay exponentially until the numerical noise. 

The absence of a clear-cut observation for the polynomial structure via the Chebyshev coefficients at higher $m$-modes does not imply that the functions are no longer polynomials in the coordinate $x^1$. In fact, since we know the puncture analytically, we can confirm it is a polynomial in $x^1$. Our numerical investigation therefore suggests that the vanishing Chebyshev coefficients are undetectable due to machine roundoff error. To see why the puncture is a polynomial in $x^1$, recall from Eq.~\eqref{eq:x1_x2_PolarParticle} that 
the coordinate $x^1$ is proportional to the cosine of the polar angle $\pfphi$ in the particle's reference frame. Section~IVB in the companion paper~\cite{Bourg2023} shows that at fixed $\varrho$, the puncture is a  polynomial in $\cos\pfphi$ of order fixed by the parameters $n_{\rm max}$ and $m$.

\subsection{Domain \texorpdfstring{$\D_2$}{D₂}: the effective source}\label{sec:results_EffSource}
\begin{figure}[tb!]
	\centering
	\includegraphics[width=\columnwidth]{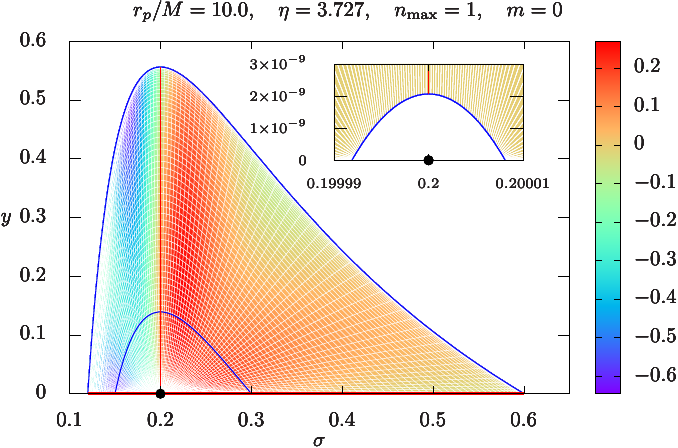}
    \caption{Effective source in the hyperboloidal coordinates $\{\sigma,y\}$ within domain $\D_2$ for $m=0$, $r_p = 10 M$, $n_{\rm max}=1$ and $\eta \approx 3.727$ --- cf.~\eqref{eq:eta_plus}. The inset zooms in on the region near the particle for a numerical grid with $N_2 = N_1 = 100$. Blue and red lines indicate, respectively, surfaces of spectral coordinates $x^1=$~constant. and $x^2=$~constant.} 
    \label{fig:Seff_cheb1}    
\end{figure}
\begin{figure}[tb!]
	\centering
	\includegraphics[width=\columnwidth]{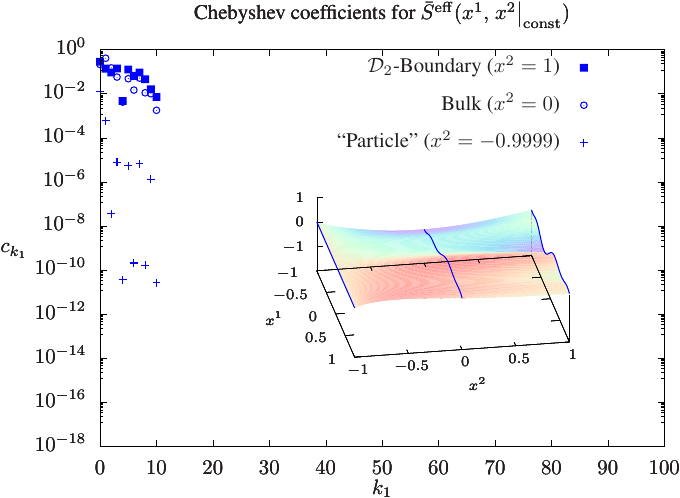}
    \caption{The effective source's Chebyshev coefficients $c_{k_1}$ for the same parameters as in Fig.~\ref{fig:Seff_cheb1}. The coefficients are given along the $x^1$-direction at the domain's boundary $x^2=1$ (blue squares), bulk $x^2=0$ (blue circles) and `at the particle' $x^2\approx -1$ (blue crosses); these lines are shown in blue in the inset, which displays the effective source in the spectral coordinates ${x^1,x^2}$. The vanishing of $c_{k_1}$ (within numerical roundoff error) for $k_1>k_1^{\rm max}$ indicates the polynomial character of these functions.} 
    \label{fig:Seff_cheb2}
\end{figure}
\begin{figure}[tb!] 
	\includegraphics[width=\columnwidth]{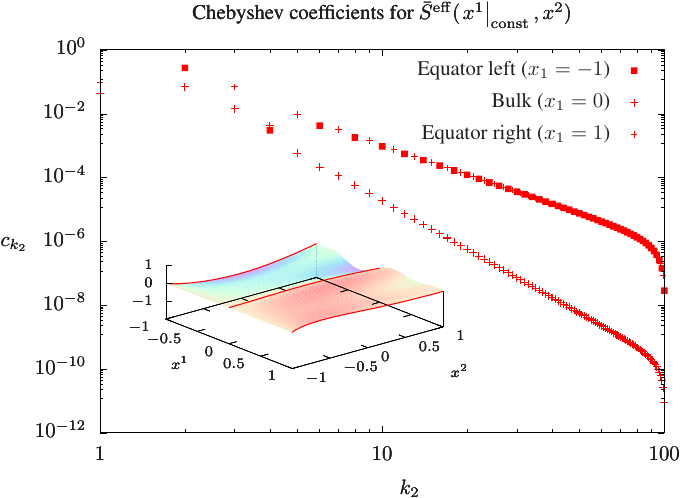}
\caption{The effective source's Chebyshev coefficients $c_{k_2}$ for the same parameters as in Fig.~\ref{fig:Seff_cheb1}. The coefficients are given along the $x^2$-direction on the equator to the left $x^1=-1$ (red squares) and right $x^1=1$ (red diamonds) of the particle, as well as in the domain's bulk $x^1=0$ (red circles); these lines are shown in red in the inset, which displays the effective source in the spectral coordinates ${x^1,x^2}$.  The coefficients' algebraic decay indicates a finite differentiability, with a degree of regularity controlled by the puncture's order $\bar{n}_{\rm max}$ (see Fig.~\ref{fig:Seff_regular_order}). 
}
	\label{fig:Seff_cheb3}
\end{figure}
We now turn our attention to the effective source, defined in the whole domain $\D_2$. Figure~\ref{fig:Seff_cheb1} displays the effective source $\bar S^{\rm eff}$ in the physical coordinates $\{\sigma, y\}$. As in the previous section, here the parameters are $r_p = 10 M$, $\eta \approx 3.727$, $n_{\rm max}=1$, and $m=0$. The puncture's order parameter $n_{\rm max}=1$ guarantees that the field is continuous at the particle. As explained in Secs.~\ref{sec:RegCond} and~\ref{sec:Chebyshev grids}, however, the algorithm does not calculate the effective source directly at the particle since the spectral grid does not include the point $(\sigma_p, 0)$. The figure's inset shows the grid point closest to the particle for a grid with $N_1= N_2 =100$, where we observe the small circumference with dimensions $\delta \sigma \sim 10^{-5}$ and $\delta y \sim 10^{-9}$ around the particle's location. 

To infer the effective source's regularity properties, we study its behavior along lines of constant coordinates $x^1$ and $x^2$. The blue lines, given by $x^2=$~constant, correspond to level sets of constant radius in the particle's frame. The red lines correspond to $x^1=$~constant, describing a surface of constant polar angle $\pfphi$ in the particle's reference frame; see Eqs.~\eqref{eq:ParticlePolarCoordinates} and \eqref{eq:x1_x2_PolarParticle}. We display those lines in the $\{\sigma,y\}$-plane in Fig.~\ref{fig:Seff_cheb1} and also in the $\{x^1, x^2\}$ parameterization of the effective source in the insets in Figs.~\ref{fig:Seff_cheb2} and \ref{fig:Seff_cheb3}.

We next study the spectral representation in the Chebyshev basis along lines $x^1=$~constant and $x^2=$~constant\footref{fn:appCheb}. Figure~\ref{fig:Seff_cheb2} depicts the coefficients $c_{k_1}$ along the lines $x^2=1$, $x^2=0$ and $x^2\approx -1$, corresponding, respectively, to the transition surface into domain ${\D_1}$, a surface in the bulk and the grid surface closest to the particle. As anticipated by the results for the puncture field, we observe a polynomial behavior of the functions $\bar{\cal S}^{\rm eff}(x^1, \left.x^2\right|_{\rm const})$, inferred from the vanishing of the coefficients $c_{k_1}$ for a given $k_1>k_1^{\rm max}$.

Figure~\ref{fig:Seff_cheb3} shows the coefficients $c_{k_2}$, associated with the function $\bar{\cal S}^{\rm eff}(\left.x^1\right|_{\rm const}, x^2)$, along the lines $x^1=0$, $x^1=1$ and $x^1=-1$. Each curve intersects the particle at $(\sigma,y)=(\sigma_p,0) \Leftrightarrow x^2=-1$ from a different direction: along the equator $y=0$ to the left ($x^1=-1$) or to the right ($x^1=1$) of the particle, and vertical $\delta r = 0$ to the particle ($x^1=0$). Despite $\bar S^{\rm eff}$ being continuous at the particle, the algebraic decay $c_{k_2} \sim k_2 ^\varkappa$ evinces the finite differentiability of $\bar S^{\rm eff}$. The order of differentiability is controlled by the parameter $n_{\rm max}$; its effect on the effective source is discussed in the next section.

\subsubsection{Puncture order \texorpdfstring{$n_{\rm max}$}{n{\_}max}}
\begin{figure*}[t!]
	\centering
    \includegraphics[width=\textwidth]{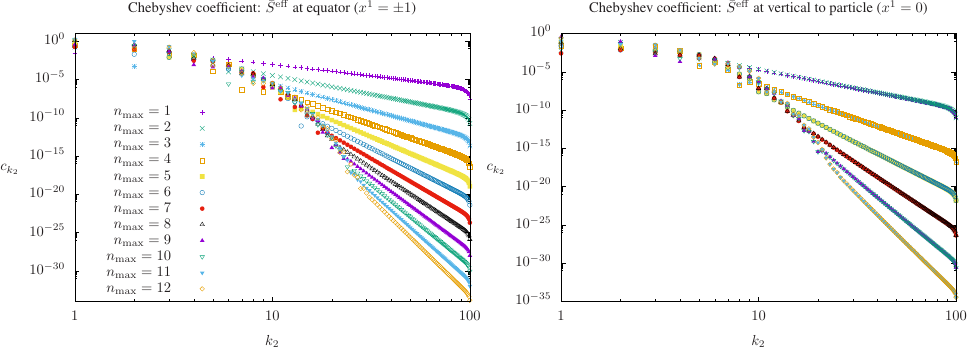}
	\caption{
 Decay rates of the effective source's Chebyshev coefficients for different puncture orders $n_{\rm max}$. The decay rates are associated with the function's regularity class as shown in Table~\ref{tab:reg_order_nbarmax}.   {\em Left Panel:} For $x^1\neq 0$, e.g. along the equator $x^1=\pm1$, the regularity class increases with each order $n_{\rm max}$. {\em Right Panel:} vertical to the particle $\delta r = 0$ ($x^1 = 0$), the regularity class increases every second order in $n_{\rm max}$. The results agree with the theoretical predictions from Ref.~\cite{Bourg2023}. Results are for mode number $m=0$, orbital radius $r_p = 10 M$, and $\eta \approx 3.727$, but the qualitative behavior is independent of these parameters.  
}
	\label{fig:Seff_regular_order}
\end{figure*}
\begin{table}[bt!]
	\begin{tabular}{c @{\quad} c @{\quad} c } 
        \toprule
		 & \multicolumn{2}{c}{ $(\varkappa;  \kappa)$}  \\
		    \cmidrule{2-3}
		$n_{\rm max}$ & $x^1\neq 0$ & $x^1=0$   \\ 
		\midrule
		$1$ & (3;1) &(5;2)\\
		$2$ & (5;2)&(5;2) \\ 
		$3$ & (7;3)& (9;4)\\
		$4$ & (9;4)& (9;4)\\
		$5$ & (11;5)&(13;6) \\
		$6$ & (13;6)& (13;6)\\ 
		$7$ & (15;7)&(17;8) \\ 
		$8$ & (17;8)& (17;8)\\
		$9$ & (19;9)& (21;10)\\ 
		$10$ & (21;10)& (21;10)\\
		$11$ & (23;11)& (25;12)\\
		$12$ & (25;12)& (25;12)\\
        \bottomrule
	\end{tabular}
    \caption{Effective source regularity properties. A singularity of order $\sim \varrho^{\kappa} \log \varrho$ as $\varrho \rightarrow 0$, cf.~\eqref{eq:ParticlePolarCoordinates2}, is inferred via a Chebyshev coefficient decay rate $c_i\sim i^{-\varkappa}$, with $\varkappa = 2\kappa+1$. The decay rates are independent of the mode number $m$.}
	\label{tab:reg_order_nbarmax}
\end{table}

As discussed in the previous section, the algebraic decay of the Chebyshev coefficients $c_{k_2}$ is a strong indication of the effective source's singular behavior at the particle. Here, we exploit the dependence of the coefficients' decay rate on the parameter $n_{\rm max}$ to infer the effective source's regularity class and compare the results against the analytical predictions from Ref.~\cite{Bourg2023}.

Figure~\ref{fig:Seff_regular_order} displays the effects of the parameter $n_{\rm max}$ on the Chebyshev coefficients. As expected, the coefficients' decay rates improve as one increases $n_{\rm max}$. However, one observes two distinct features: if one approach the particle along a direction ($x^1\neq 0$), the decay rate improves at each higher $n_{\rm max}$ order (left panel), whereas along a direction vertical to particle ($x^1=0$), the decay rate improve every two orders in $n_{\rm max}$ (right panel).

This property is in accordance with the analytical predictions from Ref.~\cite{Bourg2023}. Besides, the decay rates from Fig.~\ref{fig:Seff_regular_order} provide an important cross check between the numerical and the theoretical results. Table \ref{tab:reg_order_nbarmax} lists pairs of coefficients $(\varkappa; \kappa)$ inferred from Chebyshev coefficients. The former accounts for the coefficients' decay rate $c_{k_2} \sim k_{2}^{-\varkappa}$, from which we infer the effective source's singular behavior $~\varrho^\kappa \log \varrho$ as $\varrho \rightarrow 0$, with $\varrho$ given by eq.~\eqref{eq:ParticlePolarCoordinates2}, along different directions. The results agree with the analytical predictions~\cite{Bourg2023}, and they are independent of the angular mode $m$.

\section{Results}\label{sec:NumericalSolutions}

This section reports on the performance of the $m$-mode spectral solver. First, we benchmark the values for the scalar self-force obtained with our solver against known results in the literature. In particular, we concentrate on a particle at $r_p=10 M$. The values for the self-force were obtained with modes in the range $m\in[0,40]$. The results employ spectral truncation parameters $N_2 = N_1 = 100$, corresponding to our highest resolution used as reference for the convergence tests. Then, we discuss global properties of the solution and elaborate on the solver's efficiency by performing converge and timing tests. We end with a detailed forecast of the method's efficiency in the second-order self-force problem.

\subsection{Self-force}
\begin{figure*}[tb!]
	\centering
    \includegraphics[width=\textwidth]{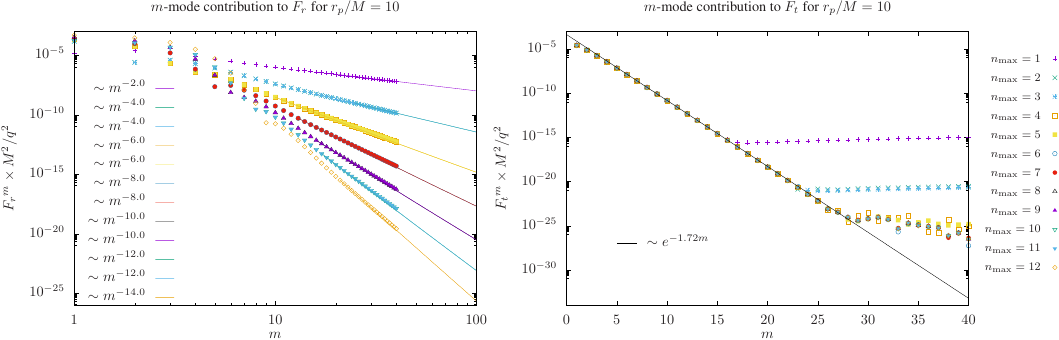}	
	\caption{Contributions from individual $m$-modes to the time (right panel) and radial component (left panel) of the scalar self-force. {\em Right Panel:} $F_t^m$ decays exponentially with respect to the azimuthal number $m$, and its values are independent of the puncture order $n_{\rm max}$. High-order punctures decrease the saturation limit for $F_t^m$. {\em Left Panel:} $F_r^m$ decays algebraically, with the decay rate increasing by two orders at every second order in $n_{\rm max}$. A fit for the large-$m$ behavior around $m\sim[15,40]$ reproduces the predictions from Ref.~\cite{Dolan:2010mt} and allows us to extrapolate the values of $F_r^m$ for $m\sim[0,100]$.} 
	\label{fig:Ft}
\end{figure*}
\begin{table*}[tb!]
	\begin{tabular}{c @{\quad} c c @{\quad} c c @{\quad}| @{\quad} c c @{\quad} c c} 
		\toprule
					   &       \multicolumn{4}{c}{Numerical}    & \multicolumn{4}{c}{High-$m$ fit}   \\
		\cmidrule{2-9}
		$n_{\rm max}$  &  $m\in[0,20]$ & (Rel. Error)  &  $m\in[0,40]$ & (Rel. Error)  & $m\in[0,100]$  &  (Rel. Error) & $m\in[0,1000]$  &  (Rel. Error)  \\
		\midrule
		$1$  & ${\color{gray}0.8777115480 }$	&  $(4\times 10^{-1})$    & $1.{\color{gray}1261013163 }$	       & $(2\times 10^{-1})$	 	& 	$1.{\color{gray}2782692807 }$	& 	$(7\times 10^{-2})$	&	$1.37{\color{gray}22512504 }$ 	& $(4\times 10^{-3})$		\\
		$2$  & $1.37{\color{gray}70545545}$	& $(1\times 10^{-3})$	&$1.378{\color{gray}2686923 }$	       & $(1\times 10^{-4})$	&  	$1.3784{\color{gray}363344 }$	& 	$(9\times 10^{-6})$		 &	$1.378448{\color{gray}1459}$	&  $(8\times 10^{-8})$		\\
		$3$  & $1.37{\color{gray}70547701 }$	& $(1\times 10^{-3})$	&$1.378{\color{gray}2691210 }$	       & $(1\times 10^{-4})$	&  	$1.3784{\color{gray}370398 }$	& 	$(8\times 10^{-6})$		  &     $1.378448{\color{gray}9523}$	 & $(5\times 10^{-7})$		\\
		\midrule
		$4$  & $1.3784{\color{gray}352533 }$ & $(9\times 10^{-6})$	&$1.37844{\color{gray}78364 }$	       & $(3\times 10^{-7})$	&  	$1.378448{\color{gray}1410 }$	& 	$(8\times 10^{-8})$	     	   &	$1.378448{\color{gray}1438}$   	& $(8\times 10^{-8})$		\\
		$5$  & $1.3784{\color{gray}352531 }$	& $(2\times 10^{-6})$	& $1.37844{\color{gray}78362 }$	       & $(3\times 10^{-7})$	&  	$1.378448{\color{gray}1408 }$	& 	$(8\times 10^{-8})$	     	   &	$1.378448{\color{gray}1436}$  	& $(8\times 10^{-8})$		\\
		$6$  & $1.37844 {\color{gray}79362 }$	& $(2\times 10^{-7})$	& $1.37844825{\color{gray}37}$	       & $(3\times 10^{-9})$	&  	$1.37844825{\color{gray}58}$	& 	$(1\times 10^{-9})$	   &   $1.37844825{\color{gray}58}$	&$(1\times 10^{-9})$		\\
		\midrule
		$7$  & $1.37844{\color{gray}79353 }$	& $(2\times 10^{-7})$	& $1.37844825{\color{gray}28 }$	       & $(3\times 10^{-9})$	&  	$1.37844825{\color{gray}48}$	& 	$(2\times 10^{-9})$	     	  &   $1.37844825{\color{gray}48}$	& $(2\times 10^{-9})$	\\
		$8$  & $1.3784482{\color{gray}485 }$	& $(7\times 10^{-9})$	& $1.37844825{\color{gray}96 }$	       & $(1\times 10^{-9})$	&  	$1.37844825{\color{gray}96 }$	& 	$(1\times 10^{-9})$	     	&  $1.3784482{\color{gray}596}$	 	& $(1\times 10^{-9})$		\\
		$9$  & $1.37844825{\color{gray}19 }$	& $(4\times 10^{-9})$	& $1.3784482{\color{gray}630 }$	       & $(4\times 10^{-9})$	&  	$1.3784482{\color{gray}631 }$     & 	$(4\times 10^{-9})$	     	& $1.3784482{\color{gray}631}$	 	& $(4\times 10^{-9})$		\\
		\midrule
		$10$  & $1.37844825{\color{gray}68 }$	& $(6\times 10^{-10})$	& $1.378448257{\color{gray}6 }$	       & $(6\times 10^{-11})$ & 	$1.378448257{\color{gray}7 }$ 	&	$(6\times 10^{-11})$	  & $1.378448257{\color{gray}7}$	& $(6\times 10^{-11})$ 		\\
		$11$  & $1.37844825{\color{gray}86 }$	& $(7\times 10^{-10})$	& $1.37844825{\color{gray}94}$	       & $(1\times 10^{-9})$	& 	$1.37844825{\color{gray}95 }$ 	&	$(1\times 10^{-9})$	  & $1.37844825{\color{gray}94}$	& $(1\times 10^{-9})$	\\
		$12$  & 1.3784482${\color{gray}467 }$	& $(8\times 10^{-9})$	& $1.3784482{\color{gray}467 }$	       & $(8\times 10^{-9})$	& 	$1.3784482{\color{gray}468 }$ 	&	$(8\times 10^{-9})$	   & $1.3784482{\color{gray}468}$	 & $(8\times 10^{-9})$									\\
		\bottomrule
	\end{tabular}
	\caption{Comparison of the radial self-force $F_r \times 10^5\, M^2/q^2$, as computed in this paper, against the highly accurate value $F^{\rm self}_r=1.3784482575667959\times 10^{-5}q^2/M^2$~\cite{Heffernan:2012su}.}
	\label{tab:Fr_values}
\end{table*}
The scalar self-force is obtained from the residual field via
\beq
F^{\rm self}_a =  \lim_{x^b\rightarrow x^b_p} q \nabla_a \Phi^{\mathcal{R}}(x^b).
\eeq
With the $m$-mode decomposition from Eq.~\eqref{eq:Freq_m_decomposition_2}, it follows for the self-force's time component
\beq
\label{eq:Ft_m}
F^{\rm self}_t = q \sum_{m=1}^\infty F_t^m, \quad F_t^m = 2 \omega_m\,  {\rm Im}\left. \bigg( \phi^{\mathcal{R}}_m \bigg)\right|_{\substack{r=r_p, \\\theta=\pi/2}}.
\eeq
The self-force's radial component reads
\beq
\label{eq:Fr_m}
F^{\rm self}_r = q \sum_{m=1}^\infty F_r^m, \quad F_r^m = \left(2-\delta_{m,0}\right) \left. {\rm Re} \bigg( \partial_r\phi^{\mathcal{R}}_m \bigg)\right |_{\substack{r=r_p\\\theta=\pi/2}}.
\eeq

The elliptic solver provides the conformal field $\bar \phi^{\mathcal{R}}(\sigma, y)$. To calculate the self-force components \eqref{eq:Ft_m} and \eqref{eq:Fr_m}, one must first map the spectral solution into the physical fields via Eq.~\eqref{eq:phi_hyperboloidal_rescaling}. In particular, the $r$-derivative follows from
\bea
\partial_r\phi^{\mathcal{R}}_m\bigg|_{\substack{r=r_p,\\ \theta=\pi/2}} &=& \left.-\dfrac{r_{\rm h}}{r_p^2}{\cal Z}\, \bigg( \partial_\sigma \bar \phi^{\mathcal{R}}_m + \bar \phi^{\mathcal{R}}_m \partial_\sigma \ln{\cal Z}   \bigg) \right|_{\substack{{\sigma=\sigma_p,}\\{y=0}}  } \nn \\
\partial_\sigma \ln{\cal Z}\bigg|_{\substack{{\sigma=\sigma_p,}\\y=0}  } &=& \dfrac{1}{\sigma_p} + s_m H'(\sigma_p).
\eea
Besides, as discussed in Sec.~\ref{sec:Chebyshev grids}, the spectral grid in the domain $\D_2$ does not include the particle location at the coordinate value $x^2=-1$. Thus, one must also interpolate the solution and its $\sigma$-derivatives into the particle's location to calculate $F_t^m$ and $F_r^m$.

The right panel of Fig.~\ref{fig:Ft} displays the results for the $m$-mode contributions to the time component of the scalar self-force. As expected, the mode contributions $F_t^m$ decay exponentially with $m$, and their values are independent of the puncture order $n_{\rm max}$; this is because the dissipative pieces of the self-force are only sensitive to the radiative piece of the scalar field, which is smooth at the particle. The parameter $n_{\rm max}$, however, does show an indirect impact in the solution. For a fixed numerical resolution (here, $N_1 = N_2 = 100$), higher puncture orders $n_{\rm max}$ decrease the saturation floor for the values $F_t^m$. Indeed, with $n_{\rm max} = 1$, the saturation occurs for $m\approx 15$ at order $F_t^m \sim 10^{-15}$. Then, $n_{\rm max} = 2$ and $n_{\rm max} = 3$ reduces 5 orders of magnitude in the saturation floor, allowing us to achieve $F_t^m \sim 10^{-20}$ at $m\approx 20$. For $n_{\rm max} \geq 4$,  $F_t^m$ stagnates at order  $\sim 10^{-25}$ at $m\approx 25$.

The left panel of Fig.~\ref{fig:Ft} depicts the $m$-mode contributions to the radial component of the scalar self-force. We observe the expected algebraic decay for $F_r^m ~ \sim m^{-\nu}$, with the decay rate $\nu$'s dependence on the puncture order $n_{\rm max}$ following the predictions from Ref.~\cite{Dolan:2010mt}: $\nu$ increases by two at every second order in $n_{\rm max}$. Indeed, by fitting a large-$m$ behavior in the range $m\sim[15,40]$, we obtain decay rates starting at $\nu=2$ for $n_{\rm max}=1$, increasing to $\nu=4$ for $n_{\rm max}=2,3$, and going up to $\nu=14$ for $n_{\rm max}=12$. 

Finally, Table \ref{tab:Fr_values} shows the total value for the radial self-force, obtained after summing over the $m$-mode contributions. We display the number of significant digits (black) and relative error by comparing our results against the high-accurate value reported in Ref.~\cite{Heffernan:2012su}. First, we consider the sum over the numerical values for $m\in[0,20]$ and $m\in[0,40]$. For punctures orders in the range $n_{\rm max} \in [1,7] $, the inclusion of higher $m$-modes enhances the results by a few orders of magnitude as expected. However, we observe a saturation in the number of significant digits for puncture orders $n_{\rm max} \geq 8$. Then, we extrapolate the data with the help of a large-$m$ fit and calculate $F_r$ by summing over $m\in[0,100]$ and $m\in[0,1000]$. This strategy increases the accuracy of the radial self-force for punctures with order $n_{\rm max} \in [1,7]$.

\subsection{Global solution}
\begin{figure}[tb!]
	\centering
	\includegraphics[width=\columnwidth]{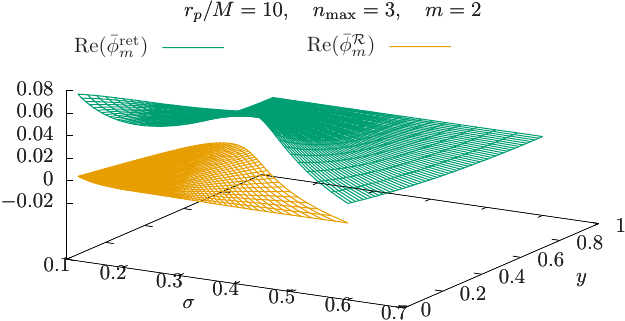}
	\caption{Global solution obtained by the elliptic hyperboloidal $m$-mode solver, with retarded field $\bar \phi^{\rm ret}(\sigma, y)$ in $\D_1$ and the residual field $\bar \phi^{\mathcal{R}}(\sigma, y)$ in $\D_2$ for $r_p=10M$, $n_{\rm max} = 3$ and $m=2$}
	\label{fig:Solution3D}
\end{figure}

We now turn our attention to the global solution in domains $\D_1$ and $\D_2$. As a representative example, Fig.~\ref{fig:Solution3D} shows the real part of the retarded field $\bar \phi^{\rm ret}(\sigma, y)$ in $\D_1$ (green) and residual field $\bar \phi^{\mathcal{R}}(\sigma, y)$ in $\D_2$ (yellow) for $r_p=10M$, $n_{\rm max} = 3$ and $m=2$. The discontinuity in the boundary $\D_1$-$\D_2$ determined by the puncture field is evident.

\begin{figure*}[tb!]
	\centering	
    \includegraphics[width=\textwidth]{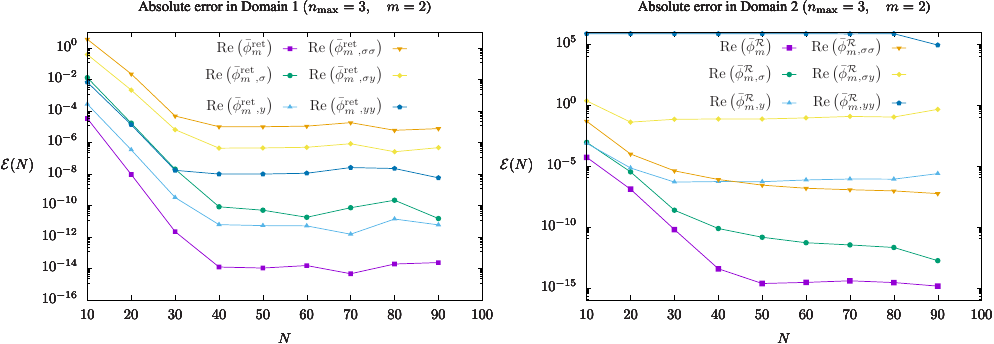}
	\caption{
	Absolute error ${\cal E}(N)$ convergence tests for $m=2$. {\em Left Panel:} The retarded field $\bar \phi^{\rm ret}(\sigma, y)$ in domain $\D_1$ exhibits exponential convergence, which does not depend on the puncture order. Accuracy is lost when calculating higher derivatives. {\em Right Panel:} With a moderate puncture order $n_{\rm max}=3$, the residual field $\bar \phi^{\mathcal{R}}(\sigma, y)$ in domain $\D_2$ quickly converges to machine roundoff error. The interpolation of higher $y-$derivatives are ill-resolved at the particle's position, as their regularity conditions \eqref{eq:cond_first_der_implications} and \eqref{eq:cond_second_der} involve up to the $5^{\rm th}$ derivatives along the spectral coordinates $(x^1, x^2)$.
}
	\label{fig:Error_doms}
\end{figure*}
We next perform convergence tests for the fields and their first and second $(\sigma,y)$-derivatives based on the infinity-norm error $\cal E$ defined in Eq.~\eqref{eq:error_infintynorm}. Apart from providing an estimate of the numerical systematic errors, this measure also allows us to infer regularity properties of the underlying solution.

Figure~\ref{fig:Error_doms} displays convergence tests for the retarded field $\bar \phi^{\rm ret}(\sigma, y)$ in domain $\D_1$ (left panel) and the residual field $\bar \phi^{\mathcal{R}}(\sigma, y)$ in domain $\D_2$ (right panel), with parameters $m=2$ and $n_{\rm max} = 3$. In the left panel, we observe the expected exponential convergence for the retarded field and its derivatives. Typical for spectral solvers, the absolute error for $\bar \phi^{\rm ret}$ saturates around the numerical roundoff  error for double float operations $\sim 10^{-14}$. We lose 3 to 4 orders of magnitudes at each higher derivative, with first and second derivatives saturating around $\sim 10^{-11}-10^{-10}$ and $\sim 10^{-8}-10^{-4}$. 

\begin{figure}[tb!]
	\centering
    \includegraphics[width=\columnwidth]{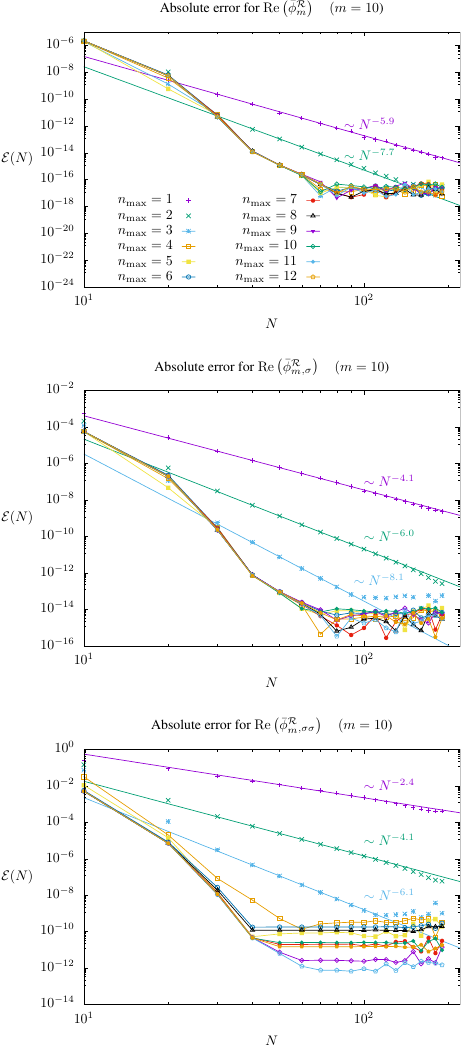}
	\caption{Puncture's order $n_{\rm max}$ influence on error measures for residual field $\bar \phi^{\mathcal{R}}(\sigma, y)$ (top panel), its first (middle panel) and second (bottom panel) $\sigma$-derivatives. The algebraic decay ${\cal E}(N)\sim N^{-\tilde \varkappa}$ is a consequence of the fields' singular behavior. The convergence rate $\tilde \varkappa$ agrees with the theoretical predictions~\cite{Bourg2023}, cf. Eqs.~\eqref{eq:converge_rate_phi_sigsig}--\eqref{eq:converge_rate_phi}. Example for $m=10$.
}
	\label{fig:Error_nbarDep}
\end{figure}

Accuracy loss is usually expected when calculating higher derivatives. However, our global solver is affected by this issue in a stronger way due to the delicate regularity conditions at the particle. Indeed, the phenomenon is even more prominent in domain $\D_2$ (right panel). Even though  the error associated with $\bar \phi^{\mathcal{R}}(\sigma, y)$ quickly drops to the numerical roundoff  error  $ \sim10^{-15}-10^{-14}$, one loses several order of magnitudes when calculating higher derivatives. Particularly problematic are the $y$-derivatives at the particle, which are responsible for the worst error measures.

Such high errors originate from the interpolation of the $y$-derivatives to the particle position. An accurate spectral interpolation to the particle relies on the data evaluated at the grid point closest to the particle, $x^2 \approx -0.999$, and such a calculation involves a delicate division of small numbers in Eqs.~\eqref{eq:map_dsigma}-\eqref{eq:map_d2sigmay}. Indeed, a regular limit to the particle position, $x^2\rightarrow -1$, requires the validity of the regularity conditions \eqref{eq:cond_first_der_implications} and \eqref{eq:cond_second_der}.

Finally, we study the role played by the puncture order $n_{\rm max}$ in the convergence rates. Figure~\ref{fig:Error_nbarDep} focuses on the error in domain $\D_2$, since the solution for the retarded field in domain $\D_1$ is not affected by the puncture's order. As a representative examples, we show results for $m=10$. The top, middle and bottom panels display, respectively, the convergence plot for the residual field $\bar \phi^{\mathcal{R}}(\sigma, y)$, its first and second $\sigma$-derivative. The algebraic decay ${\cal E}(N) \sim N^{-\tilde \varkappa}$ due to the residual field's singular structure is evident, specially for low puncture orders $n_{\rm max} = 1 \sim 3$, with decay rates in accordance with the theoretical predictions~\cite{Bourg2023}.

Indeed, as discussed in Appendix \ref{sec:app_convergence}, the error's convergence rate $\tilde \varkappa$ allows us to infer the regularity class of the underlying field.  Section~\ref{sec:results_EffSource} verified the singular behavior $\bar S_{\rm eff} \sim |\delta r|^{n_{\rm max}} \ln |\delta r|$ as one approaches the particle $\delta r \rightarrow 0$ along the equator. Thus, one expects 
\begin{align}
\label{eq:converge_rate_phi_sigsig}
\bar \phi^{\mathcal{R}}_m{}_{,\sigma \sigma} \sim |\delta r|^{n_{\rm max}} \ln |\delta r| \Rightarrow {\cal E}(N) &\sim N^{-2n_{\rm max}} , \\
\label{eq:converge_rate_phi_sig}
\bar \phi^{\mathcal{R}}_m{}_{,\sigma} \sim |\delta r|^{n_{\rm max}+1} \ln |\delta r| \Rightarrow {\cal E}(N) &\sim N^{-2(n_{\rm max}+1)},\\
\label{eq:converge_rate_phi}
\bar \phi^{\mathcal{R}}_m \sim |\delta r|^{n_{\rm max}+2} \ln |\delta r| \Rightarrow  {\cal E}(N) &\sim N^{-2(n_{\rm max}+2)},
\end{align}
which agrees with the results in Fig.~\ref{fig:Error_nbarDep}.

\subsection{Timing}

\begin{figure*}[ht!]
	\centering	
    \includegraphics[width=\textwidth]{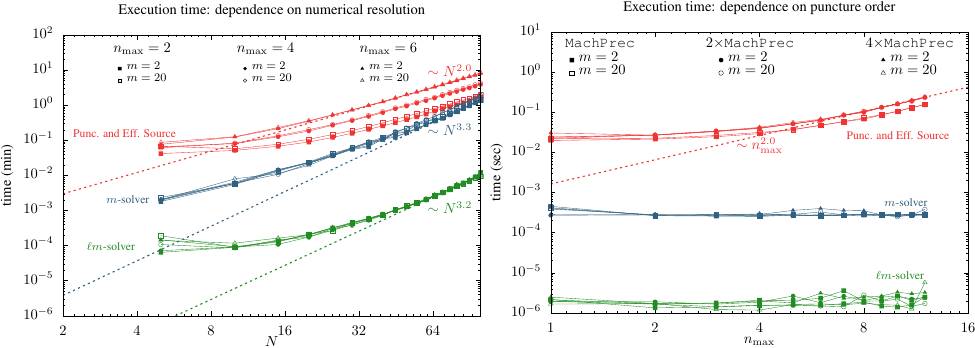}	
	\caption{CPU runtime for the  components in our scheme: evaluation of puncture field and effective source (red), $(\ell,m)$-mode at world tube (green) and $m$-mode spectral solver (blue). \emph{Left panel}: Dependence on the spectral resolution $N$. Effective source calculation scales as $\sim N^2$ with offset depending on puncture order $n_{\rm max}$, whereas $(\ell,m)$-mode and $m$-mode solver scale as $\sim N^3$ and are independent of the puncture order. 
 \emph{Right panel}: Dependence on the puncture order $n_{\rm max}$. The execution time to calculate the puncture and effective source at a single grid point scales as $n_{\rm max}^2$. Evaluation time is independent of $m$-mode.}
	\label{fig:ExectutionTime}
\end{figure*}

An $m$-mode scheme, because it requires solving PDEs rather than ODES, is inherently more expensive than schemes based on an $\ell m$-mode decomposition. In this section, we assess the execution time of our scheme. 

We can divide the timing into two main components: the calculation of the puncture field and effective source, and the application of the $m-$ mode spectral solver. In the current setup, a third component is also required: the calculation of $(\ell,m)$-modes at the world tube boundary. 

Figure~\ref{fig:ExectutionTime} plots the total execution time for each of these components: results for the calculation of the puncture field and effective source are displayed in red, for $(\ell,m)$-modes in green and $m$-mode spectral solver in blue. The left panel shows CPU time (in minutes) dependence on the spectral resolution. For that purpose, we set the underlying numerical truncation parameters in all three codes to the same value $N$. The right panel brings the CPU time (in seconds) dependence upon the the puncture order $n_{\rm max}$. 

In the left panel of Fig.~\ref{fig:ExectutionTime}, we observe the scaling $\sim N^3$ for the calculation of an individual $(\ell,m)$-mode. This scaling is expected, as it reflects the properties of the LU decomposition algorithm employed in the code ODE-HypSpecGSF \cite{PanossoMacedo:2022fdi}. If the same algorithm were also employed in the $m-$ mode spectral solver, the scaling would assume the prohibitive rate of $\sim N^6$, but the use of the iterative BICGStab with finite difference pre-conditioner reduces (see Sec.~\ref{sec:NumericalMethods}) the time scaling back to $\sim N^3$. These results are independent of the puncture order $n_{\rm max}$ and $m-$mode, as depicted in the figure for the examples 
$(n_{\rm max}, m)=(2,2)$ (solid square), $(n_{\rm max}, m)=(2,20)$ (hollow square), $(n_{\rm max}, m)=(4,2)$ (solid circle); $(n_{\rm max}, m)=(4,20)$ (hollow circle); $(n_{\rm max}, m)=(6,2)$ (solid triangle) and $(n_{\rm max}, m)=(6,20)$ (hollow triangle).

Finally, the calculation of the puncture field and effective source scales as $\sim N^2$ by construction, as it requires the evaluation of the effective source at each point of the the 2-dimensional spectral grid $\{x^1_{i_1},x^2_{i_2}\}$, $i_1,i_2 = 0  \cdots N$. Even though the $m$-mode spectral solver has a higher scaling rate than the calculation of the puncture field and effective source, it is more efficient than them for small to moderate values of $N$. Besides, though the puncture and effective source evaluation time does not depend on the particular $m$ mode, it does become more expensive as the puncture order $n_{\rm max}$ increases.

The right panel of Fig.~\ref{fig:ExectutionTime} shows the dependence of the puncture and effective source evaluation time with respect to the puncture order $n_{\rm max}$ at a given grid point, and we observe the behavior $\sim n_{\rm max} ^2$. Recall that calculations near the particle involve the division of small numbers. To avoid losing significant digits in such calculations, it is require to increase in the internal precision for the floating operations. These evaluations were performed in \texttt{Mathematica}, and the figure display the time for internal precisions set to \texttt{MachinePrecision}, $2\times$\texttt{MachinePrecision} and $4\times$\texttt{MachinePrecision}. For comparison, the figure also displays the timing for the $(\ell,m)$-mode calculation and the $m$-mode spectral solver per ``unit of grid points $N$'', i.e., dividing by their scaling $N^3$. Apart from confirming that these timings are independent of the puncture order $n_{\rm max}$, it is also evident that the puncture and effective source evaluation is the most expensive element of the calculation. However, we note that the cost associated with the beyond-machine-precision calculations can likely be avoided by re-expanding the puncture in a small neighbourhood of the particle, which would lead to analytical rather than numerical cancellations; this strategy has been employed in prior $m$-mode schemes~\cite{Dolan:2010mt,Dolan:2011dx,Dolan:2012jg,Thornburg:2016msc}.

All in all, the total CPU time will scale as
\begin{align}
t_{\rm CPU}[{\rm sec}] &\approx 10^{-3}\, m_{\rm max}\, N^2\bigl( n_{\rm max}  + 0.5 N + 10^{-3} \ell_{\rm max}N \bigr)\nn\\
&\quad +t_o
\end{align}
with $t_o \sim 10^{-2}$. The wall time, however, can be significantly improved by parallelisation. The $m$-mode calculations are independent of each other. For a given $m$, the evaluation of each $(\ell,m)$-mode out of a total $\ell_{\rm max}$ is also independent of the others, and so is the calculation of the puncture and effective source at a fixed grid point.

The analysis discussed here does not account for the underlying accuracy obtained at a given numerical resolution $N$. If the same accuracy is required for each mode, then higher modes are more expensive than low modes. However, we can typically accept significantly lower accuracy in higher modes because their contribution to any given quantity decays rapidly with $m$. This means we can consider that the expense per mode is roughly independent of the mode number. 

The next section provides an estimate to relate the required accuracy with the numerical resolution and the respective total computational time by introducing a toy-model study for second order self-force calculations.

\subsection{Calculations at second order: a toy-model exercise}
An elliptic $m$-mode solver such as the one developed here can be used for self-force calculations at both first and second order. The elliptic operator on the left-hand side of the field equations is constructed out of the background metric, and it remains the same at all orders. The only change in moving from first  to second order is that a different source appears on the right-hand side of the elliptic equation (boundary conditions can also potentially change~\cite{Pound:2015wva}, but that is not immediately relevant here). For example, in the gravitational case, the field equations for the first- and second-order Teukolsky variables take the form~\cite{Spiers:2023mor,Spiers:2023cip}
\begin{align}
{\cal O}\psi^{(1)}_{4} &= {\cal S}^{ab}\bigl(8\pi T^{(1)}_{ab}\bigr),\\
{\cal O}\psi^{(2)}_{4L} &= {\cal S}^{ab}\bigl(8\pi T^{(2)}_{ab}-\delta^2G_{ab}[h^{(1)},h^{(1)}]\bigr),
\end{align}
where ${\cal O}$ and ${\cal S}^{ab}$ are second-order linear differential operators, and $\delta^2G_{ab}$ is a second-order bilinear differential operator that acts on the first-order metric perturbation. These equations can be readily decomposed into Fourier modes and $m$ modes, reducing ${\cal O}$ to a 2D elliptic operator ${\cal O}_m$ that, after a conformal rescaling, takes the same generic form~\eqref{eq:operator_A} as the operator ${\boldsymbol A}$. The source can then be regularized using a puncture~\cite{Pound:2014xva,Spiers:2023cip}, just as we regularized the scalar field equation. Thus, an equation of the form~\eqref{eq:EqRes} is valid also at second order, with the right-hand side understood as an effective second-order source $\bar S^{(2)}_{\rm eff}$. 

The source $\bar S^{(2)}$ extends over the entire black-hole region, which can lead to numerical and theoretical challenges associated with slow falloff and infrared divergences at large distances~\cite{Miller:2023ers,Pound:2015wva}. A compactified hyperboloidal framework, combined with a spectral method, provides an ideal method of largely eliminating the numerical challenges~\cite{PanossoMacedo:2022fdi, PanossoMacedo:2023qzp}; and an appropriate choice of gauge (manifesting the physical spacetime's asymptotic flatness) largely sidesteps the theoretical challenges~\cite{Spiers:2023cip,Spiers:InPrep}.

However, as described in the Introduction, there is one major computational burden in current second-order self-force calculations: the calculation of the second-order source term in the vicinity of the particle. In this section, we study whether an $m$-mode scheme can reduce this burden.
For that purpose, we consider a toy quadratic source $\bar S^{(2)}[\bar\phi,\bar\phi]$, which will mimic $\delta^2G_{ab}[h^{(1)},h^{(1)}]$. In the effective-source region around the particle, within an $m$-mode scheme, we write it as
\beq
\bar S^{(2)}_m = \bar S^{(2), {\rm SS}}_{m} + \bar S^{(2), {\rm SR}}_{m} + \bar S^{(2), {\rm RR}}_{m},
\eeq
where $\bar S^{(2), {\rm SS}}_{m}:=\bar S^{(2)}_{m}[\bar\phi^{\mathcal{P}},\bar\phi^{\mathcal{P}}]$, $\bar S^{(2), {\rm SR}}_{m}:=\bar S^{(2)}_{m}[\bar\phi^{\mathcal{P}},\bar\phi^{\mathcal{R}}]+\bar S^{(2)}_{m}[\bar\phi^{\mathcal{R}},\bar\phi^{\mathcal{P}}]$, and $\bar S^{(2), {\rm RR}}_{m}:=\bar S^{(2)}_{m}[\bar\phi^{\mathcal{R}},\bar\phi^{\mathcal{R}}]$.
Each of these three terms is made up of a sum of products of its two arguments and their derivatives.

In current second-order calculations, the $\ell m$ modes of the ``singular-singular'' source, $\bar S^{(2), {\rm SS}}_{\ell m}$, would be calculated by performing numerical 2D integration of the 4D analytical expression $\bar S^{(2)}[\bar\phi^{\mathcal{P}},\bar\phi^{\mathcal{P}}]$ against a spherical harmonic. This represents the dominant computational expense in calculating the source. For example, in the gravitational case, calculating the modes of $\delta^2G_{ab}[h^{(1)},h^{(1)}]$ up to $\ell=10$ for a single value of the orbital radius $r_p$, on a radial grid around  $r_p$, takes approximately 15,000 CPU hours~\cite{Seff:InPrep}. As explained in the Introduction, we expect to be able to entirely sidestep this expense in an $m$-mode scheme because we expect that $\bar S^{(2), {\rm SS}}_{m}$ can be calculated analytically with relative ease~\cite{Dolan:2012jg,Thornburg:2016msc,Bourg2023}.

The singular-regular piece of the source represents a lower but still dramatic expense in current second-order calculations. The expense in this case arises from calculating the $\ell m$ modes of the first-order puncture, which are needed in order to solve the field equations for the $\ell m$ modes of the residual field. As described in Ref.~\cite{Bourg:2024vre}, the $\ell m$ modes of the puncture are calculated using a mix of analytical and numerical integration. Each integral is evaluated in $\lesssim 1\,$s~\cite{Seff:InPrep}, but the expense grows due to the large number of integrals required: several hundred $\ell m$ modes at several hundred radial grid points around the particle, amounting to $\sim 20$CPU hours for each value of $r_p$. Again, this cost is entirely avoided in an $m$-mode scheme because the integrals are easily evaluated analytically.

While an $m$-mode scheme is free of these particular costs, the gains must be measured against the cost of solving PDEs rather than ODEs. In the remainder of this section, we assess our scheme's capability to accurately and efficiently construct a quadratic second-order source. Since the singular-singular source term should be amenable to a purely analytical calculation, we focus on the singular-regular and regular-regular source terms.

\begin{figure*}[t!]
	\centering
    \includegraphics[width=\textwidth]{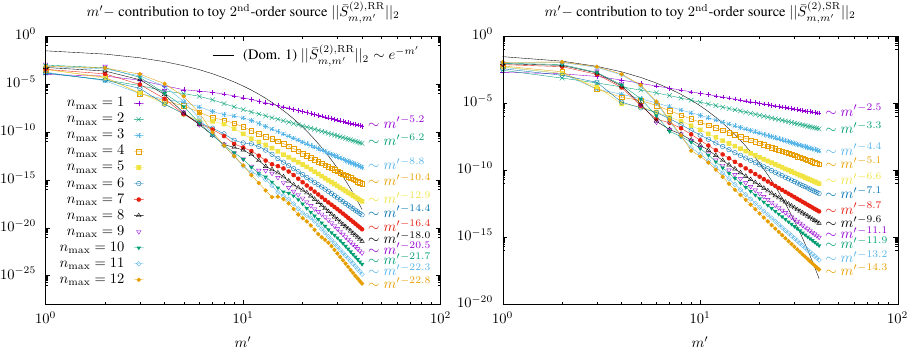}
	\caption{Contributions from individual $m'$-modes to the toy-model second-order source \eqref{eq:S_SR}--\eqref{eq:S_RR}, with the particle at $r_p = 10 M$.
 {\em Left Panel:} In domain $\D_1$, the regular-regular piece $\bar S^{(2), {\rm RR}}_{m,m'}$ is quadratic in the retarded field $\bar \phi_{m'}^{\rm ret}$, and the contribution decays exponentially with $m'$ (black line). In domain $\D_2$, $\bar S^{(2), {\rm RR}}_{m,m'}$ results from quadratic combinations of the residual field $\bar \phi_{m'}^{\rm ret}$, and the $m'$-modes contributions decay algebraically, with the decay rate improving when higher-order punctures are used. {\em Right Panel:} The singular-regular term $\bar S^{(2), {\rm SR}}_{m,m'}$ also decays algebraically with $m'$, with decay rate roughly half of its counterpart $\bar S^{(2), {\rm RR}}_{m,m'}$. 
}
	\label{fig:S_2nd_ToyModel}
\end{figure*}

To mimic the structure of $\delta^2G_{ab}$, we assume $\bar S$ has the form $t^{ab}_1\nabla_a\phi\nabla_b\phi+t^{ab}_2\phi\nabla_a\nabla_b\phi$ for some smooth tensors $t^{ab}_1$ and $t^{ab}_2$ that are independent of $\tau$ and $\phi$. Because our current implementation suffers from low accuracy in its calculation of $(\sigma,y)$-derivatives, we assume for simplicity that $t^{ab}_n$ only have $\tau$ and $\phi$ components; in the Conclusion we discuss how future implementations can calculate derivatives more accurately. We therefore assume contribution from derivatives of order $n$ have the form $\partial^n_a \bar \phi_m \sim m^n \phi$. These assumptions lead to the following  qualitative behaviour of the toy second-order source:
\beq
\bar S^{(2)}_m \sim \sum_{m'} m' \,  m'' \, \bar \phi_{m'} \bar \phi_{m''}, \quad m'' = m - m',
\eeq
where the condition $m''=m-m'$ arises from the requirement that the product of first-order modes, $\exp(m'\phi)\exp(m''\phi)$, yields the $m$ mode of $\bar S^{(2)}$,  $\exp(m\phi)$. 
We are specifically interested in how many $(m',m'')$ modes are required to calculate a given $m$ mode of the source. To assess this we can examine the limit of large $m'$ and $m''$ at fixed $m$, meaning $m''\approx -m'$. Since we are interested in the magnitude of contributions, we can also take the absolute value of the summand. Therefore, we consider
\beq
\bar S^{(2)}_m \sim \sum_{m'}\bar S^{(2)}_{m,m'}
\eeq
with 
\beq
\bar S^{(2)}_{m,m'}= m'^2|\bar\phi_{m'}\bar \phi_{-m'}| = m'^2|\bar\phi_{m'}|^2,
\eeq
where we have used $\bar \phi_{-m'}=(-1)^{m'}\bar\phi_{m'}^*$.

Outside the effective-source region, we will refer to $\bar S^{(2)}_m(\bar\phi^{\rm ret},\bar\phi^{\rm ret})$ as $\bar S^{(2),\rm RR}_m$. Thus, we introduce the following proxy function for the second-order source:
\bea
\label{eq:S_SR}
 \bar S^{(2), {\rm SR} }_{m,m'} &=& 2 m'^2 {\rm Re} \left( \phi_{m'}^{\mathcal{P}} \phi^{\mathcal{R}^*}_{m'} \right), \\
 \label{eq:S_RR}
 \bar S^{(2), {\rm RR} }_{m,m'} &=& \left\{
    \begin{array}{ccc} 
        m'^2 |\phi^{\rm ret}_{m'}|^2 & {\rm domain} &\D_1 \\
                                                                   \\
        m'^2 |\phi^{\mathcal{R}}_{m'}|^2 & {\rm domain}  &\D_2. \\
    \end{array}    
 \right.
\eea
Finally, we estimate the overall contribution of these functions to $\bar S_m$ by calculating their respective $L_2$ norms,
\beq
\label{eq:L2_2ndSource}
||\bar {\rm S}_{m,m'}||_2 = \sqrt{   \iint\limits_{\D_{\id}} || \bar {\rm S}_{m,m'}(\sigma,y) ||^2 d\sigma dy },
\eeq
with $\bar {\rm S}_{m,m'}$ either $\bar S^{(2), {\rm SR}}_{m,m'}$ or $\bar S^{(2), {\rm RR} }_{m,m'}$.
 
Figure~\ref{fig:S_2nd_ToyModel} displays the results for this measure of the contribution to the second-order source. As expected, the regular-regular term $\bar S^{(2), {\rm RR} }_{m,m'}$ decays exponentially with $m'$ in domain $\D_1$ as a consequence of the retarded field's smoothness in that region. This behaviour is displayed as a black line in both panels of Fig.~\ref{fig:S_2nd_ToyModel}, and it contrasts with the algebraic decay arising in domain $\D_2$ for the regular-regular $\bar S^{(2), {\rm RR} }_{m,m'}$ (left panel) and singular-regular terms $\bar S^{(2), {\rm SR} }_{m,m'}$ (right panel). In particular, Fig.~\ref{fig:S_2nd_ToyModel} shows the fits for the decay rates $\bar S^{(2), {\rm RR} }_{m,m'} \sim m'^{-\kappa^{\rm RR}}$ and $\bar S^{(2), {\rm SR} }_{m,m'} \sim m'^{-\kappa^{\rm SR}}$. 

It is evident that a higher-order puncture significantly enhances the corresponding decay rates. One observes the faster convergence rate for $\bar S^{(2), {\rm RR} }_{m,m'}$ when compared against $\bar S^{(2), {\rm SR} }_{m,m'}$, an expected result considering that the puncture field's modes converge slowly (as they must because the sum of the puncture's modes must reproduce the 4D field's $1/R$ divergence at the particle). As a quick rule-of-thumb, we can estimate $\kappa^{\rm RR} \sim 2 \kappa^{\rm SR}$ from the fact that $\kappa^{\rm SR}$ roughly corresponds to the convergence rate of the residual field, while $\kappa^{\rm RR}$ roughly corresponds to the convergence rate of the square of the residual field. The plots bear out this rough estimate. However, given the several assumptions underlying the construction of the second-order proxy function in Eqs.~\eqref{eq:S_SR} and \eqref{eq:S_RR}, the specific values should be considered only at a qualitative level.

Using our results for $||\bar S^{(2)}_{m,m'}||$, we can now  estimate two things: (i) an optimal choice of $m'_{\rm max}$, $n_{\rm max}$, and $N$ to achieve a given error tolerance $\epsilon\ll 1$ in our calculation of $\bar S^{(2)}_{m}$, and (ii) the timing of that calculation. We do this by setting $|| \left. \bar S^{(2), {\rm RR} }_{m,m'_{\rm max}}||_2\right|_{\D1}=\epsilon $. In this way, the calculation of the retarded-retarded source in domain ${\cal D}_1$ sets how many $m'$ modes are required to achieve a given accuracy goal. We then require the calculation of the source in ${\cal D}_2$ to match that accuracy: 
\beq
|| \bar S^{(2), {\rm SR} }_{m,m'_{\rm max}}||_2 = || \left. \bar S^{(2), {\rm RR} }_{m,m'_{\rm max}}||_2\right|_{\D1}, 
\eeq
(noting $|| \bar S^{(2), {\rm RR} }_{m,m'_{\rm max}}||_2$ in ${\cal D}_2$ can be neglected in our error measure because it will always be much smaller than $|| \bar S^{(2), {\rm SR} }_{m,m'_{\rm max}}||_2$). This allows us to read off the required $n_{\rm max}$ from the right panel of Fig.~\ref{fig:S_2nd_ToyModel}. Concretely, we read off which of the colored curves (corresponding to different values of $n_{\rm max}$) intersects the black curve (corresponding to $|| \bar S^{(2), {\rm RR} }_{m,m'_{\rm max}}||_2|_{\D1}$) at $m'=m'_{\rm max}$. One can then relate this error measure to the numerical resolution $N$ by requiring that the largest contribution (typically from the $m'=1$ mode) to the $m$-mode source is sufficiently accurate to meet the error threshold $\epsilon$.

\begin{figure}[t!]
	\centering	
	\includegraphics[width=\columnwidth]{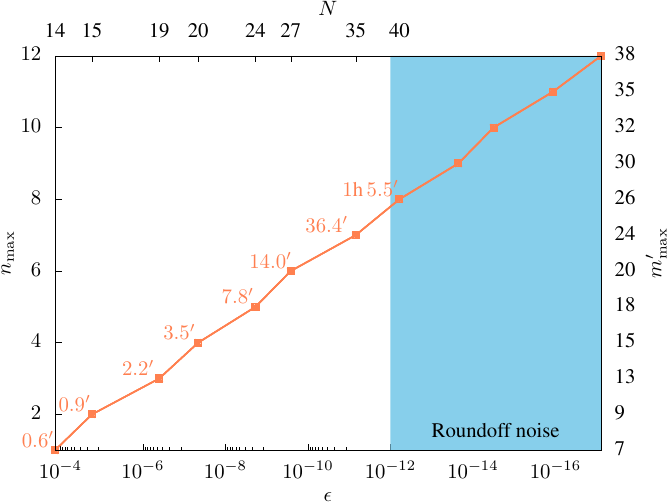}
	\caption{Numerical parameter space for particle at $r_p = 10 M$. Values along the orange line provide an estimate for the optimal puncture order $n_{\rm max}$, total number $m'_{\rm max}$ of modes, and the numerical resolution $N$ according to a given error tolerance $\epsilon$. These optimal parameter values assume fixed domains ${\cal D}_1$ and ${\cal D}_2$. Estimated
 CPU runtimes are displayed for each configuration (code executed on chip \texttt{Apple M2 Max}). 
}
	\label{fig:Rel_eps_nbar_max}
\end{figure}

Figure~\ref{fig:Rel_eps_nbar_max} summarizes this information, showing the optimal puncture order $n_{\rm max}$, the total number $m'_{\rm max}$ of first-order modes, and the numerical resolution $N$ required to achieve an  error tolerance $\epsilon$. Such a diagram allows us to restrict the numerical parameter space, and therefore optimise the computational cost of the code. 

Numbers along the orange curve in Fig.~\ref{fig:Rel_eps_nbar_max} indicate the total CPU time required to meet the target error tolerance. We observe that all of these timings are multiple orders of magnitude lower than the CPU time of current second-order $\ell m$-mode calculations.

However, there are several caveats to this analysis. First, our calibration is valid only for $r_p=10M$ and the most adequate parameter values may differ for other orbital radii. Performing an extended calibration is beyond the scope of this work, but the toy-model exercise discussed in this section may establish a standard for parameter optimisation when solving the gravitational self-force problem with the complete second-order source.

A second caveat is that we have fixed the size of the domains ${\cal D}_1$ and ${\cal D}_2$. It could be more efficient to enlarge the size of the effective source region for larger $n_{\rm max}$. By reducing the region in which we calculate the retarded field, this would reduce the number of $m'$ modes required.

A final caveat is that current $\ell m$-mode schemes could be made orders of magnitude more efficient if the $\ell m$-mode decomposition of the first-order puncture and second-order singular-singular source could be performed analytically. This would likely make $\ell m$-mode schemes at least as efficient as our $m$-mode scheme.

\section{Conclusion}

In this paper, we have successfully demonstrated a new, spectral approach to $m$-mode self-force calculations in the frequency domain. While our calculations are restricted to a scalar toy model, they provide an important proof of principle.

The punchline of our calculation is that, as hoped, we can compute the second-order source to high accuracy with only a small number of first-order $m$ modes. This conclusion applies to the ``singular times regular'' and ``regular times regular'' pieces of the second-order source. It remains to be shown that one can easily, analytically construct the $m$ modes of the ``singular times singular'' part of the source; work on that problem is underway. Assuming this ``singular times singular'' piece can be calculated analytically with relative ease, our estimates suggest that the total runtime for a second-order calculation could be reduced by several orders of magnitude (relative to current second-order self-force calculations) using an $m$-mode scheme.

We have also shown that there is significant gain in using a high-order puncture, though the gain depends on the specific quantity being calculated. In practice it appears that there are diminishing returns beyond order $n_{\rm max}\approx 6$, but we note that $n_{\rm max}=6$ is already four orders higher than the puncture used for current second-order self-force calculations.

In the future, we plan to extend our calculations to the gravitational case and a Kerr background. There are also several other avenues for future work. 

First, since $\ell$ modes are well adapted to the problem outside the near-particle shell, it would be desirable to find an efficient method of working with $\ell m$ modes in the external regions and $m$ modes within the shell. This will involve efficiently dealing with the coupling between $\ell$ modes that arises through the junction conditions at the shell boundaries, as described in Sec.~\ref{sec:NumericalMethods}. In the present paper, we have deferred that challenge by instead using a separate code to obtain the $\ell m$ modes in the external regions. 

A second avenue to explore would be a method in which the puncture is used throughout the particle's spherical shell, rather than dividing the shell into an effective-source region around the particle and a source-free region away from the particle. Using an effective source throughout the shell could be advantageous because, as shown in Fig.~\ref{fig:S_2nd_ToyModel}, the power-law convergence in the effective-source region can significantly outperform the exponential convergence in the source-free region.

The loss of of accuracy displayed in Figs.~\ref{fig:Error_doms} and \ref{fig:Error_nbarDep} is likely to jeopardise the metric reconstruction when expanding the code for the gravity case, and may impact also self-force calculation for generic orbit. Thus, a third avenue to follow would be to use a first-order reduction of the field equations. This would eliminate our loss when calculating first derivatives of the field. It would also reduce the loss of accuracy when calculating second derivatives, and one could use the higher-accuracy second $\sigma$-derivatives together with the field equations to obtain the second $y$-derivatives (which  suffer from the largest errors when calculated directly). 

\begin{acknowledgments}

We thank Leor Barack and Barry Wardell for helpful discussions. 
The authors acknowledge the use of the IRIDIS High Performance Computing Facility, and associated support services at the University of Southampton, in the completion of this work.
RPM acknowledges support from the Villum Investigator program supported by the VILLUM Foundation (grant no. VIL37766) and the DNRF Chair program (grant no. DNRF162) by the Danish National Research Foundation and the European Union's Horizon 2020 research and innovation programme under the Marie Sklodowska-Curie grant agreement No 101131233. RPM is a long-term research visitor at the STAG Research Centre, University of Southampton and this project has also received funding from the STFC Grant No.~ST/V000551/1.
PB acknowledges the support of an EPSRC Fellowship in Mathematical Sciences and support from the Dutch Research Council (NWO) (project name: Resonating with the new gravitational-wave era, project number: OCENW.M.21.119). 
AP acknowledges the support of a Royal Society University Research Fellowship, and AP and SDU acknowledge support from a Royal Society Research Grant for Research Fellows and from the ERC Consolidator/UKRI Frontier Research Grant GWModels (selected by the ERC and funded by UKRI grant number EP/Y008251/1). 
AP and PB additionally acknowledge the support of a Royal Society University Research Fellowship Enhancement Award. 
SDU additionally acknowledges the support of the fellowship Lumina Quaeruntur No.~LQ100032102 of the Czech Academy of Sciences. 

\end{acknowledgments}

\appendix

\section{Alternative coordinate mapping for effective-source region: Domain \texorpdfstring{$2$}{2}}\label{App:alternative_map_D2}

The coordinate map for the domain $\D_2$ in Sec.~\ref{sec:Map_D2} is related in a natural way to the particle's co-moving frame, but it leads to the complicated regularity conditions at the particle described in Sec.~\ref{sec:RegCond}. Here we derive an alternative map that simplifies the regularity conditions. 

As in Sec.~\ref{sec:Map_D2}, we start by selecting a circle of radius $\polarR_o$ in the polar representation for the coordinates centered at the particle, cf. Eq.~\eqref{eq:ParticlePolarCoordinates}. Combining Eqs.~\eqref{eq:CoordParticleFrame} and \eqref{eq:ParticlePolarCoordinates}, we obtain
\beq
\delta{r_o} = \polarR_o \, \cos \pfphi, \quad y_o = \dfrac{\polarR_o ^2 }{r_h ^2} \, \sin \pfphi^2.
\eeq
This surface delimits the boundary of domain $\D_2$, and it is directly parametrised by a coordinate $x^1\in[-1,1]$ as in Sec.~\ref{sec:Map_D2} via
\beq
\label{eq:x1_PolarParticle_v2}
x^1 = -\cos\pfphi.
\eeq
To cover the entire $\D_2$ with $\delta r \in [0, \delta_{r_o}]$, and $y\in[0,y_o]$ we introduce a second coordinate $x^2\in[-1,1]$ via
\beq
\label{eq:delta_r_dom2}
\delta {r} = \rho(x^2) \, \delta_{r_o}(x^1), \quad y = \rho(x^2)\, y_o(x^1),
\eeq
with $\rho(x^2)\in[0,1]$ given by
\beq
\label{eq:x2_PolarParticle}
\rho(x^2) = \dfrac{1+x^2}{2} \leftrightarrow x^2 = 2\rho -1.
\eeq
As a final step, we express Eq.~\eqref{eq:delta_r_dom2} in terms of the compact radial coordinate $\sigma$ from Eq.~\eqref{eq:HypCoord}. The result reads
\beq
\label{eq:map_dom2_v2}
\sigma = \dfrac{\sigma_p}{1 - \eta\, \sigma_p\, \sqrt{f_p}\, \rho(x^2) x^1}, \quad
y = \eta^2\, \sigma_p^2\, [1-(x^1)^2]\, \rho(x^2),
\eeq
with $\eta=\polarR_o/r_h$ the dimensionless parameter measuring the size of the ball $\polarR_o$ in units of the horizon radius $r_h$. 

With this choice, the quantity $\rho(x^2)$ contributes linearly to the mapping $y(x^1, x^2)$ in Eq.~\eqref{eq:map_dom2_v2}, which differs from the quadratic term in Eq.~\eqref{eq:map_dom2}. By construction, both maps agree at the surface $x^2=1$. Here, however, surfaces $x^2=$ constant for $x^2<1$ are no longer circles of constant radius in the polar representation for the coordinates centered at the particle.
\begin{figure*}[t!]
	\centering
    \includegraphics[width=\textwidth]{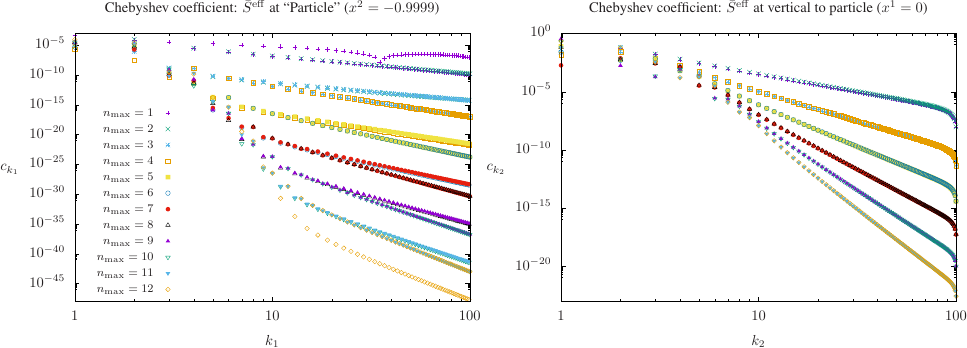}
	\caption{Coordinate map \eqref{eq:map_dom2_v2} introduces two different behaviors for the Chebyshev coefficients along $x^1=$~constant and $x^2=$~constant when compared to the map \eqref{eq:map_dom2}. {\em Left Panel:} At the surface closest to the particle $x^2\approx -1$, the function $\bar S^{\rm eff}(x^1, \left.x^2\right|_{cte.})$ is no longer a polynomal in $x^1$ and the coefficients $c_{k_1}$ decay algebraically with two branches $c_{k_1} \sim {k_1} ^{-\varkappa_{i}}$ (upper) and $c_{k_1} \sim k_1^{-\varkappa_{ii}}$ (lower). The lower branch at a given order $n_{\rm max}$ coincides with the upper branch of the next order $n_{\rm max}$. {\em Right panel:} Coordinate map \eqref{eq:map_dom2_v2} behaves as $y\sim \rho(x^2)$ as opposed to $y\sim \rho(x^2)^2$ in Eq.~\eqref{eq:map_dom2}. Thus, coefficients $c_{k_2}$ decay slower along surfaces $x^1=$~constant, with $x^1\neq \pm1$.
}
	\label{fig:Seff_regular_order_v2}
\end{figure*}

\subsection{Regularity conditions at the particle}
The advantage of this choice is that it simplifies the regularity conditions at the particle discussed in Sec.~\ref{sec:RegCond}. Indeed, given Eqs.~\eqref{eq:map_dsigma}--\eqref{eq:map_d2sigmay}, the mapping \eqref{eq:map_dom2_v2} imposes only one condition for the regularity of first ($\sigma,y$)-derivatives as $x^2\rightarrow -1$:
\beq
\label{eq:cond_first_der_v2}
\phi_{,1} \dot= 0 \rightarrow \phi_{,11} \dot= 0.
\eeq
Regularity of second ($\sigma,y$)-derivatives are simplified to 
\beq
\label{eq:cond_second_der_v2}
\phi_{,2} \dot= x^1 \phi_{,12}  - \dfrac{1+(x^1)^2}{2} \phi_{,112}.
\eeq
These conditions, then, guarantee that 
${\boldsymbol A} \bar\phi \sim {\cal O}(1)$ as $x^2 \rightarrow -1$. 

\begin{table}[b!]
	\begin{tabular}{c @{\quad} c c @{\quad} c} 
		\toprule
		 & \multicolumn{2}{c}{ $(\varkappa;  \kappa)\quad$} & $( \varkappa_{i}, \varkappa_{ii})$   \\
		    \cmidrule{2-4}
		$n_{\rm max}$ & $x^1=\pm1$ & $x^1=0$  & Closest to particle \\ 
		\midrule
		$1$ & (3;1) &(3;1) & (1,3)\\
		$2$ & (5;2)&(3;1) & (3,3)\\
		$3$ & (7;3)& (5;2) & (3,5)\\
		\midrule 
		$4$ & (9;4)& (5;2) & (5,5)\\
		$5$ & (11;5)&(7;3) & (5,7)\\
		$6$ & (13;6)& (7;3) & (7,7)\\
		\midrule
		$7$ & (15;7)&(9;4) & (7,9)\\
		$8$ & (17;8)& (9;4) & (9,9)\\
		$9$ & (19;9)& (11;5) & (9,11)\\
		\midrule
		$10$ & (21;10)& (11;5)& (11,11)\\
		$11$ & (23;11)& (13;6)& (11,13)\\
		$12$ & (25;12)& (13;6)& (13,13)\\
		\bottomrule
	\end{tabular}
	\caption{
 Equivalent of Table \ref{tab:reg_order_nbarmax} for coordinate mapping \eqref{eq:map_dom2_v2}. When evaluated closest to particle $x^2\approx -1$, the Chebyshev coefficients display two branches with rates $c_{k_1}\sim {k_1}^{-\varkappa_{i}}$ and $c_{k_1}\sim {k_1}^{-\varkappa_{ii}}$. Along $x^1=$~constant, with $x^1\neq \pm1$, coefficients $c_{k_2}$ decay slower with map \eqref{eq:map_dom2_v2}. The decay rate along the equator $x_1=\pm 1$ agrees in both maps.
	}
	\label{tab:reg_order_nbarmax_v2}
\end{table}

\subsection{Effective source's singular behavior}

The downside of working with this coordinate map is that it spoils the polynomial behavior for the puncture field and effective source along the surfaces $x^2 =$~constant. In particular, it introduces a singular behavior at the surface closest to the partice $x^2 \approx -1$, as shown in the left panel  of Fig.~\ref{fig:Seff_regular_order_v2} displaying the Chebyshev coefficients $c_{k_1}$ at $x^2=-0.9999$ for several puncture order $n_{\rm max}$. 

At every order $n_{\rm max}$, we observe a split of these coefficients into two branches, which we will refer to as the ``lower'' and the ``upper'' branch. Each branch has an overall tendency to decay algebraically with rates $c_{k_1} \sim {k_1} ^{-\varkappa_{i}}$ (upper) and $c_{k_1} \sim k_1^{-\varkappa_{ii}}$ (lower). We infer the coefficients $\varkappa_1$ and $\varkappa_2$ and display them in Table~\ref{tab:reg_order_nbarmax}. For odd puncture order $n_{\rm max}$, the coefficients behave as $\varkappa_1 = n_{\rm max}$ and $\varkappa_2 = n_{\rm max}+2$. For even order $n_{\rm max}$, the decay rates assume the same values $\varkappa_1=\varkappa_2 = n_{\rm max} +1$. Interestingly, the lower branch at a given order $n_{\rm max}$ coincides with the upper branch of the next order $n_{\rm max}$.

Another difference arises in the Chebyshev coefficients' decay rate along the surface $x^1=0$ (Fig.~\ref{fig:Seff_regular_order_v2}, right panel). Indeed, the algebraic decay for the coefficients $c_{k_2}$ are slower when using the mapping \eqref{eq:map_dom2_v2} due to the presence of the factor  $y\sim \rho(x^2)$, as opposed to the quadratic contribution $y\sim \rho(x^2)^2$ from Eq.~\eqref{eq:map_dom2}. Along the equator the decay rate coincides in the two cases because both maps coincides at $y=0$. Table~\ref{tab:reg_order_nbarmax_v2} summarises the decay rates for the coefficients $c_{k_2}$.

\section{Spectral coordinate mapping}\label{app:SpecMap}
Sec.~\ref{sec:reg_conditions} provides maps between the spectral coordinates $(x^1,x^2)\in[-1,1]^2$ and the hyperboloidal coordiantes $(\sigma,y)\in[0,1]^2$ for the different domains of interests. Given a generic map $\sigma(x^1, x^2)$ and $y(x^1,x^2)$, the relation between the derivative operators are
\bea
\label{eq:map_dsigma}
\partial_\sigma = {\cal J}^{-1}\left( y_{,2} \partial_1 - y_1 \partial_2 \,\right), \\
\label{eq:map_dy}
\partial_y = {\cal J}^{-1}\left( \sigma_{,2} \,\partial_1 - \sigma_1 \partial_2 \,\right),
\eea
for the first derivatives. Here, ${\cal J} = \sigma_{,1}\,y_{,2} - \sigma_{,2}\, y_{,1}$ is the Jacobian of the mapping. 
Second derivatives read
\begingroup
\allowdisplaybreaks
\begin{align}
\label{eq:map_d2sigma}
\partial^2_{\sigma \sigma} &= {\cal J}^{-2}\bigg[ y_{,2}^2\, \partial^2_{11}  + y_{,1}^2\, \partial^2_{22} - 2 y_{,1}y_{,2} \partial^2_{12} \nn\\*
&\qquad\  - \left( y_{,22} \, y_{,1}^2 - 2\, y_{,12}\, y_{,1} y_{,2} + y_{,11} \,y_{,2}^2\right) \partial_y  \nn\\*
&\qquad\ - \left( \sigma_{,22} \, y_{,1}^2 - 2\, \sigma_{,12}\, y_{,1} y_{,2} + \sigma_{,11} \,y_{,2}^2\right) \partial_\sigma
\bigg],  \\
\label{eq:map_d2y}
\partial^2_{y y} &= {\cal J}^{-2}\bigg[ \sigma_{,2}^2\, \partial^2_{11}  + \sigma_{,1}^2\, \partial^2_{22} - 2 \sigma_{,1}\sigma_{,2} \partial^2_{12}   \nn\\*
&\qquad\  - \left( y_{,22} \, \sigma_{,1}^2 - 2\, y_{,12}\, \sigma_{,1} \sigma_{,2} + y_{,11} \,\sigma_{,2}^2\right) \partial_y  \nn\\*
&\qquad\ - \left( \sigma_{,22} \, \sigma_{,1}^2 - 2\, \sigma_{,12}\, \sigma_{,1} \sigma_{,2} + \sigma_{,11} \,\sigma_{,2}^2\right) \partial_\sigma\bigg] \\
\label{eq:map_d2sigmay}
\partial^2_{\sigma y} &= - {\cal J}^{-2}\bigg\{ 
\sigma_{,2}y_{,2}\, \partial^2_{11} + \sigma_{,1}y_{,1}\, \partial^2_{22} \nn\\*
&\qquad\quad\   - ( \sigma_{,1}y_{,2} + y_{,1}\sigma_{,2} ) \partial^2_{12}  \nn\\*
&\qquad\quad\  - \bigl[ y_{,22} \, \sigma_{,1}y_{,1} -  y_{,12}\, \left( \sigma_{,1} y_{,2} + y_{,1} \sigma_{,2}  \right) \nn\\*
&\qquad\qquad\ + y_{,11} \,\sigma_{,2}y_{,2}\bigr] \partial_y  \nn\\
&\qquad\quad\ - \bigl[ \sigma_{,22} \, \sigma_{,1}y_{,1} - \sigma_{,12}\,\left( \sigma_{,1} y_{,2} + y_{,1} \sigma_{,2}  \right) \nn\\*
&\qquad\qquad\ + \sigma_{,11} \,\sigma_{,2}y_{,2}\bigr] \partial_\sigma
\bigg\}.
\end{align}%
\endgroup
\section{Spectral differential matrices}\label{sec:SpecMatrix}

As stated in Eq.~\eqref{eq:Disc_Derv}, differentiation of functions at the grid points follow by applying specific differentiation matrices on the functions' discrete representation~\eqref{eq:discrete_functions}~\cite{trefethen2000spectral,Boyd,canuto2007spectral,GraNov07,Ansorg2013}. For the particular grids discussed in Sec.~\eqref{sec:Chebyshev grids}, the derivative operator for the Chebyshev-Lobatto grid reads
\beq
    {D}^{x}_{i,j} = 
    \displaystyle
\begin{cases}
\displaystyle
        \frac{\kappa_i (-1)^{i-j}}{\kappa_j (x_i-x_j)} & i \neq j\\[1.25em]
\displaystyle
        -\frac{x_j}{2\left(1-\left(x_j\right)^2\right)} &0< i = j < N\\[1.75em]
\displaystyle
        \frac{2 N^2 + 1}{6} & i = j = 0\\[1.25em]
\displaystyle
        -\frac{2 N^2 + 1}{6} & i = j = N
    \end{cases},
    \label{eq:D_Lobatto}
\eeq
with
\beq
    \kappa_i =\begin{cases}
    2, & i=0,N \\
    1 & i \neq 0,N
    \end{cases}.
\eeq
For the Chebyshev-Radau grid, the differentiation matrix reads
\beq
    {D}^{x}_{i,j} = 
\begin{cases}
\displaystyle
        \dfrac{N(N + 1)}{3} & i = j = 0\\[1.25em]
\displaystyle
        (-1)^j \dfrac{\sqrt{2(1+x_j)}}{1-x_j} & i=0,\, j \neq 0\\[1.25em]
\displaystyle
        \dfrac{(-1)^{i+1}}{\sqrt{2(1+x_i)}(1-x_i)} & i\neq0, \,  j = 0\\[1.25em]
\displaystyle
       -\dfrac{1}{2\left(1-\left(x_i\right)^2\right)}& i = j \neq 0
\\[1.75em]
\displaystyle
	\dfrac{(-1)^{i-j}}{x_i-x_j} \sqrt{\dfrac{1+x_j}{1+x_i}} &  0 \neq i \neq j \neq 0  
    \end{cases}.
    \label{eq:D_radau} 
\eeq

\section{Linear Algebraic System}\label{sec:LinAlgSystem}
\begin{table*}[th!]
    \renewcommand{\arraystretch}{1.66}
    \setlength{\tabcolsep}{0pt}
	\begin{tabularx}{.75\textwidth}{
        |>{\centering\arraybackslash\hsize=.75\hsize}X
        *4{|>{\centering\arraybackslash\hsize=.75\hsize}X}
        *2{|>{\centering\arraybackslash\hsize=\hsize}X}
        |>{\centering\arraybackslash\hsize=1.25\hsize}X|
        } 
		\hline
		 	&   \multicolumn{2}{c|}{   $i_1 = 0$} & \multicolumn{2}{c|}{  $i_1 = N^{\id}_1$}   & $i_2 = 0$  &	$i_2 = N^{\id}_2$      &  	$i_1 \in (0,N^{\id}_1)$ \\
		\hline 
		\hline
		\cellcolor{gray!25} $\D_0$	& \multicolumn{4}{c|}{  \cellcolor{gray!25} {$i_2 \in [0,N_2^{\id}]$}} &   \multicolumn{2}{c|}{  \cellcolor{gray!25} {$i_1 \in (0,N_1^{\id})$}} &  \cellcolor{gray!25} $i_2 \in (0,N_2^{\id})$  \\
		\hline
		\cellcolor{gray!25} $\id = 0$ & \multicolumn{2}{c|}{\cellcolor{gray!25} Eq.~\eqref{eq:TransitionCondition_D0_D1_field} } &  \multicolumn{2}{c|}{ \cellcolor{gray!25} Eq.~\eqref{eq:EqHomo}  } & \cellcolor{gray!25} Eq.~\eqref{eq:EqHomo} & \cellcolor{gray!25} Eq.~\eqref{eq:EqHomo} & \cellcolor{gray!25} Eq.~\eqref{eq:EqHomo} \\
		\hline
		\hline
		$\D_1$	               & \multicolumn{4}{c|}{  {$i_2 \in [0,N_2^{\id}]$}} &   \multicolumn{2}{c|}{  {$i_1 \in (0,N_1^{\id})$}} & {$i_2 \in (0,N_2^{\id})$}  \\
		\hline
		$\id =1$ & \cellcolor{gray!25} Eq.~\eqref{eq:TransitionCondition_D1_D3_field}  &  Eq.~\eqref{eq:ExtData_lm_mode_plus} & \cellcolor{gray!25} Eq.~\eqref{eq:TransitionCondition_D0_D1_der} & Eq.~\eqref{eq:ExtData_lm_mode_minus} & Eq.~\eqref{eq:EqHomo}  & Eq.~\eqref{eq:TransitionCondition_D1_D2_field}   & Eq.~\eqref{eq:EqHomo}  \\
		\hline
		\hline
		$\D_2$	               & \multicolumn{4}{c|}{  {$i_2 \in (0,N_2^{\id})$}} &   \multicolumn{2}{c|}{  {$i_1 \in [0,N_1^{\id}]$}} & {$i_2 \in (0,N_2^{\id})$}  \\
		\hline
		$\id =2$ & \multicolumn{2}{c|}{ Eq.~\eqref{eq:EqRes} }& \multicolumn{2}{c|}{ Eq.~\eqref{eq:EqRes}} & Eq.~\eqref{eq:TransitionCondition_D1_D2_der} & Eq.~\eqref{eq:EqRes}    & Eq.~\eqref{eq:EqRes}\\
		\hline
		\hline
		\cellcolor{gray!25} $\D_3$	& \multicolumn{4}{c|}{  \cellcolor{gray!25} {$i_2 \in [0,N_2^{\id}]$}} &   \multicolumn{2}{c|}{  \cellcolor{gray!25} {$i_1 \in (0,N_1^{\id})$}} &  \cellcolor{gray!25} $i_2 \in (0,N_2^{\id})$  \\
		\hline
		 \cellcolor{gray!25}$\id =3$ &  \multicolumn{2}{c|}{ \cellcolor{gray!25} Eq.~\eqref{eq:EqHomo} } &  \multicolumn{2}{c|}{ \cellcolor{gray!25} Eq.~\eqref{eq:TransitionCondition_D1_D3_der} } & {\cellcolor{gray!25}  Eq.~\eqref{eq:EqHomo} }& {\cellcolor{gray!25} Eq.~\eqref{eq:EqHomo}} & {\cellcolor{gray!25} Eq.~\eqref{eq:EqHomo}}\\
		\hline

	\end{tabularx}
	\caption{Discrete implementation of the field equations and transition/boundary conditions yielding the linear system of algebraic equations $\vec F(\vec X) = 0$. Entries shaded in gray are used when considering an $n_d=4$ domain code, i.e., including the regions extending to future null infinity ($\D_0$) and to the black hole horizon ($\D_3$), together with the intermediate source-free region ($\D_1$) and effective-source region ($\D_2$). If external data is used in $\D_0$ and $\D_3$, then only the unshaded cells are needed.}
	\label{tab:F_of_X}
\end{table*}

\subsection{Self-force \texorpdfstring{$m$}{m}-mode spectral solver}
As discussed in Sec.~\ref{sec:coord_domain_decomposition}, the physical space defined in the hyperboloidal coordinates $\{\sigma,y\}=[0,1]^2$ is decomposed into a total of $n_{\rm d}=4$ domains. In this work, however, we restrict to $n_{\rm d}=2$ as explained in Secs.~\ref{sec:ext_conditions} and \ref{sec:SpecSolver}. In each domain, we discretise a total of $n_{\rm f}=2$ fields, representing the real and imaginary parts of the retarded (or residual) field.

\subsubsection{Discrete functions}
The generic notation introduced in Eq.~\eqref{eq:discrete_functions} to describe the discrete values of a function serves multiple purposes. First, it applies for the unknown retarded, $\bar \phi^{\id = 1}_m(\sigma,y)$,  and residual fields, $\bar \phi^{\id = 2}_m(\sigma,y)$, that we wish to determine numerically. Specifically, given the complex character of these fields, we define
\beq
\label{eq:complex_vec_phi}
\bar \phi^{\id}_{i_1, i_2} := f^{(\iF=0, \id)}_{i_1, i_2} + {\rm i} f^{(\iF=1, \id)}_{i_1, i_2},
\eeq
where we have explicitly used the fact that $n_{\rm f}=2$ for the scenario considered in this work.

Second, the notation easily adapts to functions of $(x^1,x^2)$ that are known a priori, such as the coordinate mappings $\sigma^{\id}(x^1,x^2)$ and $y^{\id}(x^1,x^2)$ from Sec.~\ref{sec:coord_domain_decomposition}. In this case, however, the index $\iF$ is irrelevant. Hence, the physical grid points reads
\beq
\sigma^{\id}_{i_1, i_2} := \sigma^{\id}(x^1_{i_1}, x^2_{i_2}), \quad y^{\id}_{i_1, i_2} := y^{\id}(x^1_{i_1}, x^2_{i_2}).
\eeq
Finally, it applies also to functions on $(\sigma,y)$, such as the coefficients for the differential operator in Eqs.~\eqref{eq:alpha_2}--\eqref{eq:Gamma_1}
\beq
\begin{array}{c}
\alpha_2{}^{\id}_{i_1,i_2}:=\alpha_2(\sigma^{\id}_{i_1, i_2}), \\
\\ 
\alpha_1{}^{\id}_{i_1,i_2}:=\alpha_1(\sigma^{\id}_{i_1, i_2}), \\ 
\\
\alpha_0{}^{\id}_{i_1,i_2}:=\alpha_2(\sigma^{\id}_{i_1, i_2}), \\
\\
\Gamma_2{}^{\id}_{i_1,i_2}:=\Gamma_1(y^{\id}_{i_1, i_2}), \\ 
\\
\Gamma_1{}^{\id}_{i_1,i_2}:=\Gamma_2(y^{\id}_{i_1, i_2}).
\end{array}
\eeq
Similarly, the discrete values for the puncture field and effective source read
\bea
\bar \phi^{\mathcal{P}}_{i_1,i_2} &:=& \bar \phi^{\mathcal{P}}\left( \sigma^{\id=2}_{i_1, i_2},y^{\id=2}_{i_1, i_2} \right),  \\
\nn \\
\bar S^{\rm eff}_{i_1,i_2} &:=& \bar S^{\rm eff}\left( \sigma^{\id=2}_{i_1, i_2},y^{\id=2}_{i_1, i_2} \right).
\eea
In this case, the index $\id$ is irrelevant as the puncture and effective source are defined only in domain $\D_{2}$.

With the above notation, the discrete representations of differential equations \eqref{eq:EqRes} and \eqref{eq:EqHomo} read
\beq
\label{eq:Disc_diff_eq}
(A\phi)^{\id}_{i_1,i_2} = \left\{
\begin{array}{cc}
0  & (\id = 0,1,3) \\
\\
\bar S^{\rm eff}_{i_1,i_2} & (\id = 2) \\
\end{array}
\right. ,
\eeq
with
\begin{align}
(A\phi)^{\id}_{i_1,i_2} &= \alpha_2{}^{\id}_{i_1,i_2} \, \left(\phi_{,\sigma \sigma}\right)^{\id}_{i_1, i_2} + \alpha_1{}^{\id}_{i_1,i_2} \, \left(\phi_{, \sigma}\right)^{\id}_{i_1, i_2} \nn \\
&\quad + \alpha_0{}^{\id}_{i_1,i_2} \, \phi^{\id}_{i_1, i_2} + \Gamma_2{}^{\id}_{i_1,i_2}\, \left(\phi_{, yy}\right)^{\id}_{i_1, i_2} \nn\\
&\quad + \Gamma_1{}^{\id}_{i_1,i_2}\, \left(\phi_{,y}\right)^{\id}_{i_1, i_2}.
\end{align}
Similarly, discrete representations for the transition and boundary conditions \eqref{eq:TransitionCondition_D0_D1_field}--\eqref{eq:ExtData_lm_mode_plus} follow directly in this notation.

\subsubsection{Algebraic system of linear equations \texorpdfstring{$\vec F(\vec X)$}{F(X)}}
The algebraic system of linear equations $\vec F(\vec X)=0$ follows by imposing the differential equations or boundary conditions at all grid points. These equations, however, should be understood in the form
\beq
{\cal O} \bar \phi - {\cal S} = 0.
\eeq
In this generic notation, the operator ${\cal O}$ represents either the second-order differential operator $\boldsymbol{A}$ from Eq.~\eqref{eq:operator_A} or the first-order derivative operators on the left-hand side of the boundary conditions \eqref{eq:TransitionCondition_D0_D1_field}--\eqref{eq:TransitionCondition_D1_D2_der} and \eqref{eq:ExtData_lm_mode_minus}--\eqref{eq:ExtData_lm_mode_plus}. The source ${\cal S}$ contains the right-hand side of Eqs.~\eqref{eq:EqRes}, \eqref{eq:EqHomo}, \eqref{eq:TransitionCondition_D0_D1_field}--\eqref{eq:TransitionCondition_D1_D2_der}, and \eqref{eq:ExtData_lm_mode_minus}--\eqref{eq:ExtData_lm_mode_plus}. Table \ref{tab:F_of_X} displays the employed equation according to the specific grid point, i.e., fixed by the domain index $\id$ and spectral grid $(i_1, i_2)$. 

In particular, the table displays two possible options for the total number of domains. The first option, given by table entries in white, is a two-domain code ($n_{\rm d} = 2$). It follows the discussion in Sec.~\eqref{sec:ext_conditions} and it excludes the domains $\D_0$ (the outer source-free region) and $\D_3$ (the inner source-free region). In this case, one must impose the external conditions \eqref{eq:ExtData_lm_mode_minus} and \eqref{eq:ExtData_lm_mode_plus} at domain $\D_1$'s boundaries, i.e., at the surfaces with label $i_1=0$ and $i_1 = N_1^{\id=1}$. The second option in the table considers all the $n_{\rm d}=4$ domains, and the conditions on the boundaries of $\D_1$ are replaced by the transitions into $\D_0$ and $\D_3$ via Eqs.~\eqref{eq:TransitionCondition_D0_D1_der} and \eqref{eq:TransitionCondition_D1_D3_field}. The table entries for these extra equations are shaded gray. 

All table entries amount to the components of a complex-valued vector ${\cal F}^{\id}_{i_1,i_2}$. The real-valued components of the vector $\vec F(\vec X)$ then follow from the complex values ${\cal F}^{\id}_{i_1,i_2}$ via
\bea
F^{(\iF,\id)}_{i_1, i_2} = \left\{
\begin{array}{cc}
{\rm Re}\left( {\cal F}^{\id}_{i_1,i_2} \right) & {\rm for} \quad \iF = 0 \\
\\
{\rm Im}\left( {\cal F}^{\id}_{i_1,i_2} \right) & {\rm for} \quad \iF = 1
\end{array}
\right. .
\eea

\section{Spectral approximation, error measures and convergence rates}\label{sec:SpecApprox_error_conv}
\subsection{\texorpdfstring{$1D$}{1D} and \texorpdfstring{$2D$}{2D} Chebyshev spectral approximation}
In Sec.~\ref{sec:NumericalMethods}, Eq.~\eqref{eq:f_approximation} introduced the approximation in terms of Chebyshev polynomials for a  real-valued function\footnote{To ease the notation, we omit the indices $\id$ and $\iF$ specifying the domain and individual field under consideration.} $f(x^1,x^2)$ defined in the domain $\{x^1,x^2\}\in[-1,1]^2$, which we reproduce here for convenience:
\beq
\label{eq:f_approximation_app}
f_{N_1, N_2}(x^1,x^2) = \sum_{k_1=0}^{N_1} \sum_{k_2=0}^{N_2} c_{k_1, k_2} T_{k_1}(x^1) T_{k_2}(x^2).
\eeq
Along surfaces $x^1=$~constant, or $x^2=$~constant, one obtains a similar representation for the one-dimensional functions $f(x^1,\left.x^2\right|_{\rm const})$ and $f(\left.x^1\right|_{\rm const}, x^2)$, respectively,
\beq
\label{eq:f_approximation_x1}
f_{N_1}(x^1) = \sum_{k_1=0}^{N_1}  \tilde c_{k_1} T_{k_1}(x^1), \quad f_{N_2}(x^2) = \sum_{k_2=0}^{N_2}  \tilde c_{k_2} T_{k_2}(x^2).
\eeq
Strictly speaking, the Chebyshev coefficients $\tilde c_{k_1}$, $\tilde c_{k_2}$ in the above expressions do not have the same values as $c_{k_1}$ and $c_{k_2}$ in Eq.~\eqref{eq:f_approximation_app}. A direct comparison between Eqs.~\eqref{eq:f_approximation_x1} and \eqref{eq:f_approximation_app} reveals (for $i\neq j = 1,2$)
\beq
\tilde c_{k_i} =   \sum_{k_j=0}^{N_j} c_{k_i,k_j} T_{k_j}(\left. x^i \right|_{\rm const}).
\eeq
In our studies, however, we are mainly interested in their asymptotic behavior, which allows us to infer the function's regularity class. Since $\tilde c_{k_i} \propto  c_{k_i}$, we abuse the notation in the main text and do not introduce the tilde notation to improve readability.

\subsection{Error Measures}

Assuming that  the numerical resolutions scale as $N_1 = N_2 = N$, we employ the infinity norm as the error measure for our solutions. For a given reference solution with high resolution $N_{\rm ref}$, the absolute error within a given domain $\D_{\id}$ reads
\beq
\label{eq:error_infintynorm}
{\cal E}(N) =  || f_{N} - f_{N_{\rm ref}}  ||_{\infty}, \quad || f ||_{\infty} = \max_{\D_{\id}} |f|.
\eeq
This quantities provides the highest error of the numerical solution within the domain $\D_{\id}$, and it helps infer the regularity class of the underlying solution. We have used $N_{\rm ref}=100$ throughout the work, see sec.~\ref{sec:NumericalSolutions}.

\subsection{Convergence}\label{sec:app_convergence}
Spectral methods are known to provide a fast convergence rate for the approximated solution towards the exact results as the numerical resolution $N$ increases~\cite{trefethen2000spectral,Boyd,canuto2007spectral,GraNov07}. This property is captured by the behavior of the Chebyshev coefficients $c_k$ or an error measure ${\cal E}(N)$ via
\beq
c_k \sim {\cal C}^{-k} \Rightarrow {\cal E}(N) \sim \bar{\cal C}^{-N},
\eeq
with some constants ${\cal C}$ and $\bar{\cal C}$. This result, however, is only valid when the underlying function is analytic in its domain of validity. Specifically, if the function is a polynomial of order $P$, then $c_k =0$ for $k>P$.

For singular functions belonging to a regularity class $C^{\kappa-1}$, the convergence rate is merely algebraic, i.e.
\beq
c_k \sim k^{- \varkappa} \Rightarrow {\cal E}(N) \sim N^{- \tilde \varkappa}.
\eeq
With $\varkappa$ and $\tilde \varkappa=1+ \varkappa$ integers associated with $\kappa$. For instance, for functions with logarithmic singularities of the form $\sim \varrho^\kappa \log \varrho$ (as $\varrho \rightarrow 0$), one obtains $\varkappa = 2\kappa+1$ and $\tilde \varkappa = 2 \kappa$. Despite the loss of efficiency in the numerical solver, this behavior is useful to infer the regularity properties of the solution.

\section{Legendre and Chebyshev spectral representations}\label{app:LegChebPoly}
\begin{figure*}[tb!]
	\centering
 \includegraphics[width=\textwidth]{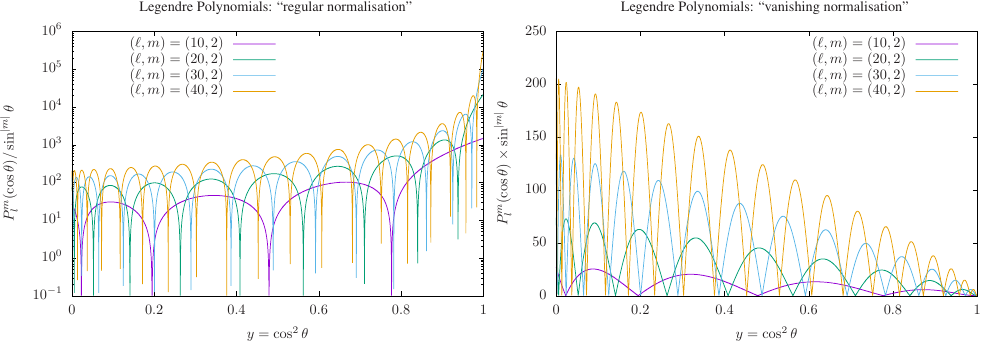}
	\caption{Adapting the associated Legendre functions to best represent them in the Chebyshev basis. {\em Left Panel:} In the so-called ``regular normalisation'' \eqref{eq:reg_norm_Plm}, the functions $\tilde P_{\ell}^m(y)$ assume regular, nonvanishing values as $y\rightarrow 1$. However, one observes steep gradients around $x =1$.  {\em Right Panel:} For the so-called  ``vanishing normalisation'' \eqref{eq:van_norm_Plm}, the functions $\bar P_{\ell}^m(y)$ vanish at $x=1$, amending the steep gradients there.}
	\label{fig:LegPolyn_Normalisation}
\end{figure*}
\begin{figure*}[tb!]
	\centering
    \includegraphics[width=\textwidth]{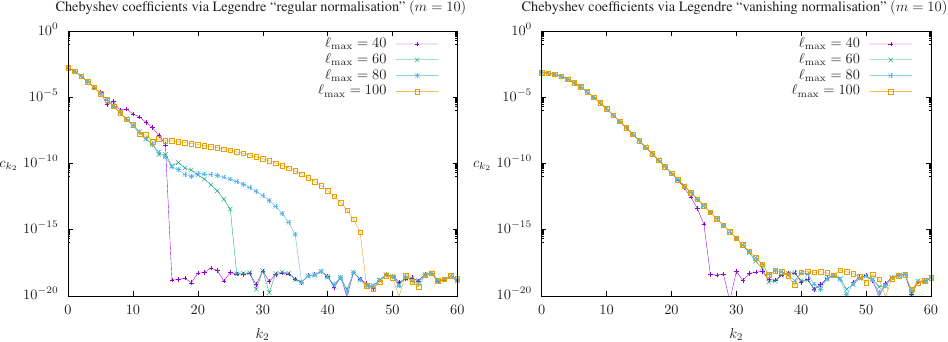}
	\caption{
	Chebyshev coefficients associated with the retarded field yielding the world tube boundary data at the interface between domains $\D_1-\D_3$. {\em Left Panel}: When using the ``regular normalisation'' to decompose the Legendre polynomials into the Chebyshev basis, the corresponding Chebyshev coefficients do not show a systematic behavior as one increases the truncation parameter $\ell_{\rm max}$. Thus, higher Chebyshev nodes are required to properly represent the boundary data. {\em Right Panel:} With the ``vanishing normalisation'', the Chebyshev coefficients decay exponentially, quickly approaching the numerical saturation limit.
	}
	\label{fig:ChebCoeff_LegNormalisation}
\end{figure*}

In this appendix, we discuss further results when relating  functions decomposed into $(\ell,m)$-modes via Legendre functions and the Chebyshev representation within the $m$-mode decomposition. We begin by pointing out that, in terms of the variable $x=\cos\theta$,  the function $\sin\theta = \sqrt{1-x^2}$ is poorly represented in terms of the Chebyshev polynomials via an expansion as in Eq.~\eqref{eq:f_approximation_x1}. Indeed, the derivatives of the function $f(x) = \sqrt{1-x^2}$ are singular at the boundary points $x=\pm 1$, spoiling the exponential convergence of the spectral decomposition. The associated Legendre functions have the structure 
\beq
\label{eq:LegPoly_decomp}
P_\ell^m(x) = \left(1-x^2\right)^{m/2} {\cal P}_{\ell-m}(x),
\eeq 
with ${\cal P}_{\ell-m}(x)$ a polynomial of order $\ell-m$ in $x$ and parity $(-1)^{\ell-m}$. Thus, a careful treatment of the term $\left(1-x^2\right)^{m/2}$ is needed for representing the function in terms of the Chebyshev basis. The delicate term $\sin^m\theta = \left(1-x^2\right)^{m/2}$ accounts for the regularity conditions at the symmetry axis $x=\pm 1$, cf. Eq.~\eqref{eq:Axis_RegCond}. 

In the scenarios considered in this work, only terms with $\ell+m=$ even contribute to the retarded and residual fields in the $(\ell,m)$-mode decomposition. With this property, the polynomials ${\cal P}_{\ell-m}(x)$ in Eq.~\eqref{eq:LegPoly_decomp} are even functions in $x$, and this result is a direct manifestation of the system's equatorial symmetry. Recall that we exploit this symmetry by introducing the variable $y = x^2$.

Thus, there are two options for adapting \eqref{eq:LegPoly_decomp} to be used in the Chebyshev representation. The first strategy is to work directly with the polynomial ${\cal P}_{\ell-m}(x)$, by defining
\beq
\label{eq:reg_norm_Plm}
\tilde P_{\ell}^m(y) = \dfrac{P_\ell^m(\sqrt{y})}{\left(1-y\right)^{m/2}}.
\eeq
We refer to this option as the the ``regular normalisation'', since $\tilde P_{\ell}^m(y)$ assumes regular, non-vanshing values as $y\rightarrow 1$. Despite being a rather intuitive approach, this option is not ideal. Apart from being highly oscillatory for $\ell\gg m$, the polynomials $\tilde P_{\ell}^m(y)$ quickly develop very steep gradients around $y=1$. The left panel of Fig.~\ref{fig:LegPolyn_Normalisation} shows examples for $m=2$ and $\ell = 10, 20, 30$ and $40$, and one observes the function increases by several orders of magnitude for $y \gtrsim 0.95$.

An alternative strategy is to damp the steep gradients by imposing that the functions vanish at $y=1$ via
\beq
\label{eq:van_norm_Plm}
\bar P_{\ell}^m(y) = P_\ell^m(\sqrt{y}) \, \left(1-y\right)^{m/2}.
\eeq
We refer to this strategy as the ``vanishing normalisation'', and it is the one employed in the definition of the hyperboloidal re-scaling function \eqref{eq:phi_hyperboloidal_rescaling}. Examples of this function are presented in the right panel of Fig.~\ref{fig:LegPolyn_Normalisation} for the same set of parameters as before. It is evident that, despite the oscillations for $\ell\gg m$, the amplitudes do not show steep gradients.

The steep gradient in the ``regular normalisation'' strategy becomes even more pronounced for higher values of $m$, and its effect directly impacts the Chebyshev representation of the boundary data \eqref{eq:ExtData_lm_mode_minus} and \eqref{eq:ExtData_lm_mode_plus}. To observe the effect of these steep functions, we study the Chebyshev representation of the worldtube boundary data constructed within the $\ell m$-mode scheme via \eqref{eq:ExtData_lm_mode_minus} and \eqref{eq:ExtData_lm_mode_plus}. In practice, the upper limit of the sum in Eqs. \eqref{eq:ExtData_lm_mode_minus} and \eqref{eq:ExtData_lm_mode_plus} is truncated at a given value $\ell_{\rm max}$, and the corresponding boundary data becomes a polynomial in the angular coordinate $y$. 

The left panel of Fig.~\ref{fig:ChebCoeff_LegNormalisation} displays the Chebyshev coefficients $c_{k_2}$ for the boundary data $\bar \phi_-(y)$ with $m=10$, constructed according to the ``regular normalisation'' strategy for different choices of truncation parameter $\ell_{\rm max}$. As expected, the coefficients $c_{k_2}$ drop to zero (numerical noise at $\sim 10^{-20}$) after a given $k_{\rm max}$ as a consequence of the $\bar \phi_-(y)$ being a polynomial. However, we do not observe a systematic convergence for the Chebyshev coefficients associated with the function $\bar \phi_-(y)$ as the parameter $\ell_{\rm max}$ increases. Thus, the spectral resolution $N_2$ to accurately represent the boundary data increases significantly with $\ell_{\rm max}$. The right panel of Fig.~\ref{fig:ChebCoeff_LegNormalisation} shows the equivalent results when we employ the ``vanishing normalisation''. In this case, the Chebyshev coefficients decay exponentially, quickly approaching the numerical saturation limit, and with self-consistent values as $\ell_{\rm max}$ increases.
\newpage
\bibliography{bibitems}
\end{document}